\documentclass[a4paper,11pt,oneside]{article}

\usepackage[a4paper,left=2cm,right=2cm,top=2.5cm,bottom=3cm]{geometry}
\usepackage{amssymb, mathrsfs, amsthm}
\usepackage[centertags]{amsmath}
\usepackage{graphicx}
\usepackage{appendix}
\usepackage{subcaption}
\usepackage{comment}
\usepackage{booktabs}
\usepackage{multicol}
\usepackage{xcolor}
\usepackage{csquotes}
\usepackage{multirow}
\usepackage{mathtools}
\usepackage{bbm}
\usepackage{bm}
\usepackage{mwe}
\usepackage[noblocks,affil-sl]{authblk}
\usepackage{caption}
\usepackage{hhline}
\usepackage{epstopdf}
\usepackage{adjustbox}
\usepackage{dsfont}
\usepackage{colortbl}
\usepackage[authoryear]{natbib}
\usepackage{chngcntr}
\usepackage{pdflscape}
\usepackage{cases}
\usepackage{cancel}
\usepackage{float}
\usepackage{endnotes}
\usepackage{hyperref}
\usepackage[most]{tcolorbox}
\usepackage{enumitem}

\numberwithin{equation}{section}
\counterwithin{figure}{section}
\counterwithin{table}{section}

\theoremstyle{plain}
\newtheorem{thm}{Theorem}[section]
\newtheorem{lem}[thm]{Lemma}
\newtheorem{prop}[thm]{Proposition}
\newtheorem{cor}[thm]{Corollary}

\newtheorem{remark}[thm]{Remark}

\renewcommand{\P}{\mathbb{P}}

\DeclareMathOperator*{\argmax}{arg\,max}

\newcommand*\diff{\mathop{}\!\mathrm{d}}

\numberwithin{equation}{section}

\definecolor{OliveGreen}{rgb}{0,0.6,0}

\begin{document}

\author[1]{An Chen\thanks{an.chen@uni-ulm.de}}
\author[2]{Christian Laudagé\thanks{christian.laudage@rptu.de}}
\author[3]{Shihao Zhu\thanks{shihao.zhu@uni-ulm.de}}

\affil[1,3]{Institute of Insurance Science, Ulm University, Germany}
\affil[2]{Department of Mathematics, RPTU University Kaiserslautern-Landau, Germany} 

\title{Optimal collective investment under regulatory uncertainty}
\date{\today}
\maketitle

\begin{abstract}

We study optimal collective investment subject to a Value-at-Risk (VaR)-type constraint that accounts for uncertainty about future regulatory requirements. The constraint incorporates several possible wealth thresholds and imposes a minimum weighted average probability of meeting them. We characterize the optimal terminal wealth, which combines regions of constant wealth at regulatory thresholds with Merton-type investment regions. Unlike a classical VaR constraint with only one threshold, our formulation generates protection against several thresholds. For two possible thresholds, we identify when both are protected and when protection concentrates on the larger one. Numerical results show that the regulatory constraint can substantially reshape the optimal payoff and generate large welfare losses relative to unconstrained investment when the required compliance probability is high, even the smaller wealth threshold is high, or initial capital is limited. In heterogeneous investor pools, welfare losses depend strongly on the composition of the risk aversions and the Pareto weights governing risk sharing. Losses are particularly pronounced when highly risk-averse investors dominate the pool. Moreover, investors with smaller capital contributions can bear disproportionately large shares of the total welfare loss, while larger contributors may bear less than their capital-proportional share. Thus, regulatory uncertainty not only reduces aggregate welfare but can also create substantial disparities in how its costs are distributed among investors.

\smallskip


\smallskip


\end{abstract}

\section{Introduction}

Insurance companies, pension funds, and other delegated investment institutions manage wealth on behalf of beneficiaries with heterogeneous preferences, different contract histories, and long-dated liabilities. Their investment decisions must account for financial-market risk as well as regulatory requirements that may change before these liabilities are settled. Uncertainty about the applicable solvency benchmark can affect how much wealth an institution seeks to protect and in which market states it provides that protection. This raises a central question: how do optimal investment and investor welfare change when the regulatory wealth requirement is not known in advance?

For institutions managing a common investment pool, this question also has a distributional dimension. Investors differ in their risk aversions and capital contributions, while a single investment policy serves the collective as a whole. The consequences of that policy depend both on the aggregate investment outcome and on the rule used to distribute wealth among investors. A regulatory constraint imposed on the pool can therefore affect its members very differently. Even when the aggregate welfare cost is moderate, particular investors may experience substantial losses or bear a disproportionate share of the total cost. Understanding investment under regulatory uncertainty consequently requires examining both the optimal investment policy and the allocation of its welfare costs within the collective.

A second aspect is that Solvency regulations constrain investment decisions by imposing requirements on the distribution of future assets relative to liabilities. Such requirements affect admissible portfolios, capital needs, and the allocation of risk among beneficiaries. Value-at-Risk (VaR) and Expected Shortfall (ES) constraints are prominent examples of downside-risk restrictions studied in this context. However, the effective regulatory benchmark need not be known with certainty at the time of investment. Its future application may depend on revisions of regulatory rules, supervisory interpretation, internal model approval, the classification of liabilities, or the treatment of different business lines. Institutions with legacy guarantees and heterogeneous groups of policyholders may therefore face several  future requirements. Investment decisions made today must account for the possibility that different wealth thresholds will become relevant at the end of the investment horizon.

We address the aforementioned issues by studying optimal collective investment subject to a VaR-type constraint that incorporates several possible wealth thresholds. The constraint requires the weighted average probability of meeting these thresholds to exceed a prescribed compliance level. Collective welfare is evaluated through a Pareto aggregator of individual utilities, which accommodates heterogeneous risk preferences and determines how terminal wealth is shared among investors. Our analysis focuses on two questions. First, how do the possible thresholds and the required compliance level shape optimal terminal wealth and its welfare cost? Second, how does the combination of this constraint and collective risk sharing distribute that cost among investors, and which investors bear a disproportionate burden?

Our study connects several strands of the literature. Classical continuous-time portfolio choice begins with \cite{merton1969lifetime,merton1971optimum} and is further developed through martingale and duality methods, for example in \cite{cox1989optimal}. Within the literature on downside-average compliance constraints, \cite{basak2001value} analyze optimal trading strategies under VaR-based risk management and show that such constraints can generate nonstandard risk-taking incentives. Related contributions study shortfall, VaR, ES, and more general downside-risk restrictions; see, among others, \cite{gabih2005dynamic}, \cite{chen2018optimal}, \cite{chen2018a}, \cite{pengyuWei2}, \cite{RedekerWunderlich2018DynamicRiskConstraints}, \cite{bauerle2023optimal}, \cite{chen2024risk}, \cite{chen2024equivalence}, and \cite{baeuerleGoell2025varConstraint}. These studies examine how risk constraints affect optimal terminal wealth and how probability-based restrictions introduce non-concavities into the investment problem. For the axiomatic foundations of risk measures, we refer to \cite{Artzner1999risk} and \cite{follmer2011stochastic}.

Collective investment with heterogeneous beneficiaries and sharing rules has been studied by \cite{jensen2016how}, \cite{chen2021optimal}, \cite{chen2021collectiveAOR}, \cite{branger2023optimal}, and \cite{WangWangZhang2023CollectiveDC}, among others. These contributions demonstrate that the sharing rule is central to welfare comparisons and to the allocation of risk among investors with different preferences. Empirical evidence also supports the relevance of preference heterogeneity in pension settings; see \cite{alserda2019individual}. This literature provides the basis for analyzing individual welfare within a common investment pool, but leaves open how uncertainty about the applicable regulatory wealth requirement interacts with collective risk sharing.

Uncertainty and ambiguity have long been recognized as relevant to investment decisions; see \cite{ellsberg1961risk} and \cite{klibanoff2005smooth}. Uncertainty about public policy and regulation affects asset prices and risk premia, see~\cite{pastor2012uncertainty,pastor2013political}. In optimal portfolio selection, parameter uncertainty, partial information, and ambiguity have been examined in various dynamic settings; see, among others, \cite{cvitanic2006dynamic}, \cite{bjork2010optimal}, \cite{balter2021consistency}, \cite{bauerle2024learning}, and \cite{chen2025optimal}. Our focus differs from these approaches in that the uncertainty concerns the regulatory wealth requirement rather than the drift, volatility, or probability model of the financial market. We incorporate this uncertainty directly into the admissible set through a weighted compliance requirement across possible thresholds. The classical VaR constraint with a single fixed threshold is recovered as a special case.

Our contribution has two parts. First, we characterize the optimal collective terminal wealth under the proposed regulatory constraint. Using the martingale approach, we obtain a piecewise payoff structure combining regions of constant wealth at regulatory thresholds with Merton-type investment regions. Unlike a classical VaR constraint, our formulation can generate protection at several wealth levels. For two possible thresholds, we characterize when both levels are protected, when an intermediate Merton-type region appears, and when protection concentrates on the higher threshold. The boundaries of these regions depend jointly on collective preferences, the budget, and the regulatory requirements. We derive the equations determining these boundaries and the associated Lagrange multipliers, as well as the dynamic replication strategy and the individual allocations induced by the Pareto sharing rule.

Second, we examine the welfare consequences of the constraint from the investors' perspective. We begin with a single investor to study how the regulatory parameters affect the optimal payoff and the welfare loss relative to unconstrained investment. We then consider heterogeneous investor pools to examine how differences in risk aversion and capital contributions affect both the magnitude and the distribution of regulatory welfare losses. To this end, we compare individual certainty equivalents under constrained and unconstrained collective investment. At the individual level, we distinguish percentage reductions in certainty equivalents from each investor's share of the total welfare loss. The former measures the severity of the loss for an individual investor; the latter allows us to assess whether that investor bears more or less than a share proportional to the initial capital contribution.

Our numerical results show that the regulatory constraint can substantially change the shape of optimal terminal wealth while imposing moderate welfare costs in many parameter configurations. Large losses arise when the required compliance probability is high, even the smaller regulatory wealth threshold is high, or initial capital is limited. In these cases, providing protection absorbs a substantial part of the available budget and leaves fewer resources for utility smoothing and upside exposure. In heterogeneous investor pools, welfare losses are particularly pronounced when highly risk-averse investors dominate the pool. While risk aversion strongly influences the magnitude of individual welfare losses, unequal Pareto weights can lead to substantial disparities in their distribution. In particular, investors with smaller capital contributions can bear disproportionately large shares of the total welfare loss, whereas larger contributors may bear considerably less than their capital-proportional share. These disparities are especially pronounced when a large investor is less risk-averse than the remaining pool members. These findings show why aggregate welfare alone is insufficient to assess the consequences of regulation for a collective investment pool. We also compare collective investment with constrained stand-alone investment and find that the presence of less risk-averse investors can make participation attractive to more risk-averse investors.

The remainder of the paper is organized as follows. Section~\ref{sec:collective} introduces the financial market, the collective preference aggregator, and the regulatory constraint. Section~\ref{sec:solution} characterizes the optimal solutions. Appendix~\ref{sec:numerical} explains the numerical implementation, which is used in Sections~\ref{sec:impactRegulatoryUncertainty} and~\ref{sec:impactCollectiveInvestment} to examine the investment and welfare consequences of the constraint, first for a single investor and then for heterogeneous investor pools. Section~\ref{sec:conclusion} concludes and discusses directions for future research. All proofs are collected in the appendix.

\section{Model setup and optimization problem}
\label{sec:collective}

Section~\ref{subsec:market} introduces the financial market model.
Section~\ref{sec:collectiveUtilities} defines the collective utility
criterion and the associated sharing rule.
Section~\ref{subsec:uncertain_reg} formulates the investment problem
under regulatory uncertainty.

\subsection{Financial market model}
\label{subsec:market}

We assume a filtered probability space
$(\Omega,\mathcal F,\{\mathcal F_t\}_{t\in[0,T]},\mathbb P)$,
where the filtration is generated by a standard Brownian motion \(W\).
The financial market consists of a risk-free asset \(B\) and one risky
asset~\(S\), whose dynamics are given by
\begin{align}
    \diff B_t = rB_t\,\diff t
    \quad \text{and}\quad
    \diff S_t = \mu S_t\,\diff t+\sigma S_t\,\diff W_t,
    \label{eq:dS_t}
\end{align}
with initial values $B_0=1$ and $S_0=s>0$.
Here, \(r \in\mathbb R\) denotes the risk-free interest rate,
\(\mu >r\) the drift, and
\(\sigma>0\) the volatility of the risky asset. The wealth process \(X\) generated
by a self-financing investment strategy with initial wealth \(x_0>0\)
satisfies
\begin{align}
    \diff X_t
    &=(rX_t+\pi_t(\mu-r))\,\diff t
      +\pi_t\sigma\,\diff W_t,
    \quad X_0=x_0,
    \label{eq:dynamicWealthProcess}
\end{align}
where $\pi_t$ is the amount of money invested in the risky asset at
time $t$.
For brevity, the market price of risk is denoted by
$\eta:=\frac{\mu-r}{\sigma}$.
In this complete market, the unique equivalent martingale
measure \(\mathbb Q\) determines the state-price density
\begin{align}
    \xi_t
    &:=
    e^{-rt}Z_t,
    \quad \xi_0=1,
    \quad \text{where}\quad
    Z_t
    :=
    \frac{\diff\mathbb Q}{\diff\mathbb P}
    \bigg|_{\mathcal F_t}
    =
    \exp\left(-\frac12\eta^2t-\eta W_t\right).
    \label{eq:xi_t}
\end{align}
The state-price density satisfies
$\diff\xi_t=-r\xi_t\,\diff t-\eta\xi_t\,\diff W_t$
and has the following lognormal distribution under \(\mathbb P\):
\[
    \xi_t
    \sim
    LN\left(
        \left(-r-\frac12\eta^2\right)t,
        \eta^2t
    \right).
\]
For further details on the martingale approach and state-price densities
in complete markets, see
\cite{basak1995general} or~\cite{jensen2001paying}.

\subsection{Collective utility function}
\label{sec:collectiveUtilities}

We consider a group of \(n\) investors, such as pension beneficiaries
or financial institutions, including banks and insurance companies.
These investors do not trade directly on the financial market.
Instead, their aggregated wealth $x_0>0$ is delegated to a single
manager, who invests it collectively on their behalf.
Delegation is motivated by professional expertise, economies of scale,
and administrative convenience; see, for example,~\cite{kim2016time}.
Differences in investors' risk attitudes are represented
by CRRA utility functions
\[
    U_i(x)
    =
    \frac{x^{1-\gamma_i}}{1-\gamma_i},
    \quad x>0,\quad
    \gamma_i>0,\quad \gamma_i\neq1,
    \quad i=1,\dots,n.
\]
The parameter \(\gamma_i\) measures the risk aversion of investor \(i\).
CRRA preferences are standard in the retirement and asset-allocation
literature; see, for instance, the empirical evidence on preference
heterogeneity in \cite{alserda2019individual}.

To capture heterogeneous preferences within a single welfare criterion,
we use the following Pareto utility aggregator:
\begin{equation}
    U(x)
    =
    \max_{\substack{x_1,\dots,x_n\ge0\\
                     \sum_{i=1}^n x_i=x}}
    \sum_{i=1}^n\beta_iU_i(x_i),
    \quad x>0,
    \label{eq:U_B}
\end{equation}
where $\sum_{i=1}^n\beta_i=1$ and
$\beta_i>0$, $i=1,\dots,n$ are the the Pareto weights, which specify the relative importance assigned to
investors' utilities in the collective objective. One possible choice
is \(\beta_i=x_{0,i}/x_0\), where \(x_{0,i}>0\) is investor \(i\)'s
initial capital contribution and
\(x_0=\sum_{i=1}^n x_{0,i}\).
This choice links welfare priorities to capital contributions, but
does not generally imply that individual terminal allocations are
proportional to those contributions or that their time-zero market
values equal \(x_{0,i}\). The sharing rule therefore permits
redistribution within the pool.
The aggregator \(U\) is standard in heterogeneous-agent portfolio choice
and risk-sharing problems; see
\cite{dumas1989two,chen2021optimal,branger2023optimal}.
Technically, \(U\) is the negative infimal convolution of the functions~\(-\beta_1U_1,\dots,-\beta_nU_n\).

Since each \(U_i\) is strictly increasing and strictly concave, the
collective utility \(U\) is also strictly increasing and strictly
concave on \((0,\infty)\).
Moreover, \(U\) inherits the usual Inada-type behavior from the
individual CRRA utilities.

\begin{lem}
    \label{lem:Inada}
    \(U\) satisfies the Inada conditions, i.e.,
    $\lim\limits_{x\downarrow0}U'(x)=\infty$ and
    $\lim\limits_{x\to\infty}U'(x)=0$.
\end{lem}

Strict concavity and the Inada condition at zero imply that
the maximizer in~\eqref{eq:U_B} is unique and has strictly positive
components for every \(x>0\).
For a fixed \(x>0\), denote this solution by
$(x_1^\star,\dots,x_n^\star)$.
The first-order conditions imply that there exists
\(\lambda_x>0\) such that
\[
    \beta_iU_i'(x_i^\star)=\lambda_x,
    \quad i=1,\dots,n.
\]

By the envelope theorem, $U'(x)=\lambda_x$, so the Pareto sharing rule
can be written as
\begin{align}
    x_i^\star
    &=
    I_i\!\left(\frac{U'(x)}{\beta_i}\right),
    \quad i=1,\dots,n,
    \label{ind_XT}
\end{align}
where \(I_i=(U_i')^{-1}\).
Equivalently, the inverse marginal utility of the collective utility
function satisfies
\begin{equation}
    I(y)
    :=
    (U')^{-1}(y)
    =
    \sum_{i=1}^n
    I_i\!\left(\frac{y}{\beta_i}\right),
    \quad y>0.
    \label{eq:collective_inverse_marginal}
\end{equation}

We apply this sharing rule pointwise to obtain individual terminal
payoffs from the optimal collective terminal wealth \(X_T^\star\):
\begin{equation}
    X_{T,i}^\star
    =
    I_i\!\left(\frac{U'(X_T^\star)}{\beta_i}\right),
    \quad i=1,\dots,n,
    \label{eq:sharing_rule_collective}
\end{equation}
with $\sum_{i=1}^n X_{T,i}^\star=X_T^\star$, see \cite{branger2023optimal}.

Concluding, the manager's problem can be formulated using the collective
utility function \(U\). Investor heterogeneity affects both the
optimal aggregate payoff through \(U\) and its distribution through
the Pareto sharing rule.

\subsection{Regulatory uncertainty and the compliance constraint}
\label{subsec:uncertain_reg}

Risk-based regulation, such as Solvency II, can impose downside-average compliance constraints that are formulated
in terms of VaR. In practice, however, the regulatory benchmark, against which a firm’s risk exposure
is assessed, may itself be subject to uncertainty. For instance, the relevant VaR threshold need not
be known ex ante, as it can depend on the business line, the classification of liabilities, supervisory
interpretation, or future regulatory adjustments. For simplicity, we refer to this situation as regulatory
uncertainty and assume that a firm wants to protect against it by considering the corresponding
multiple thresholds rather than a single, unambiguous threshold. To model this regulatory uncertainty
in a parsimonious way, we consider a finite collection of deterministic thresholds
\[
    0<L_1<\cdots<L_J<\infty.
\]
Each \(L_j\) is a possible regulatory wealth requirement
at time \(T\). The thresholds describe possible regulatory benchmarks.
We assign them the following weights
\[
    \alpha_j>0,
    \quad
    \sum_{j=1}^J\alpha_j=1.
\]

The manager chooses a nonnegative,
\(\mathcal F_T\)-measurable terminal wealth \(X_T\) to maximize
collective expected utility subject to a budget constraint and a
weighted average compliance constraint:
\begin{align}
    \sup_{X_T\ge0}\;
    \mathbb E\!\left[U(X_T)\right]
    \quad \text{subject to}\quad
    \mathbb E\!\left[\xi_TX_T\right]\le x_0,
    \quad
    \sum_{j=1}^J\alpha_j
    \mathbb P\!\left(X_T\ge L_j\right)\ge a,
    \label{eq:policyholder-problem}
\end{align}
where \(a\in(0,1)\) is the minimum overall compliance level.
The quantity $\mathbb E[\xi_TX_T]$ is the time-zero market value of
the terminal payoff \(X_T\).
The second constraint requires the weighted average
probability of meeting the regulatory thresholds to be at least \(a\).
It permits trade-offs between compliance at different thresholds:
a higher probability of meeting one threshold can compensate for
a lower probability of meeting another. It does not require
\(\mathbb P(X_T\ge L_j)\ge a\) for every \(j\).

The average compliance constraint can represent an
assessment of solvency across business lines.
Suppose an insurer has \(J\) business lines.
Line \(j\) has a terminal liability \(\ell_j>0\) at time \(T\).
Assets are managed in one pool, producing aggregate terminal wealth
\(X_T\).
For internal accounting or regulatory purposes, the pooled
assets are notionally allocated across the lines in fractions
\(c_j>0\), with \({\sum_{j=1}^Jc_j=1}\).
The terminal assets attributed to line \(j\) are
$A_{j,T}=c_jX_T$.
Line \(j\) is solvent at time \(T\) if $A_{j,T}\ge\ell_j$.
Equivalently,
\[
    c_jX_T\ge\ell_j
    \quad\Longleftrightarrow\quad
    X_T\ge L_j,
    \quad
    L_j:=\frac{\ell_j}{c_j}.
\]
The lines can be relabeled in increasing order of their
implied thresholds; lines with identical thresholds can be combined
by adding their weights.

Thus, $\mathbb P(X_T\ge L_j)$ is the solvency probability of business
line \(j\) under pooled asset management.
In this interpretation, the weights \(\alpha_j>0\)
describe the relative importance of the lines.
They may represent premium shares, shares of technical provisions,
allocated risk-capital shares, or supervisory importance weights.
The compliance constraint then imposes a minimum weighted average
of line-level solvency probabilities, without guaranteeing the same
solvency probability for every line.

The formulation nests the standard VaR constraint as a special case.
Indeed, if \(J=1\), then~\eqref{eq:policyholder-problem} reduces to
$\mathbb P(X_T\ge L_1)\ge a$, or equivalently
$\mathbb P(X_T<L_1)\le1-a$.
For \(J>1\), the constraint can be written in terms of the monetary
risk measure
\[
    \rho(X)
    :=
    \inf\left\{
        m\in\mathbb R
        \,\middle|\,
        \sum_{j=1}^J\alpha_j
        \mathbb P(X+m\ge L_j)\ge a
    \right\}.
\]
This functional is monotone and translation invariant, in the sense
of \cite{Artzner1999risk}.
The average compliance requirement is equivalent to $\rho(X_T)\le0$.

For comparison, the Weighted VaR constraint considered
in~\cite{pengyuWei1} takes the form
\begin{align}
    \int_0^1q_X(u)\,\diff\zeta(u)\ge x,
    \label{eq:WVaR_Constraint}
\end{align}
where \(x>0\), \(q_X\) is a quantile function of \(X\),
and \(\zeta\) is a fixed probability measure on \([0,1]\).
Our average compliance constraint instead has the quantile representation
\[
    \int_0^1g\big(q_X(u)\big)\,\diff u\ge a,
\]
where
\(g(z)=\sum_{j=1}^J\alpha_j\mathbf 1_{[L_j,\infty)}(z)\), which can only be rewritten in the form in~\eqref{eq:WVaR_Constraint}, if $\zeta$ were allowed to depend on $X$, which is not the case in the setup of~\cite{pengyuWei1}. 

We now use~\eqref{eq:policyholder-problem} to study both the
aggregate investment decision and its consequences for individual
investors. The regulatory parameters determine the required
protection, while risk aversions and the Pareto weights determine
collective preferences and the allocation of terminal wealth.
This allows us to examine how the regulatory constraint changes
investor welfare and whether its costs are distributed
disproportionately relative to initial capital contributions.

To distinguish a nonbinding constraint from an unattainable
compliance requirement, we introduce two benchmark levels.
Let \(\lambda^M>0\) be the unique solution of
$\mathbb E[\xi_TI(\lambda^M\xi_T)]=x_0$, and let
$X_T^M:=I(\lambda^M\xi_T)$ be the corresponding Merton payoff.
We define
\begin{align}
    a^M
    &:=
    \sum_{j=1}^J
    \alpha_j\mathbb P\!\left(X_T^M\ge L_j\right)
    =
    \sum_{j=1}^J
    \alpha_j\mathbb P\!\left(I(\lambda^M\xi_T)\ge L_j\right).
    \label{eq:riskConstraintMerton}
\end{align}
Moreover, we set
\begin{align}
    \overline a
    &:=
    \sup_{X\ge0}
    \left(
        \sum_{j=1}^J\alpha_j\mathbb P(X\ge L_j)
    \right),
    \quad \text{subject to}\quad
    \mathbb E[\xi_TX]\le x_0.
    \label{eq:MaximumValueRiskConstraint}
\end{align}
Here, the supremum is taken over nonnegative,
\(\mathcal F_T\)-measurable payoffs satisfying the budget constraint.
The level \(a^M\) is the compliance achieved by unconstrained optimal
investment, whereas \(\overline a\) is the supremum of compliance
levels achievable with the available capital.

Clearly, $a^M\le\overline a$.
If \(e^{-rT}L_J\le x_0\), the constant payoff \(X=L_J\)
is affordable and meets every threshold, so \(\overline a=1\).
If \(a\le a^M\), the Merton payoff already satisfies the compliance
constraint. If \(a>\overline a\), no payoff satisfies both constraints.

\begin{cor}
    \label{cor:benchmark-cases}
    The following statements hold:
    \begin{enumerate}
        \item[(i)]
        If \(a\le a^M\), then the Merton payoff
        $X_T^M=I(\lambda^M\xi_T)$ is the unique solution
        to~\eqref{eq:policyholder-problem}.
        If \(a<a^M\), the average compliance constraint is slack.

        \item[(ii)]
        If \(a>\overline a\), then
        the feasible set
        of~\eqref{eq:policyholder-problem} is empty.
    \end{enumerate}
\end{cor}

Corollary~\ref{cor:benchmark-cases} directs attention to
\(a^M<a\le\overline a\), where the Merton payoff does not satisfy
the compliance requirement.
Feasibility and the existence of an optimum with finite expected
utility are distinct questions; in particular, a payoff attaining
the maximum compliance level may assign zero wealth to some states.
For \(J=2\), we analyze the boundary case
\(a^M<a=\overline a\) in
Proposition~\ref{prop:optimalPayoffBoundaryCasePI}
and the interior case \(a^M<a<\overline a\) in
Theorem~\ref{thm:2bounds}.

\section{Solution of the portfolio optimization problem}
\label{sec:solution}

This section characterizes the manager's optimal investment
policy under the compliance constraint in
\eqref{eq:policyholder-problem}.
Section~\ref{sec:structureOfOptimalPayoff} establishes existence and
reduces the problem to terminal payoffs that are nonincreasing functions
of the state-price density.
Section~\ref{subsec:two-threshold} characterizes the optimal payoff
for two possible regulatory thresholds, identifying the regions
in which wealth equals a threshold and those in which it follows
a Merton-type allocation.
Section~\ref{subsec:strategies-two-threshold} derives the corresponding
dynamic investment strategy.
Together with the Pareto sharing rule, these results provide the
individual payoffs used in the subsequent welfare analysis.
Throughout, we maintain the assumption \(\eta>0\).

\subsection{Existence and general structure of the optimal payoff}
\label{sec:structureOfOptimalPayoff}

We first show that it is sufficient to consider terminal
payoffs that are nonincreasing functions of the state-price density.
This reduction preserves expected utility and compliance probabilities
while weakly reducing the initial cost of the payoff.

Let \(\mathcal G\) denote the set of Borel-measurable nonincreasing
functions $g:(0,\infty)\to[0,\infty)$.
Then, we obtain the following reduced portfolio optimization problem:
\begin{align}
    \sup_{g\in\mathcal G}\;
    \mathbb E[U(g(\xi_T))]
    \quad \text{subject to}\quad
    \mathbb E[\xi_Tg(\xi_T)]\le x_0,
    \quad
    \sum_{j=1}^J\alpha_j
    \mathbb P(g(\xi_T)\ge L_j)\ge a.
    \label{eq:reduced-problem}
\end{align}
The admissibility and expected-utility conventions introduced
in Section~\ref{subsec:uncertain_reg} also apply to the reduced problem. The following result shows that this reduction is without loss
of generality.

\begin{lem}
\label{lem:cost-efficient-reduction}
The following statements hold:
\begin{enumerate}
    \item[(i)]
    If \(g\in\mathcal G\) is feasible
    for~\eqref{eq:reduced-problem}, then \(g(\xi_T)\) is feasible
    for~\eqref{eq:policyholder-problem}.

    \item[(ii)]
    If \(X_T\) is feasible for~\eqref{eq:policyholder-problem},
    then there exists \(g\in\mathcal G\) such that \(g(\xi_T)\)
    is feasible for~\eqref{eq:reduced-problem} and
    \[
    \begin{aligned}
        \mathbb E[U(g(\xi_T))]
        &=\mathbb E[U(X_T)],\quad 
        \sum_{j=1}^J\alpha_j\mathbb P(g(\xi_T)\ge L_j)
        =\sum_{j=1}^J\alpha_j\mathbb P(X_T\ge L_j),\quad 
        \mathbb E[\xi_Tg(\xi_T)]
        \le\mathbb E[\xi_TX_T].
    \end{aligned}
    \]
\end{enumerate}
Consequently, the sets of optimal payoffs of~\eqref{eq:policyholder-problem} and~\eqref{eq:reduced-problem} are equal.
In particular, if~\eqref{eq:reduced-problem} admits an optimizer \(g^\star\), then
\(g^\star(\xi_T)\) is an optimizer of~\eqref{eq:policyholder-problem}.
\end{lem}

Lemma~\ref{lem:cost-efficient-reduction} allows us to search
for an optimal payoff in the cost-efficient class \(X_T=g(\xi_T)\).
Such payoffs allocate more wealth to states with lower state prices
and they are anti-comonotone with the state-price density.
The following theorem establishes existence for compliance
levels strictly below the maximum attainable level.

\begin{thm}
\label{thm:existence-policyholder}
If \(a<\overline a\), then the optimization
problem~\eqref{eq:policyholder-problem} admits a maximizer.
\end{thm}

\begin{remark}
\label{rem:innovationFormSolution}
The proof of Theorem~\ref{thm:existence-policyholder}
establishes existence without first assuming a Lagrangian
characterization or a particular payoff structure.
By Lemma~\ref{lem:cost-efficient-reduction}, the problem can be
restricted to nonincreasing functions of the state-price density.
Helly's selection theorem then provides the compactness argument
for a maximizing sequence.
This separates the existence argument from the multiplier-based
characterization used below.
\end{remark}

Although existence does not rely on Lagrange multipliers,
a Lagrangian argument provides a useful verification principle.
Consider multipliers \(\lambda>0\) and \(\nu\ge0\), and an
\(\mathcal F_T\)-measurable candidate payoff
\(X_T^{\lambda,\nu}\) with finite expected utility, satisfying
\[
    X_T^{\lambda,\nu}(\omega)
    \in
    \argmax_{x\ge0}
    \left\{
        U(x)-\lambda\xi_T(\omega)x
        +\nu\sum_{j=1}^J
        \alpha_j\mathbf 1_{\{x\ge L_j\}}
    \right\},
\]
for $\mathbb P$-almost all $\omega\in\Omega$. Suppose that \(X_T^{\lambda,\nu}\) satisfies the budget and
average compliance constraints with equality:
\[
    \mathbb E[\xi_TX_T^{\lambda,\nu}]=x_0
    \quad \text{and}\quad
    \sum_{j=1}^J\alpha_j
    \mathbb P\!\left(
        X_T^{\lambda,\nu}\ge L_j
    \right)=a.
\]
For any other feasible payoff \(Y\), pointwise
optimality implies
\[
\begin{aligned}
    &U(X_T^{\lambda,\nu})
    -\lambda\xi_TX_T^{\lambda,\nu}
    +\nu\sum_{j=1}^J
    \alpha_j\mathbf 1_{\{X_T^{\lambda,\nu}\ge L_j\}}\ge
    U(Y)-\lambda\xi_TY
    +\nu\sum_{j=1}^J
    \alpha_j\mathbf 1_{\{Y\ge L_j\}}
    \quad\text{a.s.}
\end{aligned}
\]
Taking expectations and using feasibility of \(Y\)
therefore yields
\[
    \mathbb E[U(X_T^{\lambda,\nu})]
    \ge
    \mathbb E[U(Y)]
    +\lambda\big(x_0-\mathbb E[\xi_TY]\big)
    +\nu\left(
        \sum_{j=1}^J\alpha_j\mathbb P(Y\ge L_j)-a
    \right)
    \ge
    \mathbb E[U(Y)].
\]
Thus, once suitable multipliers and a pointwise maximizer
satisfying the two constraints are found, optimality follows.

For general \(J\), the pointwise comparison involves
the Merton-type candidate \(I(\lambda\xi_T)\) and the regulatory
thresholds \(L_1,\dots,L_J\).
We now specialize to \(J=2\), for which this comparison gives an
explicit payoff characterization.
The parameters determining its regions are obtained from a system
of equations and generally require numerical calculation.

\subsection{Optimal payoffs in the two-threshold case}
\label{subsec:two-threshold}

Unless stated otherwise, we assume \(J=2\) throughout
the remainder of the paper.
We first characterize investment at the maximum attainable
compliance level and then turn to the interior case
\(a^M<a<\overline a\), where the manager trades off expected
utility against additional regulatory protection.

We start with the  boundary case of $a=\overline{a}$. 

\begin{prop}\label{prop:optimalPayoffBoundaryCasePI}
    For $a=\overline{a}$, we obtain the following statements:
    \begin{itemize}
        \item[(i)] If $\alpha_1\leq \frac{L_1}{L_2}$, then the unique optimal payoff of~\eqref{eq:policyholder-problem} is
        \[
            X_T^{\star} = \begin{cases}
                L_2, & \lambda^{\star}\xi_T\leq \frac{1}{L_2},\\
                0, & \text{otherwise},
            \end{cases}
        \]
        where $\lambda^{\star}>0$ is the unique solution to $\mathbb{E}[\xi_T X_T^{\star}] = x_0$. Then, $\overline{a} = \P\left(\lambda^{\star}\xi_T\leq \frac{1}{L_2}\right)$.
        \item[(ii)] If $\alpha_1> \frac{L_1}{L_2}$, then the unique optimal payoff of~\eqref{eq:policyholder-problem} is
        \[
            X_T^{\star} = \begin{cases}
                L_2, & \lambda^{\star}\xi_T\leq \frac{\alpha_2}{L_2-L_1},\\
                L_1, & \frac{\alpha_2}{L_2-L_1}<\lambda^{\star}\xi_T\leq \frac{\alpha_1}{L_1},\\
                0, & \text{otherwise},
            \end{cases}
        \]
        where $\lambda^{\star}>0$ is the unique solution to $\mathbb{E}[\xi_T X_T^{\star}] = x_0$. Then, 
        \[
            \overline{a} = \alpha_1\P\left(\lambda^{\star}\xi_T\leq \frac{\alpha_1}{L_1}\right)+\alpha_2 \P\left(\lambda^{\star}\xi_T\leq \frac{\alpha_2}{L_2-L_1}\right).
        \]
    \end{itemize}
\end{prop}

These payoffs concentrate the available capital on
meeting regulatory thresholds and assign zero wealth to the
remaining states.
If at least one investor has \(\gamma_i>1\), their collective
expected utility is~\(-\infty\).
Their uniqueness follows from the maximum-compliance problem,
rather than from a comparison of finite expected utilities.
This boundary case is therefore economically distinct from
the strictly positive interior solution below.

For completeness, the next result treats the case
in which the higher threshold can be financed in every state.
Although \(a\) cannot attain $1$, since \(a\in(0,1)\),
these payoffs characterize the full-compliance boundary.

\begin{prop}
\label{prop:riskMaximizer}
Let \(e^{-rT}L_2=x_0\).
Then the unique optimal payoff
solving~\eqref{eq:MaximumValueRiskConstraint} is
\(X_T^\star=L_2\) almost surely.
If \(e^{-rT}L_2<x_0\), then an optimal payoff
solving~\eqref{eq:MaximumValueRiskConstraint} is
\begin{align}
    X_T^\star
    =
    \begin{cases}
        I(\lambda^\star\xi_T),
        &\xi_T\le\dfrac{U'(L_2)}{\lambda^\star},\\[1.2ex]
        L_2,
        &\text{otherwise},
    \end{cases}
    \label{eq:riskMaximizer}
\end{align}
where \(\lambda^\star>0\) is the unique solution of
\(\mathbb E[\xi_TX_T^\star]=x_0\).
In both cases, \(\overline a=\alpha_1+\alpha_2=1\).
\end{prop}

We now consider \(a^M<a<\overline a\).
Here, the unconstrained payoff fails the compliance requirement,
but the manager need not devote the entire budget to maximizing
compliance.
The optimal payoff balances threshold protection against
utility-maximizing investment.
Our derivation uses a Lagrangian verification argument, drawing
on the approach of~\cite{gabih2009utility} for an expected-loss
constraint.
For brevity, write
\[
    l(\nu)
    :=
    \frac{U(L_2)-U(L_1)+\nu\alpha_2}{L_2-L_1},
    \quad \nu>0.
\]

\begin{thm}\label{thm:2bounds}

If $a^{M}<a<\overline{a}$, then the optimal payoff is given by\footnote{If the regulator specifies ES-style limits on shortfall levels $L_i$ instead of probability constraints, the same cutoff/plateau logic persists, with the calibration equations using $\mathbb{E}[(L_i-X_T)^+]$ or $\mathbb{E}[\xi_T(L_i-X_T)^+]$ in place of probabilities, see~\cite{chen2021optimal}.}
 \begin{align}\label{eq:optimalPayoffTwoBounds}
        X_T^\star=
        \begin{cases}
            I(\lambda \xi_T), & \xi_T \leq \frac{U^{\prime}(L_2)}{\lambda},\\
            L_{2}, & \frac{U^{\prime}(L_2)}{\lambda} <  \xi_T \leq \frac{q^{\star}}{\lambda},\\
            I(\lambda \xi_T), & \frac{q^{\star}}{\lambda}<\xi_T \leq \frac{U^{\prime}(L_1)}{\lambda},\\
            L_{1}, &\frac{\max\{U^{\prime}(L_1),q^{\star}\}}{\lambda} < \xi_T\leq \frac{q^{\star\star}}{\lambda},\\
            I(\lambda \xi_T),  &  \xi_T >\frac{\max\{q^{\star},q^{\star\star}\}}{\lambda},
         \end{cases}
    \end{align}
where $\lambda > 0$, $q^{\star}> U^{\prime}(L_2)$ and $q^{\star\star}\in\big(U^{\prime}(L_1),\infty\big)$ satisfy
\begin{align*}
    \mathbb{E}\!\left[\xi_T X_T^\star\right]=x_0,\quad  \alpha_1\,\mathbb{P}\!\left(X_T^\star \ge L_1\right)+\alpha_2\,\mathbb{P}\!\left(X_T^\star \ge L_2\right)= a,
\end{align*}
and simultaneously, there exists $\nu>0$ such that
\begin{align}\label{eq:firstDualityGapCondition}
U(L_1)- q^{\star\star}L_1
-\Bigl[
U\big(I(q^{\star\star})\big)-q^{\star\star}I(q^{\star\star})
\Bigr] =-\nu \alpha_1,
\end{align}
and, if $l(\nu)< U^{\prime}(L_1)$, then $q^{\star}\in\big(U^{\prime}(L_2),U^{\prime}(L_1)\big)$ satisfies
\begin{align*}
    &U(L_2)- q^{\star}L_2
-\Bigl[
U\big(I(q^{\star})\big)-q^{\star}I(q^{\star})
\Bigr] =-\nu \alpha_2,
\end{align*}
else if, $U^{\prime}(L_1)\leq l(\nu)<q^{\star\star}$, then 
\[
    q^{\star} = l(\nu)\in\big[U^{\prime}(L_1),q^{\star\star}),
\] 
and otherwise, $q^{\star}$ satisfies
\[
    U(L_2) - q^{\star}L_2 -U(I(q^{\star}))+q^{\star}I(q^{\star}) = -\nu, \quad q^{\star}> U^{\prime}(L_1).
\]

\end{thm}

The payoff provides regulatory protection by holding
wealth constant at a threshold over a range of market states.
Outside these regions, wealth follows \(I(\lambda\xi_T)\).
This has the same functional form as unconstrained optimal
investment, but generally uses a different budget multiplier.
Thus, financing threshold protection also changes investment
in the Merton-type regions.

\begin{figure}
    \centering

    \begin{subfigure}{0.31\textwidth}
        \centering
        \includegraphics[width=\linewidth]
        {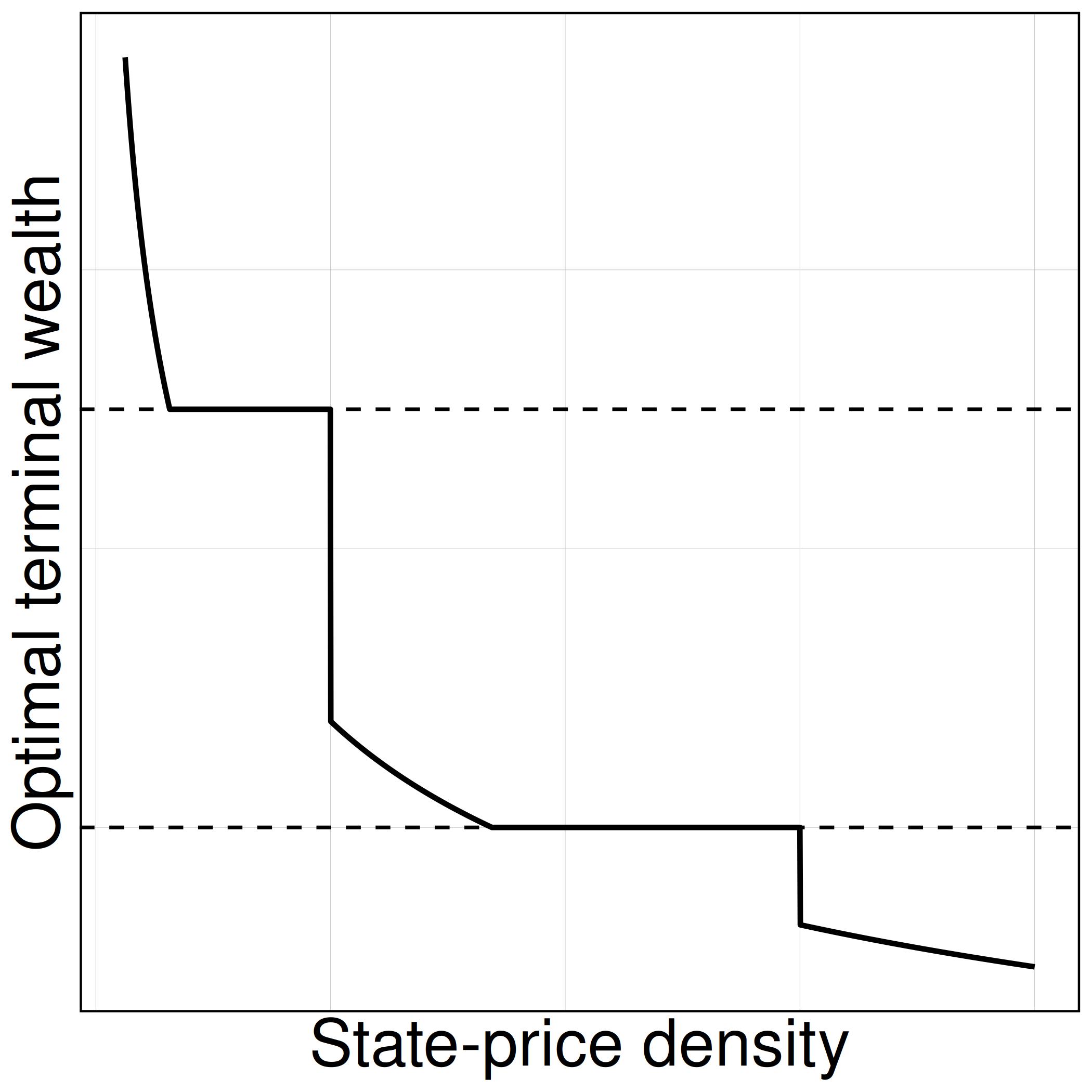}
        \caption{$l(\nu)<U'(L_1)$}
    \end{subfigure}
    \begin{subfigure}{0.31\textwidth}
        \centering
        \includegraphics[width=\linewidth]
        {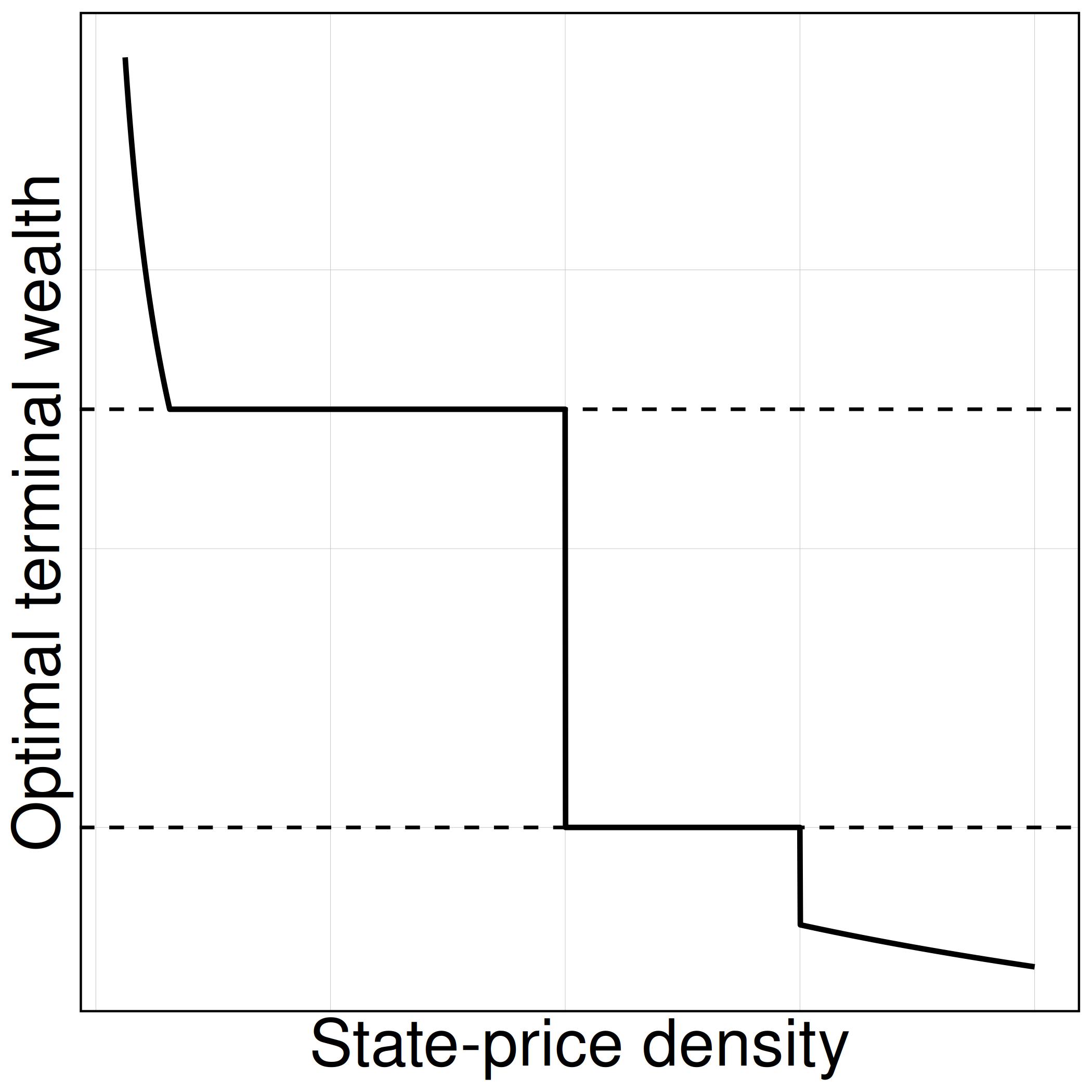}
        \caption{$U'(L_1)\le l(\nu)<q^{\star\star}$}
    \end{subfigure}
    \begin{subfigure}{0.31\textwidth}
        \centering
        \includegraphics[width=\linewidth]
        {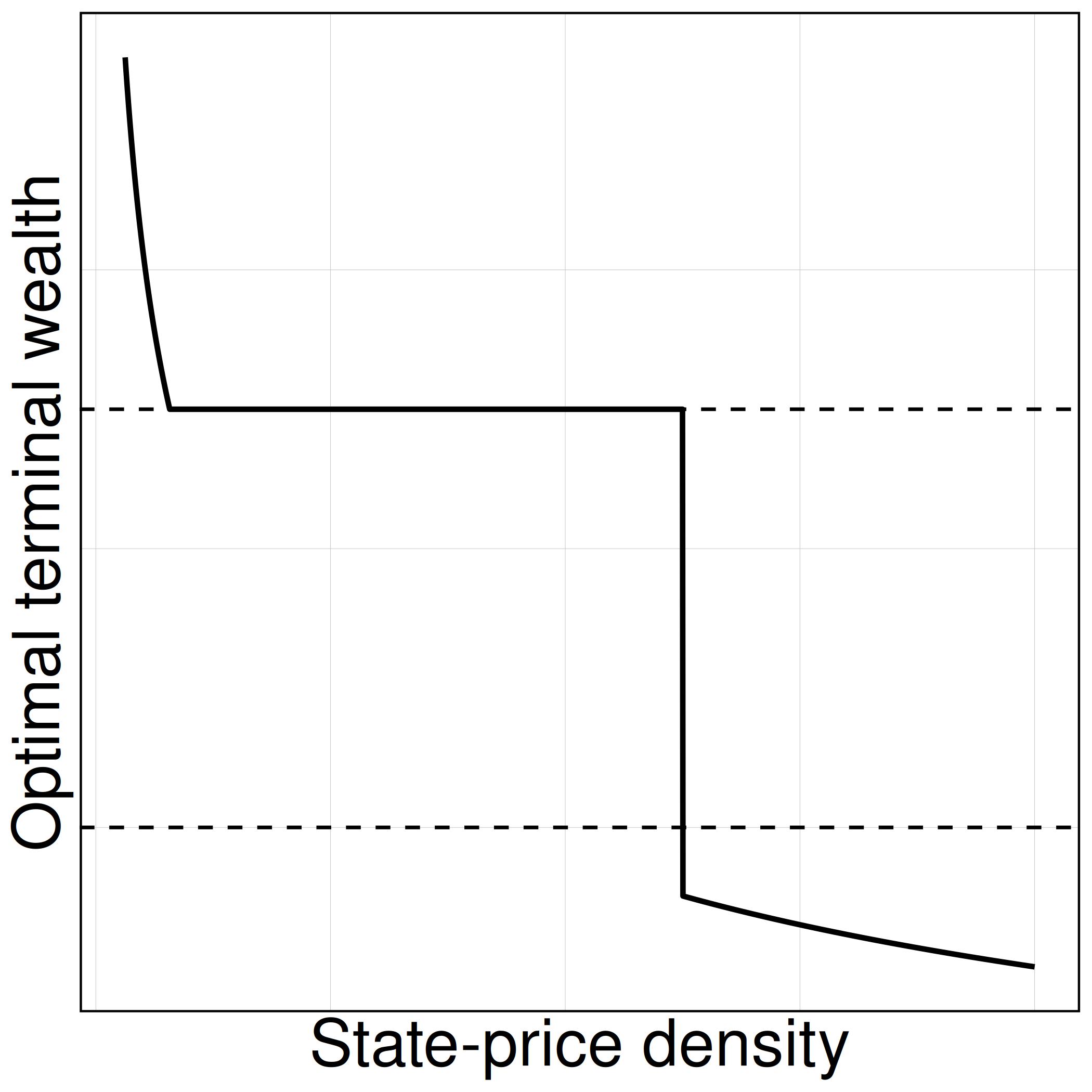}
        \caption{$l(\nu)\ge q^{\star\star}$}
    \end{subfigure}

    \caption{The three possible structures of optimal
    terminal wealth as a function of the state-price density,
    as characterized in Theorem~\ref{thm:2bounds}.}
    \label{fig:possibilities_optimal_payoff}
\end{figure}

\begin{remark}
\label{rem:calculateLagrangeParameters}
The parameters can be calculated by treating
\(\lambda\), \(q^\star\), and \(q^{\star\star}\) as functions of
\(\nu>0\).
For a fixed \(\nu\), first solve
\eqref{eq:firstDualityGapCondition} to get \(q^{\star\star}(\nu)\).
Next, determine \(q^\star(\nu)\) using the relevant case in
Theorem~\ref{thm:2bounds}.
Finally, choose \(\lambda(\nu)\) so that the resulting payoff
satisfies the budget constraint. Substituting this payoff into the weighted compliance
constraint gives a scalar equation in \(\nu\).
Solving this equation so that compliance equals \(a\) determine the corresponding $\nu$ used in the optimal payoff.

The payoff has three possible forms, illustrated in
Figure~\ref{fig:possibilities_optimal_payoff}.
If \(l(\nu)<U'(L_1)\), we have two plateaus at the two thresholds, separated
by a Merton-type region.
If \(U'(L_1)\le l(\nu)<q^{\star\star}\), both plateaus remain,
but the intermediate Merton-type region disappears.
If \(l(\nu)\ge q^{\star\star}\), the lower-threshold plateau
also disappears, and protection concentrates on \(L_2\).
These cases describe how the manager allocates protection
across the two possible regulatory requirements.
\end{remark}

\begin{remark}
\label{rem:dualGap}
Focusing on $q^{\star\star}$, note that 
    \[
        q^{\star\star}I(q^{\star\star})-U\big(I(q^{\star\star})\big) = \sup\limits_{x\in (0,\infty)}\big(q^{\star\star}x - U(x)\big) =:U^{\star}(q^{\star\star}),
    \]
    i.e.,~we evaluate the Fenchel-Legendre transform of $U$ in the point $q^{\star\star}$. Condition~\eqref{eq:firstDualityGapCondition}
can therefore be written as
    \[
        U{^\star}(q^{\star\star}) = \Big(q^{\star\star}\,L_1-U(L_1)\Big)+\nu\alpha_1.
    \]
Equivalently,
\[
    \big[
        U(I(q^{\star\star}))
        -q^{\star\star}I(q^{\star\star})
    \big]
    -
    \big[
        U(L_1)-q^{\star\star}L_1
    \big]
    =
    \nu\alpha_1.
\]
The left-hand side is the loss in the utility-minus-cost
criterion from choosing \(L_1\) instead of its unconstrained
maximizer \(I(q^{\star\star})\).
At the switching boundary, this loss is exactly offset by the
compliance reward \(\nu\alpha_1\).
In particular, the positive gap implies
\(q^{\star\star}\neq U'(L_1)\).
The left-hand side of
\eqref{eq:firstDualityGapCondition}, viewed as a function of
\(q^{\star\star}>0\), is concave, since its second derivative
is \(I^{\prime}(q^{\star\star})<0\).
\end{remark}

The next remark explains the formal connection with
the classical single-threshold VaR solution.

\begin{remark}
Consider \(\alpha_1\downarrow0\), with
\(\alpha_2=1-\alpha_1\uparrow1\).
The compliance constraint then approaches
\(\mathbb P(X_T\ge L_2)\ge a\).
For fixed \(\nu>0\),
\eqref{eq:firstDualityGapCondition} formally reduces to
\[
    U(L_1)-q^{\star\star}L_1
    -\left[
        U(I(q^{\star\star}))
        -q^{\star\star}I(q^{\star\star})
    \right]
    =
    0.
\]
Strict concavity implies that this zero-gap condition
holds at \(q^{\star\star}=U'(L_1)\).
The lower-threshold plateau therefore disappears in the
formal limiting characterization.
For an active and feasible limiting VaR constraint, the
single-threshold payoff takes the form
\[
    \lim_{\alpha_1\downarrow 0} X_T^{\star}
    =
    \begin{cases}
        I(\lambda\xi_T),
        &\xi_T\le\dfrac{U'(L_2)}{\lambda},\\
        L_2,
        &\dfrac{U'(L_2)}{\lambda}
         <\xi_T\le\dfrac{q^\star}{\lambda},\\
        I(\lambda\xi_T),
        &\xi_T>\dfrac{q^\star}{\lambda},
    \end{cases}
\]
where the parameters \(\lambda\) and \(q^\star\) satisfy
the budget constraint and the average compliance constraint.
The latter is equivalent to
\[
    \mathbb P\!\left(
        \xi_T\le\frac{q^\star}{\lambda}
    \right)
    =
    a.
\]
Thus, \(q^\star/\lambda\) is the \(a\)-quantile
of the state-price density, as in the classical solution
of~\cite{basak2001value}.
The remaining equation
\[
    U(L_2)-q^\star L_2
    -\left[
        U(I(q^\star))-q^\star I(q^\star)
    \right]
    =
    -\nu
\]
determines \(\nu\) after \(\lambda\) and \(q^\star\)
have been found. Note that, for a single threshold, determining $\nu$ is unnecessary to obtain 
the payoff itself.
In contrast, with two thresholds, the corresponding switching conditions
jointly determine how protection is allocated between them.
\end{remark}

The solution in Theorem~\ref{thm:2bounds} also admits a natural interpretation
in terms of regulatory uncertainty.

\begin{remark}\label{rem:discussionVaR-typeRiskConstraint}
    The optimal solution \(X_T^{\star}\) from
    Theorem~\ref{thm:2bounds} also protects the collective against the following
    form of regulatory uncertainty: suppose the regulator (or rating agency) focuses on one business line at time $T$ due to targeted audits, political salience, limited supervisory capacity, or shifting consumer protection priorities, which could be described by a random index $J^\star$ with
    \[
    \mathbb{P}(J^\star=j)=\alpha_j,\quad j = 1,2.
    \]
    Depending on the realized value of the random index \(J^\star\), the threshold is
    chosen. If the regulator's choice is not driven by financial-market states,
    in the sense that \(J^\star\) is independent of the state-price density
    \(\xi_T\), then, by the law of total probability,
    \[
        \mathbb{P}(X_T^{\star}\ge L_{J^\star})
        =
        \alpha_1\,\mathbb{P}(X_T^\star\ge L_1) + \alpha_2\,\mathbb{P}(X_T^\star\ge L_2)
        \geq a.
    \]
    This means that the optimal payoff protects investors against potential regulatory uncertainty. This situation can also be modeled by a random threshold $\widetilde{L}$ such that for all $i$ it holds $\mathbb{P}\big(\widetilde{L}=L_i\big)=\alpha_i$ and~$\widetilde{L}$ is independent of $X_T^{\star}$.
\end{remark}

\subsection{Optimal investment strategies in the two-threshold case}
\label{subsec:strategies-two-threshold}

We now derive the dynamic strategy that replicates
the optimal terminal payoff.
Market completeness and the finite state-price cost imply
that the payoff can be replicated by a self-financing strategy,
unique up to the usual almost-everywhere identification
of trading strategies. Its wealth process is
\begin{align}
    X_t^\star
    =
    \mathbb E\left[
        \frac{\xi_T}{\xi_t}X_T^\star
        \,\middle|\,
        \mathcal F_t
    \right].
    \label{eq:wealth-process-general}
\end{align}

\noindent Define
\[
    F(t,z)
    :=
    \mathbb E\left[
        \frac{\xi_T}{z}X_T^\star
        \,\middle|\,
        \xi_t=z
    \right],
    \quad z\in(0,\infty).
\]
Then \(X_t^\star=F(t,\xi_t)\).
For \(t<T\), applying Ito's formula and comparing
diffusion coefficients with
\eqref{eq:dynamicWealthProcess} yields
\begin{align}
    \pi_t^\star
    =
    -\frac{\mu-r}{\sigma^2}
    \xi_t\partial_zF(t,\xi_t).
    \label{eq:general-pi}
\end{align}

For \(0\le b<c\le\infty\), define
\[
    A_\gamma(t,z;b,c)
    :=
    \mathbb E\left[
        \frac{\xi_T^{1-1/\gamma}}{z}
        \mathbf 1_{\{b<\xi_T\le c\}}
        \,\middle|\,
        \xi_t=z
    \right].
\]
To accommodate payoff regions that disappear, we set
\(A_\gamma(t,z;b,c)=0\) whenever \(b\ge c\).

Let \(\tau:=T-t\), and denote the density and cumulative distribution
function of a standard normal distribution by \(\varphi\) and
\(\Phi\), respectively.
Conditionally on \(\xi_t=z\), under \(\mathbb P\),
\[
    \log(\xi_T)
    \sim
    N\left(
        \log(z)+\left(-r-\frac12\eta^2\right)\tau,
        \eta^2\tau
    \right).
\]
Consequently,
\begin{align}
    A_\gamma(t,z;b,c)
    &=
    z^{-1/\gamma}
    \exp\left(
        \left(1-\frac1\gamma\right)
        \left(-r-\frac12\eta^2\right)\tau
        +
        \frac12\left(1-\frac1\gamma\right)^2
        \eta^2\tau
    \right)
    \nonumber
    \Big[
        \Phi(d_\gamma(t,z;c))
        -
        \Phi(d_\gamma(t,z;b))
    \Big],
    \label{eq:Ap}
\end{align}
where
\[
    d_\gamma(t,z;c)
    :=
    \frac{
        \log(c/z)+r\tau+
        \left(\frac1\gamma-\frac12\right)\eta^2\tau
    }{
        \eta\sqrt{\tau},
    }.
\]
with the conventions
\(\Phi(d_\gamma(t,z;0))=0\) and
\(\Phi(d_\gamma(t,z;\infty))=1\). Note that
\begin{align}\label{eq:derivativeAuxiliaryTradingStrategy}
    \partial_z A_{\gamma}(t,z;b,c) = -z^{-1}A_{\gamma}(t,z;b,c)\left(\frac{1}{\gamma}+\frac{1}{\eta\sqrt{\tau}}\cdot \frac{\varphi\big(d_\gamma(t,z;c)\big)
	-
	\varphi\big(d_\gamma(t,z;b)\big)}{\Phi\big(d_\gamma(t,z;c)\big)
	-
	\Phi\big(d_\gamma(t,z;b)\big)}\right).
\end{align}

\noindent Similarly, define
\[
    B(t,z;b,c)
    :=
    \mathbb E\left[
        \frac{\xi_T}{z}
        \mathbf 1_{\{b<\xi_T\le c\}}
        \,\middle|\,
        \xi_t=z
    \right].
\]
Again, set \(B(t,z;b,c)=0\) for \(b\ge c\).
For \(b<c\), this is the limit of \(A_\gamma(t,z;b,c)\)
as \(\gamma\to\infty\), and
\[
    B(t,z;b,c)
    =
    e^{-r\tau}
    \left[
        \Phi(d_\infty(t,z;c))
        -
        \Phi(d_\infty(t,z;b))
    \right],
\]
where
\[
    d_\infty(t,z;c)
    :=
    \frac{\log(c/z)+r\tau-\frac12\eta^2\tau}
         {\eta\sqrt{\tau}}.
\]

\noindent Recall that the inverse marginal utility
in~\eqref{eq:collective_inverse_marginal} is
\[
    I(y)
    =
    \sum_{i=1}^n
    \beta_i^{1/\gamma_i}y^{-1/\gamma_i}.
\]

We apply these building blocks to the following cases.

\noindent\textbf{Case 1 \((a\le a^M)\):}
By Corollary~\ref{cor:benchmark-cases},
$X_T^\star=I(\lambda^M\xi_T)$.
Therefore,
\[
    X_t^\star
    =
    \sum_{i=1}^n
    \beta_i^{1/\gamma_i}
    (\lambda^M)^{-1/\gamma_i}
    A_{\gamma_i}(t,\xi_t;0,\infty).
\]
The untruncated conditional moments simplify to
\[
    A_{\gamma_i}(t,z;0,\infty)
    =
    z^{-1/\gamma_i}
    \exp\left[
        \left(
            -r\left(1-\frac1{\gamma_i}\right)
            +
            \frac12
            \left(\frac1{\gamma_i^2}-\frac1{\gamma_i}\right)
            \eta^2
        \right)\tau
    \right].
\]
Thus, the intermediate wealth generally differs from
\(I(\lambda^M\xi_t)\).
Using~\eqref{eq:general-pi}, the optimal strategy is
\[
    \pi_t^\star
    =
    \frac{\mu-r}{\sigma^2}
    \sum_{i=1}^n
    \frac1{\gamma_i}
    \beta_i^{1/\gamma_i}
    (\lambda^M)^{-1/\gamma_i}
    A_{\gamma_i}(t,\xi_t;0,\infty).
\]

\noindent\textbf{Case 2 \((a=\overline a)\):}
First suppose \(x_0<e^{-rT}L_2\), so that
Proposition~\ref{prop:optimalPayoffBoundaryCasePI} applies.
If $\alpha_1\le L_1/L_2$, the terminal payoff is
\[
    X_T^\star
    =
    L_2\mathbf 1_{\{\xi_T\le c_2\}},
    \quad
    c_2=(\lambda^\star L_2)^{-1}.
\]
Hence,
\[
    X_t^\star=L_2B(t,\xi_t;0,c_2).
\]
The replicating strategy is
\[
    \pi_t^\star
    =
    -\frac{\mu-r}{\sigma^2}
    \xi_t
    \partial_z[L_2B(t,z;0,c_2)]
    \big|_{z=\xi_t}
    =
    \frac{e^{-r\tau}L_2}{\sigma\sqrt{\tau}}
    \varphi(d_\infty(t,\xi_t;c_2)).
\]
This strategy replicates a payment of \(L_2\) in
the states selected for regulatory protection.
Under \(\eta>0\), the position in the risky asset is nonnegative.

If $\alpha_1>L_1/L_2$, set
\[
    c_1=\frac{\alpha_1}{\lambda^\star L_1},
    \quad \text{and}
\quad     c_2=\frac{\alpha_2}{\lambda^\star(L_2-L_1)}.
\]
The wealth process is
\[
    X_t^\star
    =
    L_2B(t,\xi_t;0,c_2)
    +
    L_1B(t,\xi_t;c_2,c_1).
\]
The replicating strategy is
\begin{align*}
    \pi_t^\star
    &=
    -\frac{\mu-r}{\sigma^2}\xi_t
    \partial_z
    \left[
        L_2B(t,z;0,c_2)
        +
        L_1B(t,z;c_2,c_1)
    \right]\big|_{z=\xi_t}\\
    &=
    \frac{e^{-r\tau}}{\sigma\sqrt{\tau}}
    \left[
        (L_2-L_1)\varphi(d_\infty(t,\xi_t;c_2))
        +
        L_1\varphi(d_\infty(t,\xi_t;c_1))
    \right].
\end{align*}

For the full-compliance boundary,
if \(x_0=e^{-rT}L_2\), the constant terminal payoff is
replicated by \(X_t^\star=e^{-r\tau}L_2\) and
\(\pi_t^\star=0\).
If \(x_0>e^{-rT}L_2\), the payoff selected in
Proposition~\ref{prop:riskMaximizer} has wealth process
\[
    X_t^\star
    =
    L_2B\left(t,\xi_t;\frac{U'(L_2)}{\lambda^\star},\infty\right)
    +
    \sum_{i=1}^n
    \beta_i^{1/\gamma_i}
    (\lambda^\star)^{-1/\gamma_i}
    A_{\gamma_i}\left(t,\xi_t;0,\frac{U'(L_2)}{\lambda^\star}\right).
\]

Its replicating strategy follows by differentiating
this expression according to~\eqref{eq:general-pi}.

\noindent\textbf{Case 3 \((a^M<a<\overline a)\):}
For the parameters from Theorem~\ref{thm:2bounds}, define
\[
\begin{aligned}
    c_2^0&:=\frac{U'(L_2)}{\lambda},
    &c_2^1&:=\frac{q^\star}{\lambda},\\
    c_1^0&:=\frac{U'(L_1)}{\lambda},
    &c_1^1&:=\frac{\max\{U'(L_1),q^\star\}}{\lambda},\\
    c_1^2&:=\frac{q^{\star\star}}{\lambda},
    &c_{\mathrm{tail}}
    &:=\frac{\max\{q^\star,q^{\star\star}\}}{\lambda}.
\end{aligned}
\]
Then \(X_t^\star=F(t,\xi_t)\) is simplified to
\begin{align*}
    F(t,z)
    &=
    L_2B(t,z;c_2^0,c_2^1)
    +
    L_1B(t,z;c_1^1,c_1^2)\\
    &\quad+
    \sum_{i=1}^n
    \beta_i^{1/\gamma_i}\lambda^{-1/\gamma_i}
    \Big[
        A_{\gamma_i}(t,z;0,c_2^0)
        +
        A_{\gamma_i}(t,z;c_2^1,c_1^0)
        +
        A_{\gamma_i}
        (t,z;c_{\mathrm{tail}},\infty)
    \Big].
\end{align*}
The zero convention for empty intervals makes this
formula valid for all three payoff structures in
Figure~\ref{fig:possibilities_optimal_payoff}.
In particular, the final Merton-type region begins at the
maximum of \(q^\star/\lambda\) and
\(q^{\star\star}/\lambda\).

The corresponding optimal investment strategy is
\begin{align}
    \pi_t^\star
    =
    -\frac{\mu-r}{\sigma^2}
    \xi_t
    \partial_zF(t,\xi_t).
    \label{eq:tradingTwoBounds}
\end{align}

Applying the Pareto sharing rule to the replicated
terminal wealth yields the individual terminal allocations.
The dynamic strategy determines the aggregate investment outcome,
while the sharing rule determines how its benefits and regulatory
costs are distributed among investors.

\section{Numerical study}\label{sec:numericalStudy}

We now combine the previous findings to perform a numerical study. For this, we apply the numerical procedure described in Appendix~\ref{sec:numerical}. This allows us to illustrate the economic implications from two perspectives. First, in
Section~\ref{sec:impactRegulatoryUncertainty}, we isolate the effect of
regulatory uncertainty by considering only one investor, i.e.,~$n=1$. This allows us to compare the constrained solution with the unconstrained Merton payoff and to
identify how the different parameters of the average compliance constraint
affect the shape of the optimal terminal payoff. Second, in
Section~\ref{sec:impactCollectiveInvestment}, we turn to the collective setting with four investors, i.e.,~$n=4$, 
and analyze how heterogeneous risk aversions and Pareto weights affect the
results. Unless stated otherwise, we assume the following model parameters:
\[
    T = 1,\ r = 0.01,\ \mu = 0.05,\ \sigma = 0.2,\
    L_1 = 3,\ L_2 = 5,\ \alpha_1 = 1-\alpha_2 = 0.7,\
    a = 0.4,\ x_0 = 2.5.
\]
Thus, the initial wealth is below the discounted upper regulatory threshold, $x_0<e^{-rT}L_2$.
Consequently, the average compliance constraint cannot be satisfied by investing all
wealth in the risk-free asset. Instead, the manager has to allocate wealth
state-dependently: wealth is shifted towards those states in which meeting one
of the regulatory thresholds is most cost efficient, while the remaining states
are treated according to the utility-maximization trade-off, recall the optimal payoff in Theorem~\ref{thm:2bounds}.

\subsection{Impact of average compliance constraint without collective investment}
\label{sec:impactRegulatoryUncertainty}

\begin{figure}
\centering

\begin{subfigure}{0.425\textwidth}
\centering
\includegraphics[width=\linewidth]{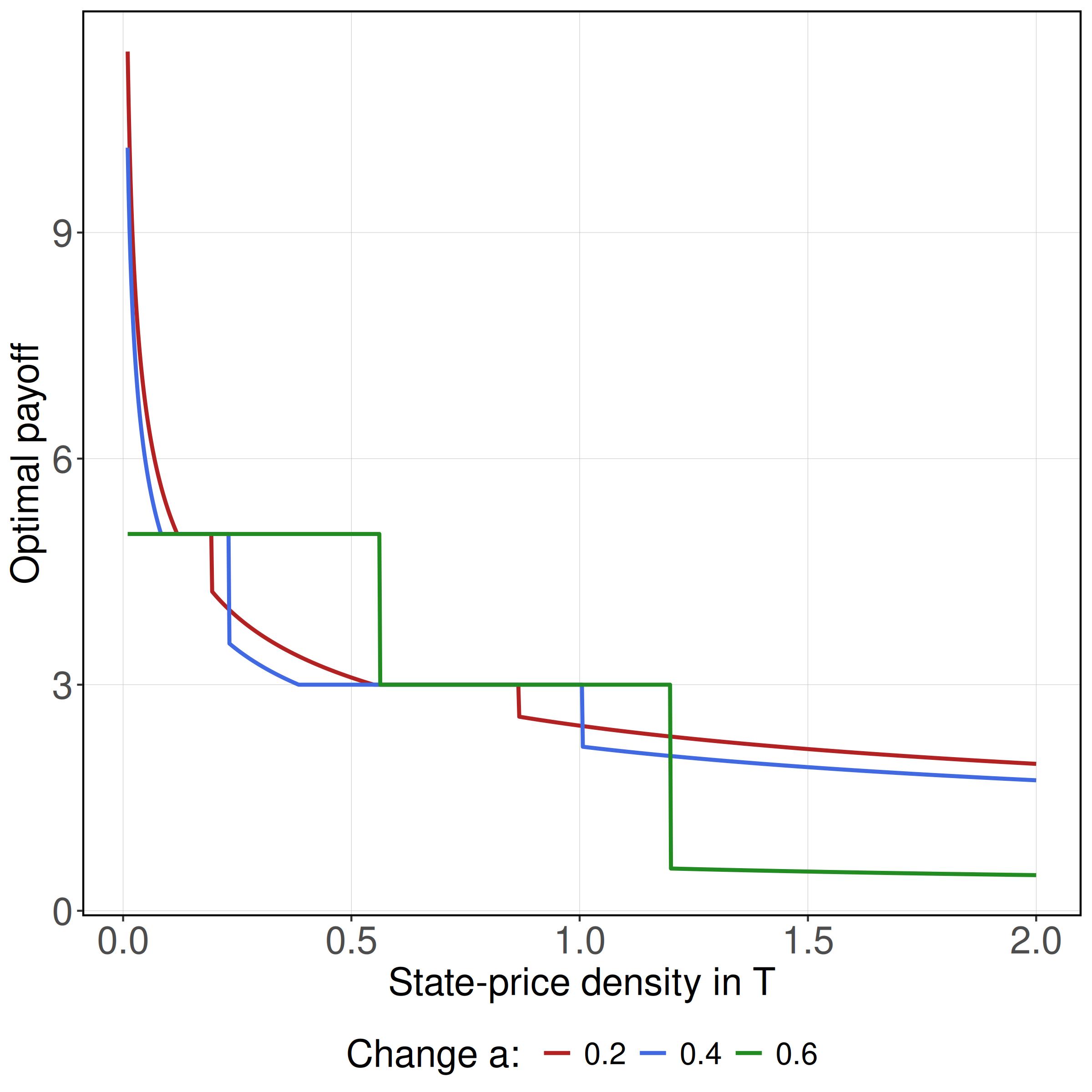}
\caption{Change $a$}
\end{subfigure}
\hspace{1cm}
\begin{subfigure}{0.425\textwidth}
\centering
\includegraphics[width=\linewidth]{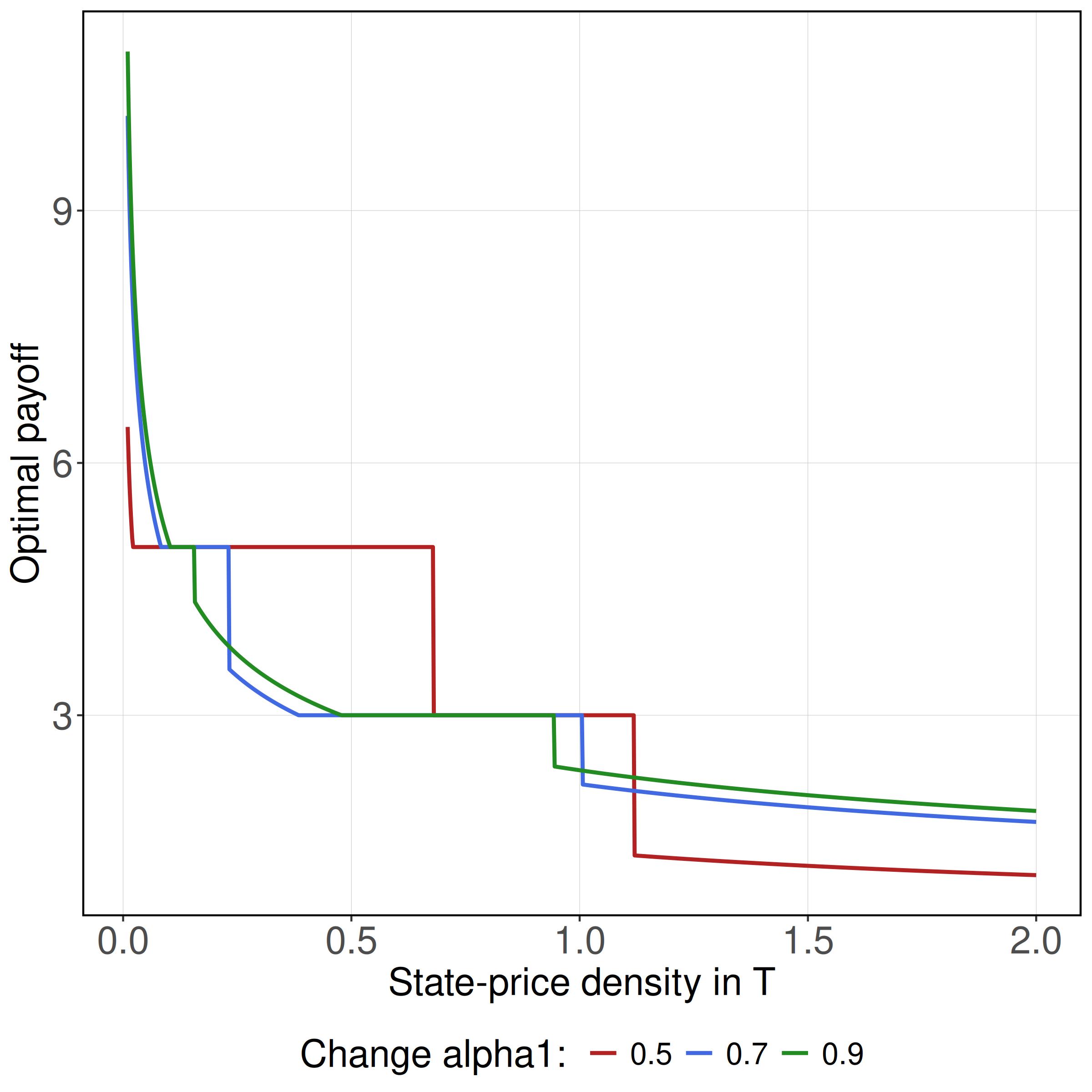}
\caption{Change $\alpha_1$}
\end{subfigure}

\vspace{0.2cm}

\begin{subfigure}{0.425\textwidth}
\centering
\includegraphics[width=\linewidth]{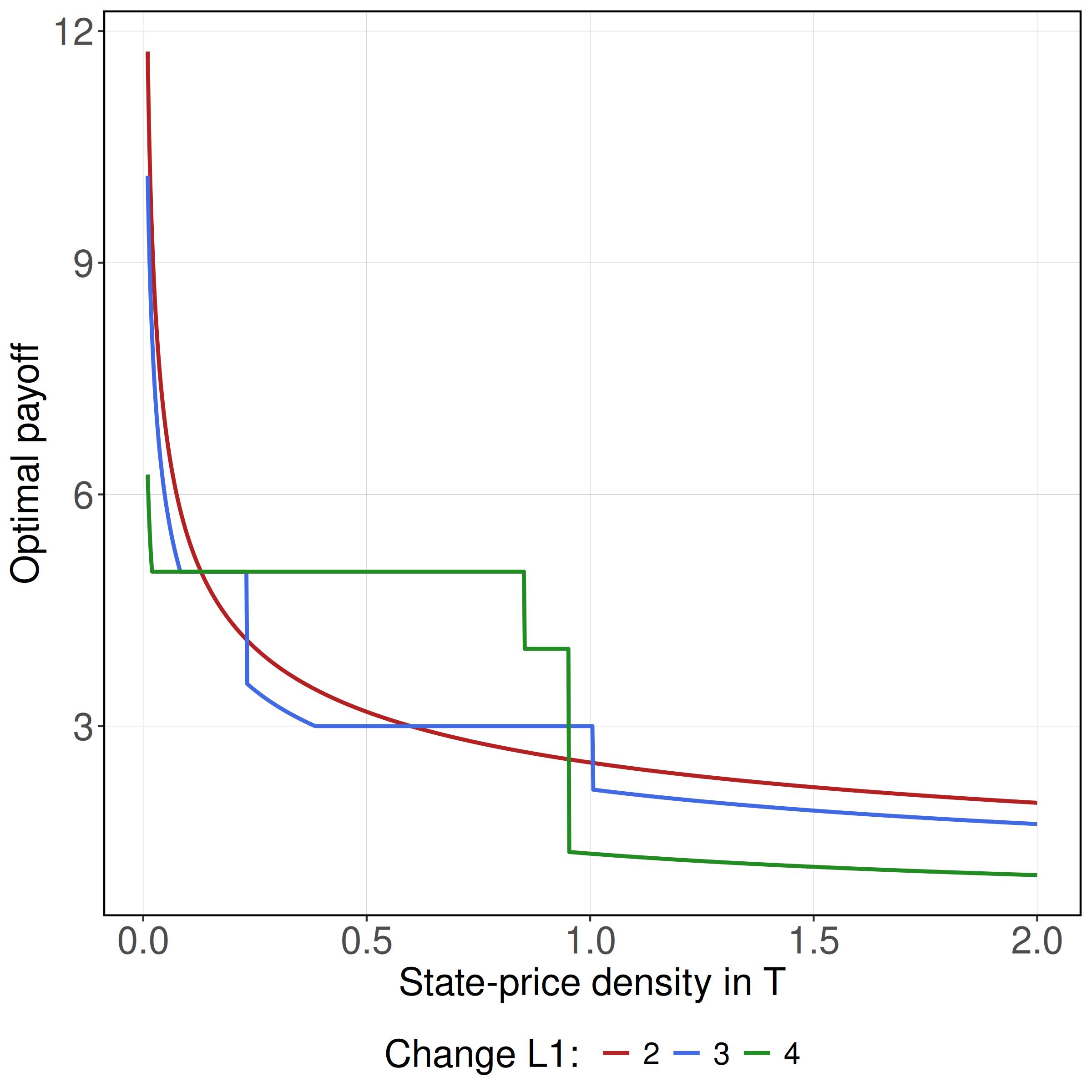}
\caption{Change $L_1$}
\end{subfigure}
\hspace{1cm}
\begin{subfigure}{0.425\textwidth}
\centering
\includegraphics[width=\linewidth]{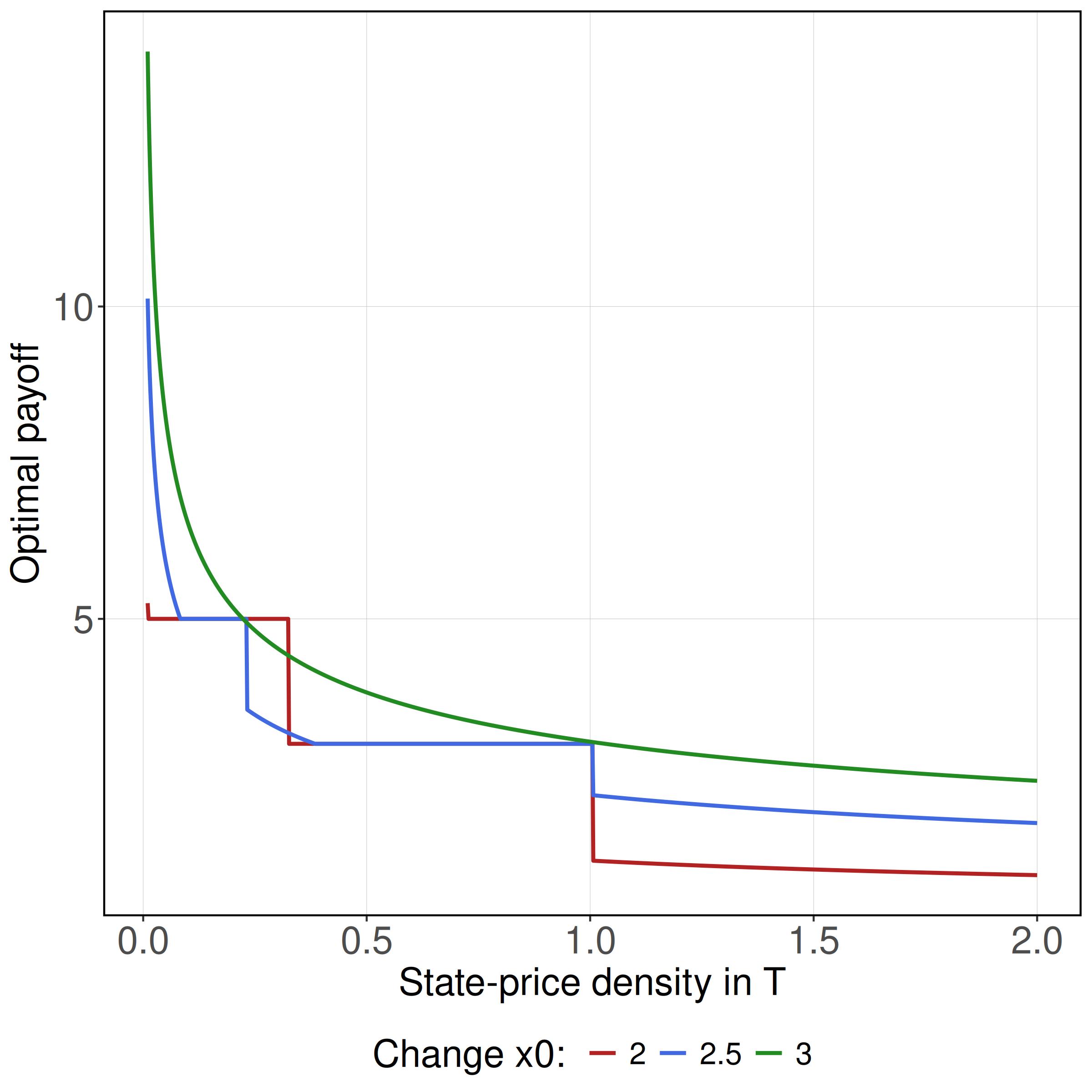}
\caption{Change $x_0$}
\end{subfigure}

\caption{Optimal payoff $X_T^{\star}$ depending on the state-price density
$\xi_T$ by varying parameter values individually.}
\label{fig:optimal_payoff}
\end{figure}

We begin by considering a single investor with risk aversion 
$\gamma:=\gamma_1=3$. This removes the redistribution channel from the model and allows
us to focus on the differences between the Merton payoff and the optimal payoff under the average compliance constraint. Figure~\ref{fig:optimal_payoff} shows the optimal terminal
payoff $X_T^\star$ as a function of the state-price density $\xi_T$ when one
parameter is varied at a time.

Several common features are visible across the panels. First, the optimal 
payoff remains decreasing in the state-price density. Hence, wealth is high in
good states and low in bad states, as it holds for the Merton payoff.
Second, the average compliance constraint introduces flat regions at the regulatory thresholds
$L_1$ and~$L_2$, recall also the cutoff 
structures derived in Section~\ref{subsec:two-threshold}. In those regions, the
Merton payoff would decrease slightly below a regulatory threshold and instead, it is optimal to increase the terminal wealth exactly to that threshold in
order to increase the average compliance probability. In other states, the payoff follows a
Merton-type branch. The optimal payoff therefore combines utility smoothing
with targeted threshold protection.

Panel~(a) varies the required compliance level $a$. For the low value
$a=0.2$, the constraint is only mildly binding and the optimal payoff remains
close to the Merton payoff. For the baseline value~$a=0.4$, the average compliance constraint
yields more visible plateaus at $L_1$ and $L_2$. Increasing the
required compliance level further to $a=0.6$ changes the solution
substantially. The payoff is then mainly designed to satisfy the average compliance constraint, and even in relatively good states the payoff does not exceed the
upper threshold $L_2$ over a large range of state prices. Thus, the
manager sacrifices upside potential in order to finance a larger
probability mass at the protected levels.

The location of the cutoffs is important for the previous interpretation, because, by the state-price density being lognormally distributed, not all parts of the x-axis carry the same probability mass. For our parameters,
$\mathbb{P}(\xi_T\leq 0.5)$ is very small, whereas
$\mathbb{P}(\xi_T\leq 1)$ is above one half. Hence, movements of cutoffs in
regions around typical values of $\xi_T$ have a much larger effect on the
average compliance probability than movements in far-left regions. This explains why the plateau related to $q^{\star\star}/\lambda$ is
often economically more important than changes in the very-good-state region
associated with $q^\star/\lambda$.

Panel~(b) varies the weight $\alpha_1$ attached to the lower threshold. A
smaller value of $\alpha_1$ increases the relative importance of the upper
threshold $L_2$. Hence, the optimizer expands the region in which
$X_T^\star=L_2$, so that the event $\{X_T^\star\geq L_2\}$ contributes more
strongly to satisfying the average compliance constraint. Conversely, when
$\alpha_1$ is large, i.e.,~the lower threshold $L_1$ receives more attention, then it is more attractive to generate compliance by protecting the lower
threshold over a wider set of states rather than by financing the more costly
upper threshold. Thus, the weights $\alpha_1$ and $\alpha_2$ determine not only the
severity of regulation, but also which threshold shapes the payoff most
strongly.

Panel~(c) changes $L_1$. This has a
large effect on the optimal payoff. When $L_1$ is reduced, the average compliance constraint
becomes easier to satisfy and the constrained solution can coincide with the
Merton solution. When $L_1$ is increased, the plateau at the lower threshold
moves upward and absorbs more resources, which reduces the payoff in other
states. In Appendix~\ref{sec:auxiliaryFiguresTables}, we demonstrate that changing $L_2$ has only limited effects, i.e.,~the manager therefore satisfies much of the
requirement through the lower threshold. This illustrates the following important feature of the average compliance constraint: the effective severity of a threshold
depends jointly on its level, its weight, and the probability mass of the
states in which it can be reached cost efficiently.

Panel~(d) varies the initial capital $x_0$. The effect is monotone and
economically intuitive. With less initial capital, the manager has fewer
resources to finance compliance, and the optimal payoff becomes more
distorted relative to the Merton payoff. In particular, when $x_0=2$, the
payoff is much more concentrated around the regulatory thresholds. In contrast,
when $x_0=3$, the initial capital is sufficiently large such that the Merton payoff satisfies the average compliance constraint.

In addition, in Appendix~\ref{sec:auxiliaryFiguresTables}, we also present the optimal payoff under changes in the risk aversion parameter $\gamma$. 

\begin{table}[ht]
\centering
\begin{tabular}{lrrr}
  \hline
\textbf{Parameter} & \multicolumn{1}{c}{\textbf{$CE^M$}} & \multicolumn{1}{c}{\textbf{$CE^{\star}$}} & \multicolumn{1}{c}{\textbf{$rc$}} \\ 
  \hline
  $a = 0.2$ & 2.5423 & 2.5325 & 0.38\% \\ 
  $a = 0.4$ & 2.5423 & 2.4731 & 2.72\% \\ 
  $a= 0.6$ & 2.5423 & 1.3054 & \textbf{48.65}\% \\ 
   \hline
  $\alpha_1= 0.5$ & 2.5423 & 2.0994 & \textbf{17.42}\% \\ 
  $\alpha_1 = 0.7$ & 2.5423 & 2.4731 & 2.72\% \\ 
  $\alpha_1 = 0.9$ & 2.5423 & 2.5116 & 1.21\% \\ 
   \hline
  $L_1 = 2.0$ & 2.5423 & 2.5423 & 0.00\% \\ 
  $L_1 = 3.0$ & 2.5423 & 2.4731 & 2.72\% \\ 
  $L_1 = 4.0$ & 2.5423 & 1.7021 & \textbf{33.05}\% \\
  \hline
  $x_0 = 2.0$ & 2.0335 & 1.5154 & \textbf{25.47}\% \\ 
  $x_0 = 2.5$ & 2.5423 & 2.4731 & 2.72\% \\ 
  $x_0 = 3.0$ & 3.0505 & 3.0505 & 0.00\% \\
   \hline
\end{tabular}
\caption{Certainty equivalents (CE) and resulting welfare loss $rc$.
The numbers marked in bold are the cases in which the welfare loss is
larger than $10\%$.} 
\label{tab:certainty_equivalents}
\end{table}

Further, as shown in Table~\ref{tab:certainty_equivalents}, we can compare the certainty equivalent of the constrained
solution with the certainty equivalent of the Merton
solution. The last column reports the relative difference between these certainty equivalents, which describes the welfare loss induced by the
regulatory uncertainty. To be precise, let us denote the certainty equivalents by 
\begin{align}\label{eq:certainty_equivalent}
    CE^b
    :=
    U^{-1}\left(\mathbb E[U(X_{T}^b)]\right),
\end{align}
where $b=M$ refers to the Merton payoff and $b=\star$ to the optimal payoff subject to the average compliance constraint. Note that in the one investor case, it holds that $CE^b = \mathbb{E}\big[(X_T^b)^{1-\gamma}\big]^{\frac{1}{1-\gamma}}$.
The (regulatory) cost by introducing the average compliance constraint is then given by $RC
    :=
    CE^M-CE^\star$, and the corresponding relative (regulatory) cost, also called welfare loss, is
\begin{align}\label{eq:certainty_equivalent_2}
    rc
    :=
    \frac{CE^M-CE^\star}{CE^M}.
\end{align}
In the case of only one investor, it holds that 
\[
rc = 1 - \Bigg(\frac{\mathbb{E}\big[(X_T^\star)^{(1-\gamma)}\big]}{\mathbb{E}\big[(X_T^M)^{(1-\gamma)}\big]}\Bigg)^{\frac{1}{1-\gamma}}.
\]

Table~\ref{tab:certainty_equivalents} confirms that our average compliance constraint can be severe, but only in
specific parameter constellations. For our initial parameter setting, the welfare loss is
$2.72\%$, which indicates that the average compliance constraint changes the shape of the optimal
payoff but does not impose a large utility cost. Similarly, reducing the
compliance requirement to $a=0.2$ leads to a negligible welfare loss of
$0.38\%$, and increasing the weight of the lower threshold to $\alpha_1=0.9$
reduces the welfare loss to $1.21\%$. In these cases, compliance can be
achieved relatively cheaply by reallocating wealth around the lower threshold.

The most pronounced losses occur when the average compliance constraint requires protection that
is difficult to finance. Increasing the compliance level to $a=0.6$ reduces the
certainty equivalent from $2.54$ to~$1.31$, corresponding to a welfare loss of
$48.65\%$. Increasing the lower threshold to $L_1=4$ leads to a welfare loss of
$33.05\%$. Reducing the initial capital to $x_0=2$ also has a large effect, with a
loss of $25.47\%$. These cases share the same mechanism: the manager
must allocate a substantial amount of wealth to threshold regions,
which leaves fewer resources for utility smoothing and for upside exposure. 

In contrast to the relatively
low weight and high cost of satisfying $L_2$, the lower threshold $L_1$ becomes the dominant driver
to satisfy the average compliance constraint. Table~\ref{tab:certainty_equivalents} therefore reinforces the graphical
evidence from Figure~\ref{fig:optimal_payoff}: regulatory uncertainty matters
most when probability mass has to be moved toward a costly and highly weighted threshold.

Overall, the numerical results show that the average compliance constraint
has two distinct effects. It changes the qualitative shape of the optimal
payoff by introducing threshold plateaus, and it may reduce welfare by forcing
wealth into regulatory compliance regions. However, the welfare loss is not
uniformly large. In many parameter constellations, the payoff shape changes
visibly while the welfare loss remains economically moderate.
Large welfare losses arise mainly when the compliance level is high, the lower threshold is high, or the available initial capital is low. An important consequence is the following: while many parameters cannot be changed by the manager -- like~the compliance level~$a$, the initial capital $x_0$ or the thresholds $L_1$ and $L_2$ -- the weight $\alpha_1$ has to be chosen carefully by the manager. Our study implies that a manager should be aware that smaller values of $\alpha_1$ may increase the welfare loss significantly.

\subsection{Impact of average compliance constraint and collective investment}
\label{sec:impactCollectiveInvestment}

After studying the effect of the average compliance constraint for only one investor, we now turn to the collective framework. The purpose of this subsection
is to analyze in which way the same average compliance constraint affects different
investors when they have heterogeneous risk aversions. Moreover, we test different Pareto weights, that is, applying different Pareto aggregators as introduced in
Section~\ref{sec:collectiveUtilities}. 

The collective setting adds a redistribution channel that is absent in the
single investor case, while the average compliance constraint is imposed on aggregate
terminal wealth. It will turn out that the average compliance constraint can affect investors in very different ways. We aim to clarify such effects in dependence of the risk aversion parameters $\gamma_i$ and the Pareto weights $\beta_i$. To do so, we use the same parameter values as in the previous section, beside the value of $\alpha_1$. We adjust it from $0.7$ to $0.5$. In this situation, the differences between the certainty equivalents of the Merton payoff and the optimal payoff subject to the average compliance constraint are more pronounced, recall also the values in Table~\ref{tab:certainty_equivalents}. Moreover, we assume $n=4$ investors.

Overall statistics are collected in Table~\ref{tab:certainty_equivalents_collective}, analogously to the ones in Table~\ref{tab:certainty_equivalents}. In total, we combine four possible choices of the risk aversion parameters and three possible choices of Pareto weights. Note that, risk aversion scenario 1 refers to the case of equal risk aversions for all four investors, which means that also the Pareto aggregator is a power utility function and hence, the choice of the Pareto weights is inconsequential. Therefore, we use risk aversion scenario 1 as baseline scenario and by a comparison with the other risk aversion scenarios, we can measure the impact of the collective investment under the average compliance constraint. The corresponding statistic is denoted by $\Delta_{\text{Hom}}$, measuring the relative difference between the certainty equivalents of the optimal payoffs~($CE^{\star}$).

Here, we see that only in risk aversion scenario 4 -- three investors are significantly risk-averse, i.e.,~risk aversion is $4.5$ -- the collective investment leads to a reduction of the $CE^{\star}$-values. On the opposite, if the majority of investors are less risk-averse (risk aversion scenario 3), then we can observe a positive effect of collective investment, i.e.,~the value of certainty equivalent increases compared to risk aversion scenario 1. As a  consequence, the gap towards the Merton solution, measured by the $rc$-value, is smaller compared to the remaining risk aversion scenarios. The largest gap is obtained for risk aversion scenario 4.

These observations hold for all choices of Pareto weights. However, the effects are either weakened or strengthened in an economically intuitive manner, depending on the allocation of the Pareto weights. 

\begin{table}[ht]
\centering
\begin{tabular}{l|rrrr}
  \hline\hline
     \multicolumn{5}{c}{\textbf{Equal weights ($\beta_1 = \beta_2 = \beta_3 = \beta_4 = 0.25$)}}\\
    \hline\hline
\textbf{Risk aversion scenario} &  \multicolumn{1}{c}{\textbf{$CE^M$}} & \multicolumn{1}{c}{\textbf{$CE^{\star}$}} & \multicolumn{1}{c}{\textbf{$rc$}} & \multicolumn{1}{c}{\textbf{$\Delta_{\text{Hom}}$}} \\ 
 \hline
1\quad $(\gamma_1 = \gamma_2 = \gamma_3 = \gamma_4 = 3)$ &  2.5420 & 2.0993 & 17.42\% & 0.00\% \\ 
  2\quad $(\gamma_1 = 1.5,\gamma_2 = \gamma_3 = 3,\gamma_4 = 4.5)$ & 2.5433 & 2.1167 & 16.77\% & 0.83\%\\ 
  3\quad $(\gamma_1 = \gamma_2 = \gamma_3 = 1.5,\gamma_4 = 4.5)$ & 2.5516 & 2.3041 & 9.70\% & 9.76\% \\ 
  4\quad $(\gamma_1 = 1.5, \gamma_2 = \gamma_3 = \gamma_4 = 4.5)$ & 2.5397 & 1.9672 & 22.54\% & -6.29\% \\ 
  \hline\hline
     \multicolumn{5}{c}{\textbf{One large investment ($\beta_1 = 0.7,\beta_2 = \beta_3 = \beta_4 = 0.1$)}}\\
    \hline\hline
    \textbf{Risk aversion scenario} &  \multicolumn{1}{c}{\textbf{$CE^M$}} & \multicolumn{1}{c}{\textbf{$CE^{\star}$}} & \multicolumn{1}{c}{\textbf{$rc$}} & \multicolumn{1}{c}{\textbf{$\Delta_{\text{Hom}}$}} \\ 
 \hline
    1\quad $(\gamma_1 = \gamma_2 = \gamma_3 = \gamma_4 = 3)$ & 2.5420 & 2.0994 & 17.41\% & 0.00\% \\ 
  2\quad $(\gamma_1 = 1.5,\gamma_2 = \gamma_3 = 3,\gamma_4 = 4.5)$ & 2.5467 & 2.2073 & 13.33\% & 5.14\%\\ 
  3\quad $(\gamma_1 = \gamma_2 = \gamma_3 = 1.5,\gamma_4 = 4.5)$ & 2.5529 & 2.3255 & 8.91\% & 10.77\% \\ 
  4\quad $(\gamma_1 = 1.5, \gamma_2 = \gamma_3 = \gamma_4 = 4.5)$ & 2.5432 & 2.0869 & 17.94\% & -0.59\% \\ 
  \hline\hline
     \multicolumn{5}{c}{\textbf{Two medium and two small investments ($\beta_1 = \beta_2 = 0.4,\beta_3 = \beta_4 = 0.1$)}}\\
    \hline\hline
    \textbf{Risk aversion scenario} &  \multicolumn{1}{c}{\textbf{$CE^M$}} & \multicolumn{1}{c}{\textbf{$CE^{\star}$}} & \multicolumn{1}{c}{\textbf{$rc$}} & \multicolumn{1}{c}{\textbf{$\Delta_{\text{Hom}}$}}\\ 
 \hline
    1\quad $(\gamma_1 = \gamma_2 = \gamma_3 = \gamma_4 = 3)$ & 2.5420 & 2.0994 & 17.41\% & 0.00\% \\ 
  2\quad $(\gamma_1 = 1.5,\gamma_2 = \gamma_3 = 3,\gamma_4 = 4.5)$ & 2.5448 & 2.1596 & 15.14\% & 2.87\% \\ 
  3\quad $(\gamma_1 = \gamma_2 = \gamma_3 = 1.5,\gamma_4 = 4.5)$ & 2.5530 & 2.3274 & 8.83\% & 10.86\% \\ 
  4\quad $(\gamma_1 = 1.5, \gamma_2 = \gamma_3 = \gamma_4 = 4.5)$ & 2.5411 & 2.0174 & 20.61\% & -3.91\% \\ 
   \hline
\end{tabular}
\caption{Certainty equivalents (CE) and resulting welfare loss $rc$ in case of collective investments with four insurers. We use the risk aversion scenario 1 as baseline scenario. The relative change to the homogeneous-preference benchmark~$\Delta_{\text{Hom}}$ is the relative difference between the certainty equivalents of the optimal payoffs for the risk aversion scenario of interest and risk aversion scenario 1.} 
\label{tab:certainty_equivalents_collective}
\end{table}

To quantify how the average compliance constraint affects different members of the collective, we measure welfare losses at the individual level  analogously to the formulas in~\eqref{eq:certainty_equivalent} and~\eqref{eq:certainty_equivalent_2}. Recall that \(X_T^M\) denotes the unconstrained collective Merton payoff and \(X_T^\star\) denotes the optimal payoff under the average compliance constraint. Then, for each benchmark ($b=M$ or $b=\star$), the individual payoff of investor \(i\) is recovered from the Pareto sharing rule,
\[
    X_{T,i}^b
    =
    I_i\left(\frac{U'(X_T^b)}{\beta_i}\right),
\]
where \(I_i=(U_i^{\prime})^{-1}\). The individual certainty equivalent under benchmark \(b\) is denoted by
\begin{align}\label{eq:certainty_equivalent_per_insurer}
    CE_i^b
    :=
    U_i^{-1}\Big(\mathbb E\big[U_i\big(X_{T,i}^b\big)\big]\Big) = \beta_i^{1/\gamma_i}\mathbb{E}\Big[U^{\prime}\big(X_T^b\big)^{(1-1/\gamma_i)}\Big]^{\frac{1}{1-\gamma_i}}.
\end{align}
We then define the individual regulatory cost by $RC_i
    :=
    CE_i^M-CE_i^\star$, and the corresponding relative regulatory cost by
\begin{align}\label{eq:regulatory_costs}
    rc_i
    :=
    \frac{RC_i}{CE_i^M}
    =
    1-\frac{CE_i^\star}{CE_i^M} = 1 - \Bigg(\frac{\mathbb{E}\big[U^{\prime}(X_T^\star)^{(1-1/\gamma_i)}\big]}{\mathbb{E}\big[U^{\prime}(X_T^M)^{(1-1/\gamma_i)}\big]}\Bigg)^{\frac{1}{1-\gamma_i}}.
\end{align}
The relative cost \(rc_i\) is our main measure of the welfare burden borne by investor \(i\). If \(rc_i\) is larger for investors with high \(\gamma_i\), then regulatory costs are mainly borne by the more risk-averse investors. If \(rc_i\) is larger for investors with low \(\gamma_i\), then regulatory costs are mainly borne by the less risk-averse investors.

Moreover, to describe the redistribution of regulatory costs within the collective, we compute each investor's share of the total regulatory cost,
$   s_i
    :=
    RC_i /\sum\limits_{k=1}^n RC_k.$ 
This cost share can be compared with the investor's initial wealth share by assuming that $\beta_i = x_{0,i} / x_0$. The difference $ d_i:=s_i-\beta_i$
measures whether investor \(i\) bears more or less than a proportional share of the total regulatory cost. We call it the excess of regulatory costs. If \(d_i>0\), investor \(i\) bears more than a proportional share; if~\(d_i<0\), investor \(i\) bears less than a proportional share.

In addition, for a more aggregate comparison, we divide the four investors into a less risk-averse group and a more risk-averse group. To do so, we define $G_L:=\{i:\gamma_i\leq \gamma_{\mathrm{med}}\}$ and $G_H:=\{i:\gamma_i> \gamma_{\mathrm{med}}\}$, where \(\gamma_{\mathrm{med}}\) denotes the median risk-aversion parameter in the pool, for which we choose $\gamma_{\mathrm{med}}=3$. The corresponding excess of regulatory costs are $d_L:=\sum_{i\in G_L}d_i$ and $d_H:=\sum_{i\in G_H}d_i$. If~\(d_L>0\) (resp.~\(d_H>0\)), then the less (resp.~more) risk-averse investors bear more than their proportional share of the regulatory costs.

We collect this information in Table~\ref{tab:certainty_equivalents_collective_agents}. Let us start with the case of equal Pareto weights. Here we see that for all risk aversion scenarios the values of the $d_i$'s do not differ significantly from zero. Indeed, the largest difference occurs for investor $1$ for risk aversion scenario 4 with only $-5.14\%$. This changes drastically for the other constellations of Pareto weights, which means that unequal Pareto weights have a more significant impact and depending on the risk aversion scenario, it can lead to unfairly distributed values of $d_i$ among the investors. The largest difference is observed when the first investor has Pareto weight $\beta_1 = 0.7$ and all other investors have risk aversion $4.5$. Then, $d_1 = -31.25\%$, i.e.,~the large Pareto weight $\beta_1$ puts obviously the focus to investor 1, who benefits from the situation by obtaining -- in relation to its invested proportion -- a significant smaller amount of the regulatory costs (negative value of $d_1$). A similar value of $d_1$ is observed in risk aversion scenario~1. The strong influence of the Pareto weights is also visible by noting that we have $d_1 = -12.11\%$, when the Pareto weight $\beta_1$ is reduced to $0.4$ in case of two medium and two small investments and risk aversion scenario~4.

\begin{landscape}

\begin{table}[ht]
\centering
\begin{tabular}{c|rrr|rrr|rrr|rrr}
  \hline
  & \multicolumn{3}{c|}{\textbf{Risk aversion scenario 1}} & \multicolumn{3}{c|}{\textbf{Risk aversion scenario 2}} & \multicolumn{3}{c|}{\textbf{Risk aversion scenario 3}} & \multicolumn{3}{c}{\textbf{Risk aversion scenario 4}}\\
  & \multicolumn{3}{c|}{$(\gamma_1 = \gamma_2 = \gamma_3 = \gamma_4 = 3)$} & \multicolumn{3}{c|}{$(\gamma_1 = 1.5,\gamma_2 = \gamma_3 = 3,\gamma_4 = 4.5)$} & \multicolumn{3}{c|}{$(\gamma_1 = \gamma_2 = \gamma_3 = 1.5,\gamma_4 = 4.5)$} & \multicolumn{3}{c}{$(\gamma_1 = 1.5, \gamma_2 = \gamma_3 = \gamma_4 = 4.5)$}\\
  \hline\hline
    & \multicolumn{12}{c}{\textbf{Equal Pareto weights ($\beta_1 = \beta_2 = \beta_3 = \beta_4 = 0.25$)}}\\
    \hline\hline
Insurer & \multicolumn{1}{c}{$CE_i^{\star}$} & \multicolumn{1}{c}{$rc_i$} & \multicolumn{1}{c|}{$d_i$} & \multicolumn{1}{c}{$CE_i^{\star}$} & \multicolumn{1}{c}{$rc_i$} & \multicolumn{1}{c|}{$d_i$} & \multicolumn{1}{c}{$CE_i^{\star}$} & \multicolumn{1}{c}{$rc_i$} & \multicolumn{1}{c|}{$d_i$} & \multicolumn{1}{c}{$CE_i^{\star}$} & \multicolumn{1}{c}{$rc_i$} & \multicolumn{1}{c}{$d_i$} \\ 
  \hline
 1 & 0.5248 & 17.42\% & 0.00\% & 0.3608 & 19.22\% & -4.56\% & 0.5116 & \textbf{10.88\%} & 0.45\% & 0.2662 & \textbf{29.38\%} & \textbf{-5.14\%} \\ 
  2 & 0.5248 & 17.42\% & 0.00\% & 0.5513 & 17.33\% & 2.50\% & 0.5116 & \textbf{10.88\%} & 0.45\% & 0.5720 & \textbf{20.67\%} & 1.71\% \\ 
  3 & 0.5248 & 17.42\% & 0.00\% & 0.5513 & 17.33\% & 2.50\% & 0.5116 & \textbf{10.88\%} & 0.45\% & 0.5720 & \textbf{20.67\%} & 1.71\% \\ 
  4 & 0.5248 & 17.42\% & 0.00\% & 0.6597 & 13.52\% & -0.44\% & 0.7714 & \textbf{7.00\%} & -1.36\% & 0.5720 & \textbf{20.67\%} & 1.71\% \\ 
  \hline
  L & 2.0993 & 69.67\% & 0.00\% & 1.4634 & 53.87\% & 0.44\% & 1.5347 & 32.65\% & 1.36\% & 0.2662 & 29.38\% & -5.14\% \\ 
  H & 0.0000 & 0.00\% & 0.00\% & 0.6597 & 13.52\% & -0.44\% & 0.7714 & 7.00\% & -1.36\% & 1.7158 & 62.00\% & 5.14\% \\ 
   \hline\hline
   & \multicolumn{12}{c}{\textbf{One large investment ($\beta_1 = 0.7,\beta_2 = \beta_3 = \beta_4 = 0.1$)}}\\
    \hline\hline
Insurer & \multicolumn{1}{c}{$CE_i^{\star}$} & \multicolumn{1}{c}{$rc_i$} & \multicolumn{1}{c|}{$d_i$} & \multicolumn{1}{c}{$CE_i^{\star}$} & \multicolumn{1}{c}{$rc_i$} & \multicolumn{1}{c|}{$d_i$} & \multicolumn{1}{c}{$CE_i^{\star}$} & \multicolumn{1}{c}{$rc_i$} & \multicolumn{1}{c|}{$d_i$} & \multicolumn{1}{c}{$CE_i^{\star}$} & \multicolumn{1}{c}{$rc_i$} & \multicolumn{1}{c}{$d_i$} \\  
    \hline
     1 & 0.8174 & 17.41\% & \textbf{-31.06\%} & 0.7794 & 15.09\% & -28.58\% & 1.0869 & 9.81\% & -17.69\% & 0.5937 & \textbf{22.45\%} & \textbf{-31.25\%} \\ 
  2 & 0.4273 & 17.41\% & 10.35\% & 0.4339 & 13.17\% & 9.68\% & 0.2970 & 9.81\% & 4.29\% & 0.5019 & 15.28\% & 10.42\% \\ 
  3 & 0.4273 & 17.41\% & 10.35\% & 0.4339 & 13.17\% & 9.68\% & 0.2970 & 9.81\% & 4.29\% & 0.5019 & 15.28\% & 10.42\% \\ 
  4 & 0.4273 & 17.41\% & 10.35\% & 0.5652 & 10.21\% & 9.22\% & 0.6460 & 6.26\% & 9.10\% & 0.5019 & 15.28\% & 10.42\% \\ 
  \hline
  L & 2.0994 & 69.66\% & 0.00\% & 1.6472 & 41.43\% & -9.22\% & 1.6809 & 29.43\% & -9.10\% & 0.5937 & 22.45\% & \textbf{-31.25\%} \\ 
  H & 0.0000 & 0.00\% & 0.00\% & 0.5652 & 10.21\% & 9.22\% & 0.6460 & 6.26\% & 9.10\% & 1.5057 & 45.87\% & \textbf{31.25\%} \\  
   \hline\hline
   & \multicolumn{12}{c}{\textbf{Two medium and two small investments ($\beta_1 = \beta_2 = 0.4,\beta_3 = \beta_4 = 0.1$)}}\\
    \hline\hline
Insurer & \multicolumn{1}{c}{$CE_i^{\star}$} & \multicolumn{1}{c}{$rc_i$} & \multicolumn{1}{c|}{$d_i$} & \multicolumn{1}{c}{$CE_i^{\star}$} & \multicolumn{1}{c}{$rc_i$} & \multicolumn{1}{c|}{$d_i$} & \multicolumn{1}{c}{$CE_i^{\star}$} & \multicolumn{1}{c}{$rc_i$} & \multicolumn{1}{c|}{$d_i$} & \multicolumn{1}{c}{$CE_i^{\star}$} & \multicolumn{1}{c}{$rc_i$} & \multicolumn{1}{c}{$d_i$} \\ 
    \hline
   1 & 0.6440 & 17.41\% & -9.32\% & 0.5202 & 17.17\% & -11.57\% & 0.7070 & 9.71\% & -6.07\% & 0.3954 & \textbf{26.41\%} & \textbf{-12.11\%} \\ 
  2 & 0.6440 & 17.41\% & -9.32\% & 0.6700 & 15.27\% & -8.17\% & 0.7070 & 9.71\% & -6.07\% & 0.6628 & 18.31\% & \textbf{-10.80\%} \\ 
  3 & 0.4057 & 17.41\% & 9.32\% & 0.4220 & 15.27\% & 10.05\% & 0.2806 & 9.71\% & 3.46\% & 0.4871 & 18.31\% & \textbf{11.46\%} \\ 
  4 & 0.4057 & 17.41\% & 9.32\% & 0.5534 & 11.88\% & 9.68\% & 0.6341 & 6.20\% & 8.68\% & 0.4871 & 18.31\% & \textbf{11.46\%} \\ 
  \hline
  L & 2.0994 & 69.65\% & -0.00\% & 1.6121 & 47.71\% & -9.68\% & 1.6947 & 29.14\% & -8.68\% & 0.3954 & 26.41\% & -12.11\% \\ 
  H & 0.0000 & 0.00\% & 0.00\% & 0.5534 & 11.88\% & 9.68\% & 0.6341 & 6.20\% & 8.68\% & 1.6369 & 54.93\% & 12.11\% \\ 
   \hline
\end{tabular}
\caption{Certainty equivalents $CE_i^{\star}$, relative regulatory costs $rc_i$ and the relative excess of regulatory costs $d_i$ of four insurers and groups of insurers (``L'' for group of insurers with risk aversion less or equal $3$ and ``H'' for group of the remaining insurers) for varying 
Pareto weights $\beta_i$ and risk aversion parameters~$\gamma_i$. Values marked in bold are discussed in detail in the main text.} 
\label{tab:certainty_equivalents_collective_agents}
\end{table}

\end{landscape}

Concluding, in terms of the criterion $d_i$, an unfair situation arises when the market consists of a market leader admitting a risk aversion being smaller than the ones of all other investors. This is also reflected by an increase in the relative regulatory costs $rc_1$ from $22.45\%$ (one large investment) to $26.41\%$ (two medium and two small investments). Moreover, note that it always holds that $d_L = -d_H$. If this value is large, then there is a significant imbalance between the groups of investors with small and large risk aversions, see e.g.,~the value of $d_L$ for risk aversion scenario 4 in case of one large investor~($\beta_1 = 0.7$).

In terms of the relative regulatory costs $rc_i$, different risk aversion scenarios have a more significant influence than by considering $d_i$, even in case of equal weights: compared with risk aversion scenario~1, risk aversion scenario~3 (resp.~4) leads to significant smaller (resp.~larger) $rc_i$-values. Hence, if there are many investors being significantly risk-averse, then this increases the costs by introducing the average compliance  constraint. The threshold plateaus induced by the average compliance constraint can therefore generate non-uniform welfare effects across the members of the pool.

Finally, to study for which agent entering the pool is beneficial, we compare the certainty equivalents from collective investment with the corresponding stand-alone benchmark. In the stand-alone case, investor \(i\) acts separately with initial wealth \(x_{0,i}\). To make the average compliance constraint comparable with the one in the collective setting, we scale the thresholds proportionally by the initial wealth share, i.e.,~we adjust the thresholds as follows
\[
    L_{j}^{i}:=\frac{x_{0,i}}{x_0}L_j,
    \quad j=1,\ldots,J.
\]

Let \(CE_i^{SA,\star}\) denote the certainty equivalent of investor \(i\) obtained by solving the stand-alone constrained portfolio optimization problem. Then, in Table~\ref{tab:certainty_equivalents_diffs_agents}, we state the following value for investor~$i$:
\[
    \Delta_i = \frac{CE_i^{\star}-CE_{i}^{SA,\star}}{CE_{i}^{SA,\star}}.
\]
Those values describe if entering the collective investment is beneficial for investor $i$ under the criterion of an increase in the certainty equivalent, i.e,~if $\Delta_i>0$ (resp.~$\Delta_i<0$), then investor $i$ should (resp.~should not) enter the pool.

\begin{table}[ht]
\centering
\begin{tabular}{c|rrrr}
  \hline
  \textbf{Insurer} & \multicolumn{4}{c}{\textbf{Risk aversion scenario}} \\
  & \textbf{1} & \textbf{2} & \textbf{3} & \textbf{4} \\
  \hline\hline
    & \multicolumn{4}{c}{\textbf{Pareto weights: $\beta_1 = \beta_2 = \beta_3 = \beta_4 = 0.25$}}\\
    \hline\hline
    1 & \textbf{0.00\%} & -39.97\% & -14.89\% & -55.71\% \\ 
   2 & \textbf{0.00\%} & 5.05\% & -14.89\% & 23.05\% \\ 
   3 & \textbf{0.00\%} & 5.05\% & -14.89\% & 23.05\% \\ 
   4 & \textbf{0.00\%} & 41.93\% & 65.96\% & 23.05\% \\ 
   \hline\hline
   & \multicolumn{4}{c}{\textbf{Pareto weights: $\beta_1 = 0.7,\beta_2 = \beta_3 = \beta_4 = 0.1$}}\\
    \hline\hline
      \hline
 1 & \textbf{-44.38\%} & -53.69\% & -35.42\% & -64.72\% \\ 
   2 & \textbf{103.54\%} & 106.69\% & 23.54\% & 169.82\% \\ 
   3 & \textbf{103.54\%} & 106.69\% & 23.54\% & 169.82\% \\ 
   4 & \textbf{103.54\%} & 203.82\% & \textbf{247.30\%} & 169.82\% \\ 
   \hline\hline
   & \multicolumn{4}{c}{\textbf{Pareto weights: $\beta_1 = \beta_2 = 0.4,\beta_3 = \beta_4 = 0.1$}}\\
    \hline\hline
 1 & -23.31\% & -45.91\% & -26.48\% & -58.89\% \\ 
   2 & -23.31\% & -20.22\% & -26.48\% & -10.87\% \\ 
   3 & 93.24\% & 101.03\% & 16.71\% & 161.84\% \\ 
   4 & 93.24\% & 197.51\% & 240.90\% & 161.84\% \\ 
   \hline
\end{tabular}
\caption{Values of $\Delta_i$ for different choices of risk aversion scenarios and Pareto weights. For the concrete values of the risk aversion parameters, we refer to Table~\ref{tab:certainty_equivalents_collective_agents}. Numbers marked in bold are especially important for the discussion in the main text.} 
\label{tab:certainty_equivalents_diffs_agents}
\end{table}

It is a standard task to prove that in case of risk aversion scenario 1 (equal risk aversion parameters) the following analytical formula holds:
\[
    \Delta_i = \frac{\beta_i^{\frac{1}{\gamma_1}-1}}{\sum\limits_{j=1}^{n} \beta_j^{\frac{1}{\gamma_1}}}-1,\quad i=1,\dots,n.
\]
So, for equal Pareto weights and equal risk aversion parameters, all investors are indifferent in the choice of entering the pool or not. Hence, a homogeneous situation in Pareto weights and risk aversion is a  fair situation for all investors. 

However, by changing the Pareto weights, we see that a significant imbalance occurs, which gives the most risk-averse investors a strong incentive to enter the pool. This effect is more pronounced in the case of only one large investor (case of $\beta_1 = 0.7$). 

By changing the risk aversion parameters, we see that the effect becomes stronger, if the distance between small and large risk aversions is significant. This is the case for risk aversion scenario 3, because here we only have one investor with risk aversion $4.5$, for which $\Delta_4 = 247.30\%$. 

Concluding, a significant heterogeneous family of investors and/or significant differences in the invested capital can lead to large discrepancies in the motivations to enter the pool. For example, in the heterogeneous-risk-aversion scenarios considered numerically, less risk-averse investors in the pool can create particularly strong incentives for more risk-averse investors to join, see risk aversion scenario 3 in Table~\ref{tab:certainty_equivalents_diffs_agents}. 

\section{Conclusion and outlook}\label{sec:conclusion}

We study the collective portfolio optimization problem under a VaR-type  constraint that accounts for multiple VaR thresholds, thereby representing regulatory uncertainty. Then, following the classical martingale approach and after proving the existence of an optimal payoff, we solve the corresponding portfolio optimization problem in the two-threshold case. Surprisingly, in comparison to the solution with respect to a classical VaR constraint, we obtain two additional equations to calculate the Lagrange multipliers. Those additional equations describe specific duality gaps with respect to the collective utility function. 

In a numerical case study, we first test the impact of the regulatory uncertainty without collective investment. This already shows that the shape of the optimal payoff can change drastically, while the associated cost of different thresholds is moderate in many parameter constellations. In particular, the welfare loss for our baseline parameter setting is only $2.72\%$. However, a large welfare loss occurs, if we change single parameters in a way such that protection against regulatory uncertainty becomes difficult to finance. This is the case for a small initial wealth (welfare loss of $25.47\%$ for $x_0=2$), a large compliance level (welfare loss of $48.65\%$ for $a=0.6$) or large values of the lower threshold (welfare loss of $33.05\%$ for $L_1=4$). Second, in case of collective investments, we demonstrate that different risk aversions can have a huge impact on the relative regulatory costs, while their distribution across the investors mainly depend on the Pareto weights of the collective utility function. Moreover, less risk averse investors in the pool can create strong incentives for more risk-averse investors to join.

A possible direction for future research is to allow for uncertainty in the weighting vector $\alpha=(\alpha_1,\dots,\alpha_J)$.
For this, let \(\mathcal A\) be a set of plausible weighting vectors. A robust version of the average compliance requirement is then
\[
    \inf_{\alpha\in\mathcal A}
    \sum_{j=1}^J
    \alpha_j\mathbb P(X_T\ge L_j)
    \ge a.
\]
In this formulation, the firm must achieve the overall compliance level even
under the most adverse priority profile in \(\mathcal A\). Another future research topic could be stress testing, used as a complementary technique alongside VaR. Financial institutions, such as ~banks or insurance companies, can simulate various regulatory changes (such as tax rate increases or new environmental standards) and evaluate the worst-case financial outcomes. Thus, stress testing can capture tail risks that may not be reflected in  VaR estimates. So, introducing stress testing as additional constraints in our setting is a challenging future task. 

\section*{Acknowledgments}
An Chen and Shihao Zhu gratefully acknowledge financial support from the Deutsche Forschungsgemeinschaft (DFG), Project-ID 509303834, FOR 5583 ``Asset Allocation and Asset Pricing under Regulatory Uncertainty''. 

\bibliographystyle{apalike} 
\bibliography{bibfile}

\appendix
\setcounter{subsection}{0}
\renewcommand\thesubsection{A.\arabic{subsection}}
\setcounter{equation}{0}
\renewcommand\theequation{A.\arabic{equation}}

\section{Proofs accompanying Section~\ref{sec:collectiveUtilities}}

\begin{proof}[Proof of Lemma~\ref{lem:Inada}]
    We obtain
    \[
        \lim\limits_{x\downarrow 0} U^{\prime}(x) \geq  \beta_{\min}\lim\limits_{x\downarrow 0}x^{-\gamma_{\max}} = \infty,
    \]
    where $\gamma_{\max} = \max\limits_{i\in\{1,\dots,n\}}\gamma_i$ and $\beta_{\min} = \min\limits_{i\in\{1,\dots,n\}}{\beta_i}$. An analogous argument shows $\lim\limits_{x\rightarrow \infty} U^{\prime}(x) \leq 0$. 
\end{proof}

\begin{proof}[Proof of Corollary~\ref{cor:benchmark-cases}]
    The proof of (ii) is clear. For (i) it is also clear that the Merton payoff solves~\eqref{eq:policyholder-problem}. Uniqueness follows by the fact that any solution to~\eqref{eq:policyholder-problem} also solves the classical Merton problem, for which it is known that the solution is unique. The rest of the statement is obvious.
\end{proof}

\section{Proofs accompanying Section~\ref{sec:structureOfOptimalPayoff}}  
\renewcommand\theequation{A.\arabic{equation}}
\renewcommand\thesubsection{A.\arabic{subsection}}

For brevity, we define
\[
    \Psi(X_T)
    :=
    \sum_{j=1}^J \alpha_j \mathbb P(X_T\geq L_j).
\] 

\begin{proof}[Proof of Lemma~\ref{lem:cost-efficient-reduction}]
To prove (i), note that if \(g\in\mathcal G\) is feasible for~\eqref{eq:reduced-problem}, then $
    \mathbb E[\xi_T g(\xi_T)]
    \le x_0$ and $\Psi(g(\xi_T))
    \ge a$. Moreover, $\mathbb E[U(X_T)]
    =
    \mathbb E[U(g(\xi_T))]$.
Thus \(g(\xi_T)\) is feasible for~\eqref{eq:policyholder-problem} and gives the same value as \(g\) in the problem~\eqref{eq:reduced-problem}.

We now prove (ii). Let \(X_T\ge0\) be feasible for problem~\eqref{eq:policyholder-problem}. Denote by \(F_X\) and \(F_\xi\) the distribution functions of \(X_T\)
and \(\xi_T\), respectively. Since \(\xi_T\) has a continuous distribution, the
random variable $F_\xi(\xi_T)$ is uniformly distributed on \((0,1)\). Next, let
\[
    F_X^{-1}(u):=\inf\{x\ge0:F_X(x)\ge u\},
    \quad u\in(0,1),
\]
be the generalized inverse of \(F_X\). Define
\[
    g(z):=F_X^{-1}\bigl(1-F_\xi(z)\bigr),
    \quad z>0.
\]
Since \(F_\xi\) is nondecreasing and \(F_X^{-1}\) is nondecreasing, the function
\(g\) is nonincreasing. Moreover, \(g\) is Borel measurable. Hence
\(g\in\mathcal G\). 

Then, by \(F_\xi(\xi_T)\) being uniformly distributed on \((0,1)\), $g(\xi_T)$ has the same distribution as \(X_T\), which implies that
\[
    \mathbb E[U(g(\xi_T))]
    =
    \mathbb E[U(X_T)]\quad\text{and}\quad \mathbb P(g(\xi_T)\ge L_j)
    =
    \mathbb P(X_T\ge L_j),\quad j=1,\dots,J.
\]
Consequently, $\Psi(g(\xi_T)) = \Psi(X_T)$. It remains to compare the budget costs. By construction,
\[
    g(\xi_T)
    =
    F_X^{-1}\bigl(1-F_\xi(\xi_T)\bigr),
\]
which is the cost-efficient rearrangement of \(X_T\) with respect to the
state-price density \(\xi_T\). Since \(\xi_T\) is continuously distributed,
Corollary~2 in \cite{bernard2014explicit} implies that this payoff is the cheapest one
among all payoffs with distribution \(F_X\). Hence,
\[
    \mathbb E[\xi_T g(\xi_T)]
    \le
    \mathbb E[\xi_T X_T]
    \le x_0.
\]
Therefore, \(g(\xi_T)\) is feasible for~\eqref{eq:reduced-problem}.
\end{proof}

\begin{proof}[Proof of Theorem~\ref{thm:existence-policyholder}]
By Lemma~\ref{lem:cost-efficient-reduction}, it is enough to prove existence of a maximizer for the optimization problem~\eqref{eq:reduced-problem}. To do so, let $V$ denote the optimal value of~\eqref{eq:reduced-problem}.

Note that $a<\bar{a}$ is only possible if the feasible set of~\eqref{eq:policyholder-problem} --- and so of~\eqref{eq:reduced-problem} --- is nonempty. Therefore, we are allowed to choose a maximizing sequence \((g_n)_{n\in\mathbb{N}}\subseteq\mathcal G\) for \eqref{eq:reduced-problem}, i.e.,~each \(g_n\) satisfies
$  \mathbb E[\xi_T g_n(\xi_T)]\le x_0$, 
$    \sum_{j=1}^J\alpha_j
    \mathbb P(g_n(\xi_T)\ge L_j)\ge a,$
and in addition, $\mathbb E[U(g_n(\xi_T))]\to V.$

Next, fix \(m\in\mathbb N\). 
On the event \(\{\xi_T\le 1/m\}\), monotonicity of every $g_n$ and \(\xi_T>0\) a.s.~implies that
\[
    g_n(1/m)\,
    \mathbb E\!\left[\xi_T\mathbf 1_{\{\xi_T\le 1/m\}}\right]
    \le
    \mathbb E\!\left[\xi_T g_n(\xi_T)\right]
    \le x_0,\qquad n\geq 1,
\]
where the last inequality is exactly the budget constraint. 
By $\mathbb E\!\left[\xi_T\mathbf 1_{\{\xi_T\le 1/m\}}\right]>0$, we obtain that
\[
    g_n(1/m)
    \le
    \frac{x_0}
    {\mathbb E\!\left[\xi_T\mathbf 1_{\{\xi_T\le 1/m\}}\right]},
    \qquad n\ge1.
\]
Consequently, for every \(z\in \big[m^{-1},m\big]\),
\[
    0\le g_n(z)
    \le
    \frac{x_0}
    {\mathbb E\!\left[\xi_T\mathbf 1_{\{\xi_T\le 1/m\}}\right]},
    \qquad n\ge1.
\]
Thus for every $m\in\mathbb{N}$, the sequence \((h_{n,m})_{n\in\mathbb{N}}\) with $h_{n,m}:=g_n|_{K_m}$, where $K_m:=\big[m^{-1},m\big]$, is uniformly bounded.
This allows us to apply
Helly's selection theorem, see e.g.,~\cite[Section~7]{rudin1976},  iteratively. Indeed, given a sequence $\big(g_n^{(m)}\big)_n\subseteq (g_n)_n$ with $g_{n}^{(m)}\rightarrow h^{(m)}$ pointwise, we can apply Helly's selection theorem to the sequence $\big(g_{n}^{(m+1)}\big|_{K_{m+1}}\big)_n$ to obtain a subsequence $\big(g_{n}^{(m+1)}\big|_{K_{m+1}}\big)_n$ with pointwise limit $h^{(m+1)}$. Note that  $h^{(m+1)}|_{K_m} = h^{(m)}$. The resulting subsequences of $(g_n)_n$ are nested, i.e.,~$(g_{n}^{(1)})\supseteq (g_n^{(2)})\supseteq \dots\, $.

Define now the diagonal subsequence by $    \widetilde g_m:=g_m^{(m)}$, $m\in\mathbb N$.
Then, for every fixed \(i\in\mathbb{N}\), the tail \((\widetilde g_m)_{m\ge i}\) is a
subsequence of \((g_n^{(k)})_n\). Hence \((\widetilde g_m|_{K_i})_m\) converges pointwise to $h^{(i)}$. Since two functions $h^{(n)}$ and $h^{(m)}$ are equal on overlaps of their domains, we obtain a nonincreasing function
$g^\star:(0,\infty)\to[0,\infty)$ with $\widetilde g_m(z)\to g^\star(z)$
for all \(z\in(0,\infty)\). This then implies that
\[
    \widetilde g_m(\xi_T)\to g^\star(\xi_T)
    \qquad\text{a.s.}
\]
For the sake of simplicity, we relabel the diagonal subsequence
\((\widetilde g_m)_m\) again by \((g_n)_n\).

Next, we verify that \(g^\star\) is feasible. By Fatou's lemma and the budget
constraint for \(g_n\), we obtain
\[
    \mathbb E[\xi_T g^\star(\xi_T)]
    \le
    \liminf_{n\to\infty}
    \mathbb E[\xi_T g_n(\xi_T)]
    \le x_0.
\]
Hence \(g^\star\) satisfies the budget constraint. To show that also the average compliance constraint is satisfied, note that by
\(g_n(\xi_T)\to g^\star(\xi_T)\) a.s. it holds for every \(j=1,\ldots,J\) that
\begin{align}\label{eq:compliance}
      \mathbf 1_{\{g^\star(\xi_T)\ge L_j\}}
    \ge
    \limsup_{n\to\infty}
    \mathbf 1_{\{g_n(\xi_T)\ge L_j\}}
    \qquad\text{a.s.}
    \end{align}
Indeed, if the right-hand side is equal to one, then there exists a subsequence
\((n_k)\) such that \({g_{n_k}(\xi_T)\ge L_j}\) for all \(k\), and passing to the
limit yields \(g^\star(\xi_T)\ge L_j\). Taking expectations in (\ref{eq:compliance}) and using Fatou's lemma,
we obtain that
\[
    \mathbb P(g^\star(\xi_T)\ge L_j)
    \ge
    \mathbb E\!\left[
        \limsup_{n\to\infty}
        \mathbf 1_{\{g_n(\xi_T)\ge L_j\}}
    \right]
    \ge
    \limsup_{n\to\infty}
    \mathbb P(g_n(\xi_T)\ge L_j).
\]
Multiplying by \(\alpha_j\) and summing over \(j\), we obtain
\[
    \sum_{j=1}^J\alpha_j
    \mathbb P(g^\star(\xi_T)\ge L_j)
    \ge
    \limsup_{n\to\infty}
    \sum_{j=1}^J\alpha_j
    \mathbb P(g_n(\xi_T)\ge L_j)
    \ge a.
\]
Thus \(g^\star\) is feasible for the reduced problem (\ref{eq:reduced-problem}).
Finally, we prove optimality. Since \(g_n(\xi_T)\to g^\star(\xi_T)\) a.s., by
upper semicontinuity of $X\mapsto \mathbb{E}[U(X)]$ it holds that
\[
    \mathbb E[U(g^\star(\xi_T))]
    \ge
    \limsup_{n\to\infty}
    \mathbb E[U(g_n(\xi_T))]
    =
    V.
\]
On the other hand, since \(g^\star\) is feasible and \(V\) is the optimal value we obtain $\mathbb E[U(g^\star(\xi_T))]= V$.
\end{proof}

\section{Proofs accompanying Section~\ref{subsec:two-threshold}}  

For the proof of Proposition~\ref{prop:optimalPayoffBoundaryCasePI}, we need the following auxiliary result.
\begin{lem}\label{lem:auxiliaryFunctionMaxRiskConstraint}
    Let $z\geq 0$ and $g:[0,\infty)\rightarrow \mathbb{R}$ be given by
    \begin{align*}
        g(s;z) = \begin{cases}
            -sz+\alpha_1+\alpha_2, & L_2\leq s,\\
            -sz+\alpha_1, & L_1\leq s <L_2,\\
            -sz,& s<L_1.
        \end{cases}
    \end{align*}
    Then, the following holds:
    \begin{enumerate}
        \item[(i)] If $\alpha_1\leq \frac{L_1}{L_2}$, then
        \begin{align*}
            \argmax_{s\geq 0} g(s;z) = \begin{cases}
                L_2, & z\leq \frac{1}{L_2},\\
                0, & z> \frac{1}{L_2}. 
            \end{cases}
        \end{align*}
        \item[(ii)] If $\alpha_1>\frac{L_1}{L_2}$, then 
        \begin{align*}
            \argmax_{s\geq 0} g(s;z) = \begin{cases}
                L_2, & z\leq \frac{\alpha_2}{L_2-L_1},\\
                L_1, & \frac{\alpha_2}{L_2-L_1}<z\leq \frac{\alpha_1}{L_1},\\
                0, & z> \frac{\alpha_1}{L_1}. 
            \end{cases}
        \end{align*}
    \end{enumerate}
\end{lem}

\begin{proof}
    Note that 
    \begin{align}\label{eq:auxiliaryFunctionMax}
         \max_{s\geq 0}g(s;z) = \max\{0,\alpha_1-L_1 z, \alpha_1+\alpha_2 - L_2 z\}. 
    \end{align}
    Note, the values $0$, $\alpha_1-L_1 z$ and $\alpha_1+\alpha_2-L_2 z$ are attained at $s=0$, $s=L_1$ and $s=L_2$, respectively. Hence, it is enough to compare the three cases on the right-hand side in~\eqref{eq:auxiliaryFunctionMax} in dependence of $z$. By comparing these cases pairwise, we obtain the following:
    \begin{enumerate}
        \item[(1)] \underline{$0$ versus $\alpha_1 - L_1z$:} If $z\leq \frac{\alpha_1}{L_1}$, then $\alpha_1-L_1 z\geq 0$ and $\alpha_1-L_1 z< 0$, otherwise. 
        \item[(2)] \underline{$0$ versus $\alpha_1 + \alpha_2 - L_2z$:} If $z\leq \frac{\alpha_1+\alpha_2}{L_2}$, then $\alpha_1+\alpha_2-L_2 z\geq 0$ and $\alpha_1+\alpha_2-L_2 z< 0$, otherwise. 
        \item[(3)] \underline{$\alpha_1 - L_1z$ versus $\alpha_1 + \alpha_2 - L_2z$:} If $z\geq \frac{\alpha_2}{L_2-L_1}$, then $\alpha_1-L_1 z\geq \alpha_1+\alpha_2-L_2 z$ and 
        $\alpha_1-L_1z<\alpha_1+\alpha_2-L_2 z$, otherwise. 
    \end{enumerate}
    Now, assume $\frac{\alpha_1}{\alpha_1+\alpha_2}\leq \frac{L_1}{L_2}$, which implies 
    \[
        \frac{\alpha_1}{L_1}\leq \frac{\alpha_1+\alpha_2}{L_2} \leq \frac{\alpha_2}{L_2-L_1}. 
    \]
    In this case, $\alpha_1-L_1z\leq \max\{0,\alpha_1+\alpha_2-L_2z\}$. This gives us that $L_2$ is the maximizer if 
    $z\leq \frac{\alpha_1+\alpha_2}{L_2}$ and $0$ is the maximizer otherwise. 

    To prove (ii), assume $\frac{\alpha_1}{\alpha_1+\alpha_2}> \frac{L_1}{L_2}$ and note that
    \[
         \frac{\alpha_2}{L_2-L_1}< \frac{\alpha_1+\alpha_2}{L_2} < \frac{\alpha_1}{L_1}. 
    \]
    An analogous consideration as before for case (i) gives us then the result for $\argmax\limits_{s\geq 0}g(s;z)$.
\end{proof}

\begin{proof}[Proof of Proposition~\ref{prop:optimalPayoffBoundaryCasePI}]

    In case of $a=\overline{a}$ a solution of~\eqref{eq:MaximumValueRiskConstraint} also solves~\eqref{eq:policyholder-problem}. Hence, it is enough to solve~\eqref{eq:MaximumValueRiskConstraint} and show that this solution is unique. 

    Now, assume first $e^{-rT}L_2\leq x_0$. Then, $a = \overline{a} = \alpha_1+\alpha_2$, a contradiction to our assumption $a<\alpha_1+\alpha_2$. Therefore, in the following we assume $e^{-rT}L_2>x_0$ and perform a pointwise Lagrange approach. To do so, assume a payoff $Y\geq 0$ satisfying the budget constraint. For an arbitrary $\lambda\geq 0$ and by using the same function $g$ as in Lemma~\ref{lem:auxiliaryFunctionMaxRiskConstraint}, we obtain 
    \begin{align*}
        \mathbb{E}\big[ \alpha_1 \mathbf{1}_{[L_1,\infty)}(Y) + \alpha_2\mathbf{1}_{[L_2,\infty)}(Y)\big]&\leq \mathbb{E}\big[ \alpha_1 \mathbf{1}_{[L_1,\infty)}(Y) + \alpha_2\mathbf{1}_{[L_2,\infty)}(Y)\big] -\lambda \big(\mathbb{E}[\xi_T Y]-x_0\big)\\
        &= \mathbb{E}\big[ g(Y;\lambda \xi_T)\big] +\lambda x_0\\
        &\leq \begin{cases}
            (\alpha_1+\alpha_2)\mathbb{P}\left(\lambda \xi_T\leq \frac{\alpha_1+\alpha_2}{L_2}\right)+\lambda x_0, & \frac{\alpha_1}{\alpha_1+\alpha_2}\leq \frac{L_1}{L_2},\\
            \alpha_1\mathbb{P}\left(\lambda \xi_T\leq \frac{\alpha_1}{L_1}\right)+\alpha_2\mathbb{P}\left(\lambda \xi_T\leq \frac{\alpha_2}{L_2-L_1}\right)+\lambda x_0, & \frac{\alpha_1}{\alpha_1+\alpha_2}> \frac{L_1}{L_2},
        \end{cases}
    \end{align*}
    where the first inequality follows from $\mathbb E[\xi_TY]\leq x_0$ and the second inequality follows by Lemma~\ref{lem:auxiliaryFunctionMaxRiskConstraint}.
    The second inequality holds with equality iff $Y=X_T^{\star}$, where 
    \[
        X_T(\lambda) =
        \begin{cases}
                L_2, & \lambda\xi_T\leq \frac{\alpha_1+\alpha_2}{L_2},\\
                0, & \text{otherwise},
        \end{cases}
    \]
    if $\frac{\alpha_1}{\alpha_1+\alpha_2}\leq \frac{L_1}{L_2}$, 
    and
    \[
        X_T(\lambda) =
        \begin{cases}
                L_2, & \lambda\xi_T\leq \frac{\alpha_2}{L_2-L_1},\\
                L_1, & \frac{\alpha_2}{L_2-L_1}<\lambda\xi_T
                \leq \frac{\alpha_1}{L_1},\\
                0, & \text{otherwise},
        \end{cases}
    \]
    if $\frac{\alpha_1}{\alpha_1+\alpha_2}> \frac{L_1}{L_2}$. It remains to show that there exists a unique value
    $\lambda^\star>0$ such that $\mathbb E[\xi_TX_T(\lambda^\star)]=x_0$. First, consider the case of $\frac{\alpha_1}{\alpha_1+\alpha_2}\leq \frac{L_1}{L_2}$, for which we set
    \[
        h_1(\lambda) = \mathbb{E}[\xi_T X_T(\lambda)] = L_2 \mathbb{E}\left[\xi_T \mathbf{1}_{\left[0,\frac{\alpha_1+\alpha_2}{L_2}\right]}(\lambda\xi_T)\right] = e^{-rT}L_2 \mathbb{Q}\left(\lambda\xi_T\leq \frac{\alpha_1+\alpha_2}{L_2}\right).
    \]
  
    By $\xi_T$ being lognormally distributed under $\mathbb Q$, its
    distribution function is continuous and strictly increasing and so, $h_1$ is continuous and strictly decreasing in
    $\lambda$. Moreover, we obtain the following limits:
    \[
        \lim_{\lambda\downarrow0}h_1(\lambda)=e^{-rT}L_2,
        \qquad
        \lim_{\lambda\to\infty}h_1(\lambda)=0.
    \]
    Then, together with $e^{-rT}L_2>x_0$, the intermediate value theorem yields a unique $\lambda^{\star}>0$ such that $\mathbb{E}[\xi_T X_T(\lambda^{\star})] = x_0$ as desired.
    This implies that
    \[
        \overline{a}
        =
        \alpha_1\P(X_T(\lambda^\star)\geq L_1)
        +\alpha_2\P(X_T(\lambda^\star)\geq L_2)
        =
        (\alpha_1+\alpha_2)
        \P\left(
        \lambda^\star\xi_T\leq \frac{\alpha_1+\alpha_2}{L_2}
        \right),
    \]
    which proves (i) by noting that $X_T^{\star} = X_T(\lambda^{\star})$. For (ii), assume $\frac{\alpha_1}{\alpha_1+\alpha_2}> \frac{L_1}{L_2}$ and set
    \begin{align*}
        h_2(\lambda) &= \mathbb{E}[\xi_T X_T(\lambda)]\\
        &= L_2 \mathbb{E}\left[\xi_T \mathbf{1}_{\left[0,\frac{\alpha_2}{L_2-L_1}\right]}(\lambda\xi_T)\right]+L_1 \mathbb{E}\left[\xi_T \mathbf{1}_{\left(\frac{\alpha_2}{L_2-L_1},\frac{\alpha_1}{L_1}\right]}(\lambda\xi_T)\right]\\
        &= e^{-rT}(L_2-L_1) \mathbb{Q}\left(\lambda\xi_T\leq \frac{\alpha_2}{L_2-L_1}\right)+e^{-rT}L_1 \mathbb{Q}\left(\lambda\xi_T\leq \frac{\alpha_1}{L_1}\right). 
    \end{align*}
    Then, we obtain the following limits: $ \lim\limits_{\lambda\downarrow 0} h_2(\lambda) =  e^{-rT}L_2$ and $\lim\limits_{\lambda\rightarrow \infty} h_2(\lambda) =0$.
    An analogous argumentation as before then ensures the existence of a unique value $\lambda^{\star}>0$ such that $\mathbb{E}[\xi_T X_T(\lambda^{\star})] = x_0$. Furthermore, we obtain 
    \[
        \overline{a} = \alpha_1 \P(X_T(\lambda^{\star})\geq L_1) + \alpha_2\P(X_T(\lambda^{\star})\geq L_2) = \alpha_1\P\left(\lambda^{\star}\xi_T\leq \frac{\alpha_1}{L_1}\right)+\alpha_2 \P\left(\lambda^{\star}\xi_T\leq \frac{\alpha_2}{L_2-L_1}\right). 
    \]
    By noting that $X_T^{\star} = X_T(\lambda^{\star})$, this finally proves (ii).
\end{proof}

\begin{proof}[Proof of Proposition~\ref{prop:riskMaximizer}]
    Let $e^{-rT}L_2 = x_0$ and note that $X_T^{\star} = L_2$ satisfies budget and average compliance constraint with equality. For another payoff $X\geq 0$ a.s.~satisfying the budget constraint and $\P(X<L_2)>0$, we have 
    \[
        \alpha_1 \P(X\geq L_1) + \alpha_2\P(X\geq L_2) <\alpha_1+\alpha_2
    \]
    and this payoff can therefore not be optimal. On the other side, if $X\geq L_2$ a.s.~and $\P(X>L_2)>0$, then $e^{-rT}\mathbb{E}_{\mathbb{Q}}[X]>e^{-rT} L_2 = x_0$, which contradicts that $X$ satisfies the budget constraint. Hence, $X_T^{\star}=L_2$ is the unique optimal payoff. 
    For $e^{-rT}L_2 < x_0$, note that the payoff in~\eqref{eq:riskMaximizer} satisfies budget and average compliance constraint with equality and hence, it is a solution to~\eqref{eq:riskMaximizer} such that $\overline{a} = \alpha_1+\alpha_2$.
\end{proof}

For the proof of Theorem~\ref{thm:2bounds}, we use the following well-known auxiliary result. To keep the manuscript as self-contained as possible, we state a short proof. 
\begin{lem}\label{lem:limitInadaUtility}
    Let $f:(0,\infty)\rightarrow\mathbb{R}$ be a continuously differentiable, strictly increasing and strictly concave function satisfying the Inada condition $\lim\limits_{x\downarrow 0} f^{\prime}(x) = \infty$. Then, for every $c\in(0,\infty)$ it holds that
    \[
        \lim_{x\downarrow 0} \Big(f^{\prime}(x)(x-c)-f(x)\Big) = -\infty.
    \]
\end{lem}

\begin{proof}
Fix $0<y<c$. Then, by concavity of $f$, it holds for any $x\in(0,y)$ that 
\[
    f(y)\leq f(x)+f'(x)(y-x).
\]
Therefore, we obtain for $x\in(0,y)$ that 
\begin{align*}
f^{\prime}(x)(x-c)-f(x)\leq(y-c)f^{\prime}(x)-f(y)\longrightarrow-\infty,\quad\text{as }x\downarrow 0.
\end{align*}
This proves the claim. 
\end{proof}

Furthermore, we introduce another auxiliary result regarding $I$.
\begin{lem}\label{lem:Inada_2}
    Let $f:(0,\infty)\rightarrow\mathbb{R}$ be a  differentiable, strictly increasing and strictly concave function  satisfying the Inada condition $\lim\limits_{x\downarrow 0} f^{\prime}(x) = \infty$. Then, it holds that
    \[
        \lim_{x\downarrow 0}\frac{f^{\prime}(x)}{f(x)} = 0.
    \]
    In particular, it holds that 
    \[
        \lim\limits_{x\to\infty}
        \frac{U(I(x))-xI(x)}{x}=0.
    \]
\end{lem}

\begin{proof}
    In case of $\lim_{x\downarrow 0}f(x)\in\mathbb{R}$ the statement is clear. Otherwise, fix $a>0$. Then, for an arbitrary $\varepsilon\in(0,a)$, we choose $\delta \in(0,\varepsilon)$. For any $z<\delta$ we obtain by strict concavity of $f$ that
    \begin{align*}
        \frac{\int_{z}^{a} f^{\prime}(t)\diff t}{f^{\prime}(z)} = \frac{\int_{z}^{\delta} f^{\prime}(t)\diff t +\int_{\delta}^{a} f^{\prime}(t)\diff t}{f^{\prime}(z)} 
        &\leq (\delta-z) +\frac{\int_{\delta}^{a} f^{\prime}(t)\diff t}{f^{\prime}(z)} \\
        &\leq \varepsilon + \frac{\int_{\delta}^{a} f^{\prime}(t)\diff t}{f^{\prime}(z)}\rightarrow \varepsilon,\quad \text{as }z\downarrow 0.
    \end{align*}
    Note, the Inada condition is used to obtain the convergence of $\frac{\int_{\delta}^{a} f^{\prime}(t)\diff t}{f^{\prime}(z)}\rightarrow 0$ as $z\downarrow 0$. Then, since $\varepsilon$ was chosen arbitrarily, we obtain that $\lim_{x\downarrow 0}\frac{f^{\prime}(x)}{f(x)} = 0$.    
    
    In case of $U$, it holds that
        \[
        \lim\limits_{x\rightarrow \infty}\frac{U(I(x))-x I(x)}{x} =  \lim\limits_{y\downarrow 0}\frac{U(y)-U^{\prime}(y)y}{U^{\prime}(y)}=\lim\limits_{y\downarrow 0}\Big(\frac{U(y)}{U^{\prime}(y)}-y\Big).
    \] 
    Then, the claim follows by the first part together with $U$ satisfying the Inada conditions due to Lemma~\ref{lem:Inada}. 
\end{proof}

\begin{proof}[Proof of Theorem~\ref{thm:2bounds}]
 We perform a pointwise Lagrange-approach in three main steps.\newline
 
\textbf{Step 1 (Candidate solution):} We compare the Lagrange-function of the interior point $I(\lambda\xi_T(\omega))$ for a specific $\omega\in\Omega$ and the two boundary solutions $L_1$ and $L_2$. For brevity, we write $\xi = \xi_T(\omega)$. With this, the Lagrangian for multipliers $\lambda>0$ and $\nu\ge0$ is given by
\begin{align*}
\mathcal L(X_T;\lambda,\nu)
=
\mathbb{E}\Big[ U(X_T)-\lambda \xi_T X_T
+\nu \sum_{j=1}^J \alpha_j \mathbf 1_{\{X_T\ge L_j\}}\Big]
-\nu a+\lambda x_0.
\end{align*}
Thus, for given $\lambda>0$ and $\nu> 0$, the pointwise Lagrange approach is based on the following map:
 \begin{align*}
     g:(x,y)\mapsto U(x)-yx+\nu \left(\alpha_1 \mathbf{1}_{[L_1,\infty)}(x)+ \alpha_2\mathbf{1}_{[L_2,\infty)}(x)\right).
\end{align*}
For brevity, we also make use of the following notations:
\begin{align*}
    f_1(x) &= g(L_1,x) - g(I(x),x),\\
    f_2(x) &= g(L_2,x) - g(I(x),x),\\
    f_3(x) &= g(L_1,x) - g(L_2,x).
\end{align*}

Now, we perform a case-by-case study. In doing so, assume first that \( I(\lambda\xi)\geq L_2\), or equivalently,~\(\lambda\xi\leq  {U^{\prime}(L_2)}\). Then, 
\begin{align*}
    f_2(\lambda\xi) = U(L_2)-\lambda\xi L_2
-\Bigl[
U\bigl(I(\lambda\xi) \bigr)-\lambda\xi\,I(\lambda\xi)\Bigr]\leq 0,
\end{align*}
because for the map $h(x) = U(x)-\lambda\xi x$ it holds that $h^{\prime}(x) = U^{\prime}(x)-\lambda\xi\geq 0$ for all $x\leq I(\lambda\xi)$. Hence,~$L_2$ cannot be optimal. Next, note that
\[
    f_1(x) = U(L_1)-x L_1
-\Bigl[
U\bigl(I(x) \bigr)-x\,I(x)\Bigr]-\nu\alpha_2,\quad x\leq U^{\prime}(L_2).
\]
In this situation, we have $f_1^{\prime}(x) = -L_1+I(x)>0$, which implies that
\[
    f_1\big(U^{\prime}(L_2)\big)\leq f_1\big(U^{\prime}(L_1)\big) = -\nu\alpha_2\leq 0.
\]
Hence, $L_1$ cannot be optimal. Concluding, if  \( I(\lambda \xi)\geq L_2\), then $I(\lambda\xi)$ is optimal.

The second case is the one of $L_1\leq I(\lambda \xi)< L_2$. Here, we first argue that $L_1$ cannot be optimal. For this, note that $f_1^{\prime}(x) = -L_1+I(x)\geq 0$ for all $x\in\big(U^{\prime}(L_2),U^{\prime}(L_1)\big]$. Thus, we obtain
\[
    \lim\limits_{x\downarrow U^{\prime}(L_2)}f_1(x)\leq f_1\big(U^{\prime}(L_1)\big) = 0. 
\]
Hence, $f_1(\lambda\xi) \leq 0$, which means that $L_1$ cannot be optimal. To compare $L_2$ and $I(\lambda\xi)$, note that $\lim\limits_{x\downarrow U^{\prime}(L_2)}f_2\big(x\big) = \nu\alpha_2 > 0$. Furthermore, 
\[
    f_2^{\prime}(x) = -L_2+I(x)\leq 0,\quad x\in \big(U^\prime(L_2),U^\prime(L_1)\big].
\]
So, if $f_2\big(U^{\prime}(L_1)\big)\geq 0$, then $L_2$ is optimal (implying that $q^{\star}\geq U^{\prime}(L_1)$). Otherwise, by continuity of~$f_2$, there exists $q^{\star} \in \big(U^{\prime}(L_2),U^{\prime}(L_1)\big)$ such that $f_2(q^{\star}) = 0$ and hence, $L_2$ is optimal for $\lambda\xi\in\big(U^{\prime}(L_2),q^{\star}\big]$ and~$I(\lambda\xi)$ is optimal for $ \lambda\xi\in\big(q^{\star},U^{\prime}(L_1)\big]$. 

For the case of \(I(\lambda\xi)<L_1\), let us first assume that  $f_2\big(U^{\prime}(L_1)\big)<0$. Then, 
\[
    f_3(\lambda\xi) = g(L_1,\lambda \xi) - g(L_2,\lambda\xi)> (U^{\prime}(L_1)-\lambda\xi)(L_1-L_2)>0,
\]
where we used the assumption $f_2\big(U^{\prime}(L_1)\big)<0$ to obtain the first inequality. This shows that $L_2$ cannot be optimal. To compare $L_1$ and $I(\lambda\xi)$, note that
\[
f_1(x)  = 
U(L_1)-x L_1+\nu \alpha_1
-\Bigl[
U\bigl(I(x)\bigr)-x I(x)
\Bigr],\quad x>U^{\prime}(L_1).
\]

Then $f_1^{\prime}(x)=-L_1+I(x)<0$, while $\lim\limits_{x\downarrow U^{\prime}(L_1)}f_1(x) = \nu \alpha_1> 0$ and by Lemma~\ref{lem:Inada} we are allowed to apply Lemma~\ref{lem:limitInadaUtility} for $U$, from which we obtain that 
\begin{align*}
    \lim_{x\rightarrow \infty}f_1(x) &= \nu\alpha_1 + U(L_1) + \lim_{x\rightarrow\infty}\Big(x\big(I(x)-L_1\big)-U\big(I(x)\big)\Big)\\
    &= \nu\alpha_1 + U(L_1) + \lim_{y\downarrow 0}\Big(U^{\prime}(y)\big(y-L_1\big)-U\big(y\big)\Big)\\
    &= -\infty.
\end{align*}
Hence there exists a threshold \(q^{\star\star}>  U^{\prime}(L_1)\) such that $f_1(q^{\star\star}) = 0$ and $L_1$ is optimal on ${\lambda\xi\in\bigl(U^{\prime}(L_1),q^{\star\star}\bigr]}$ and $I(\lambda\xi)$ is optimal on $\lambda\xi>q^{\star\star}$. 

It remains the case of $f_2\big(U^{\prime}(L_1)\big)\geq 0$. First, we compare $L_2$ and $I(\lambda \xi)$. In doing so, note that 
\[
    f_2(x) = U(L_2) - xL_2 +v(\alpha_1+\alpha_2) -U(I(x))+xI(x), \quad x> U^{\prime}(L_1), 
\]
and note that $f_2^{\prime}(x) = -L_2 + I(x)< 0$, $\lim\limits_{x\downarrow U^{\prime}(L_1)}f_2(x) = f_2\big(U^{\prime}(L_1)\big)+\nu \alpha_1\geq \nu\alpha_1>0$ and $\lim\limits_{x\rightarrow \infty}f_2(x) = -\infty$, where the latter follows again by Lemma~\ref{lem:limitInadaUtility}. Hence, there exists a value $x^{\star}> U^{\prime}(L_1)$ such that $f_2(x^{\star}) = 0$, meaning that $L_2$ is preferred over $I(\lambda \xi)$, if $\lambda\xi \in \big(U^{\prime}(L_1),x^{\star}\big]$, and $I(\lambda\xi)$ is preferred over~$L_2$, if $\lambda\xi>x^{\star}$. Next, we compare $L_1$ and $I(\lambda\xi)$. For this, we obtain
\[
    f_1(x) = U(L_1)-xL_1 + \nu\alpha_1 -U(I(x))+xI(x),\quad x> U^{\prime}(L_1).
\]
Then, $f_1^{\prime}(x) = -L_1+I(x)< 0$, $\lim\limits_{x\downarrow U^{\prime}(L_1)}f_1(x)=\nu\alpha_1>0$ and $\lim\limits_{x\rightarrow \infty}f_1(x) = -\infty$, where the latter follows by Lemma~\ref{lem:limitInadaUtility}. Hence, there exists $q^{\star\star}> U^{\prime}(L_1)$ such that $f_1(q^{\star\star}) = 0$, implying that $I(\lambda\xi)$ is optimal for $\lambda\xi > \max\{x^{\star},q^{\star\star}\}$ and $L_1$ is preferred over $I(\lambda\xi)$ for $\lambda\xi\in \big(U^{\prime}(L_1),q^{\star\star}\big]$. 

We now differ the two cases of $x^{\star}<q^{\star\star}$ and $x^{\star}\geq q^{\star\star}$. Let us begin with $x^{\star}<q^{\star\star}$. Then, from the previous considerations we obtain that $I(\lambda\xi)$ is optimal for $\lambda\xi > q^{\star\star}$ and $L_1$ is optimal for $\lambda\xi \in(x^{\star},q^{\star\star}]$. Thus, we have to compare $L_1$ and $L_2$ for $\lambda\xi\in (U^{\prime}(L_1),x^{\star}]$. For this, note that $f_3$ is strictly increasing on this interval and
\begin{align*}
    f_3(x^{\star}) &= U(L_1) - x^{\star} L_1 - U(L_2) + x^{\star} L_2 -\nu \alpha_2\\ 
    &= U(L_1) - x^{\star} L_1 - U(I(x^{\star})) + x^{\star} I(x^{\star}) + \nu\alpha_1\\
    &= f_1(x^{\star}) >0,
\end{align*}
where we used $f_2(x^{\star}) = 0$ to obtain the second equality and the inequality follows by $x^{\star}<q^{\star\star}$, $f_1(q^{\star\star}) = 0$ and $f_1$ being strictly decreasing on $(U^{\prime}(L_1),\infty)$. Moreover, it holds that
\[
    \lim_{x\downarrow U^{\prime}(L_1)}f_3(x) = f_3\big(U^{\prime}(L_1) \big) = \Bigg(U^{\prime}(L_1) - \frac{U(L_2)-U(L_1)+\nu\alpha_2}{L_2-L_1}\Bigg)(L_2-L_1) \leq 0,
\]
where the inequality is equivalent to our assumption of  $f_2\big(U^{\prime}(L_1)\big)\geq 0$. Hence, by setting $q^{\star} = l(\nu)\in \big[U^{\prime}(L_1),q^{\star\star}\big]$, i.e., $f_3(q^{\star}) = 0$, we obtain that $L_2$ is optimal for $\lambda\xi\in(U^{\prime}(L_1),q^{\star}]$ and $L_1$ is optimal on $\lambda\xi\in (q^{\star},q^{\star\star}]$. 

Finally, assume $x^{\star}\geq q^{\star\star}$. Here, we already know that $I(\lambda\xi)$ is optimal for $\lambda\xi > x^{\star}$ and $L_2$ is optimal for $\lambda\xi \in(q^{\star\star},x^{\star}]$. Thus, we have to compare $L_1$ and $L_2$ on $\lambda\xi\in (U^{\prime}(L_1),q^{\star\star}]$. For this, note that
\[
    f_3(q^{\star\star}) = U(L_1) - q^{\star\star} L_1 - U(L_2) + q^{\star\star} L_2 -\nu \alpha_2 = -f_2(q^{\star\star})<0,
\]
where we used for the second equality that $f_1(q^{\star\star}) = 0$ and for the inequality that $f_2(x^{\star})=0$, $f_2$ decreasing on $(U^{\prime}(L_1),\infty)$ and $x^{\star}\geq q^{\star\star}$. Hence, we obtain that $L_2$ is optimal for $\lambda\xi\in(U^{\prime}(L_1),x^{\star}]$ and $I(\lambda\xi)$ is optimal for $\lambda\xi > x^{\star}$. In this case, we set $q^{\star} = x^{\star}$. 

To obtain the desired representation of $X_T^{\star}$, it remains to show that $x^{\star}\geq q^{\star\star}$ is equivalent to $l(\nu)\geq q^{\star\star}$. Recall that 
\[
    f_2(x)-f_1(x) = U(L_2)-U(L_1) - x(L_2-L_1) + \nu\alpha_2,\quad x>U^{\prime}(L_1). 
\]
Applying this equation in case of $x=q^{\star\star}>U^{\prime}(L_1)$, we obtain
\begin{align*}
    f_2(q^{\star\star}) &= f_1(q^{\star\star}) + U(L_2)-U(L_1) - q^{\star\star}(L_2-L_1) + \nu\alpha_2\\
    &=\nu\alpha_2 + U(L_2) -U(L_1) - q^{\star\star}(L_2-L_1)\\
    &= (L_2-L_1)(l(\nu)-q^{\star\star}). 
\end{align*}
By $f_2$ being strictly decreasing on $(U^{\prime}(L_1),\infty)$ and by $f_{2}(x^{\star})=0$, we obtain from the previous equation the desired equivalence of 
\[
    x^{\star}\geq q^{\star\star} \quad \Longleftrightarrow \quad l(\nu)\geq q^{\star\star}. 
\]
Together with the earlier considerations, the desired representation of $X_T^{\star}$ follows immediately.\newline

\textbf{Step 2 (Optimality of candidate solution):} Assume that there are parameters $\lambda, \nu, q^{\star}, q^{\star\star}$, satisfying the constraints in the claim. We show that our pointwise candidate solution $X_T^{\star}$ is indeed optimal. For this, assume that $Y$ is another payoff satisfying the budget constraint as well as the average compliance constraint. Then, we have
\begin{align*}
    \mathbb{E}\!\left[U(X_T^{\star})\right] &= \mathbb{E}\!\left[U(X_T^{\star})\right] - \lambda \Big(\mathbb{E}\!\left[\xi_T X_T^\star\right]-x_0\Big) +\nu \Big( \alpha_1\,\mathbb{P}\!\left(X_T^\star \ge L_1\right)+\alpha_2\,\mathbb{P}\!\left(X_T^\star \ge L_2\right)-a\Big)\\
    &= \mathbb{E}\!\left[g(X_T^{\star},\lambda\xi_T)\right] +\lambda x_0 -\nu a\\
    &\geq \mathbb{E}\!\left[g(Y,\lambda\xi_T)\right] +\lambda x_0 -\nu a\\
    &=  \mathbb{E}\!\left[U(Y)\right] - \underbrace{\lambda}_{>0} \Big(\underbrace{\mathbb{E}\!\left[\xi_T Y\right]-x_0}_{\leq 0}\Big) +\underbrace{\nu}_{> 0} \Big( \underbrace{\alpha_1\,\mathbb{P}\!\left(Y \ge L_1\right)+\alpha_2\,\mathbb{P}\!\left(Y \ge L_2\right)-a}_{\geq 0}\Big)\\
    &\geq \mathbb{E}\!\left[U(Y)\right].
\end{align*}
Hence, $X_T^{\star}$ is optimal. \newline

\textbf{Step 3 (Existence of Lagrange multipliers):} Using the monotonicity and continuity arguments from above for the functions $f_1$ and $f_2$ it can be easily seen that for $\nu>0$ there exists $q^{\star\star}(\nu)\geq U^{\prime}(L_1)$ such that 
\[
    U(L_1)- q^{\star\star}(\nu)L_1
-\Bigl[
U\big(I(q^{\star\star}(\nu))\big)-q^{\star\star}(\nu)I(q^{\star\star}(\nu))
\Bigr] =-\nu \alpha_1
\]
and there exists $q^{\star}(\nu)\geq U^{\prime}(L_2)$ such that
\[
    U(L_2)- q^{\star}(\nu)L_2
-\Bigl[
U\big(I(q^{\star}(\nu))\big)-q^{\star}(\nu)I(q^{\star}(\nu))
\Bigr] =-\nu \alpha_2,
\]
if $l(\nu)< U^{\prime}(L_1)$, and otherwise, 
\[
    U(L_2)- q^{\star}(\nu)L_2
-\Bigl[
U\big(I(q^{\star}(\nu))\big)-q^{\star}(\nu)I(q^{\star}(\nu))
\Bigr] =-\nu,
\]
if this solution is beyond $q^{\star\star}(\nu)$ and $q^{\star}(\nu) =l(\nu)$ otherwise. Furthermore, local monotonicity and continuity of $f_1$ gives us that $q^{\star\star}$ is a strictly increasing and continuous function depending on $\nu$.  

Next, we argue that $q^{\star}$ is strictly increasing. To do so, assume $0<\nu_1<\nu_2$.  If $l(\nu_1)<U^{\prime}(L_1)$ and $l(\nu_2)<U^{\prime}(L_1)$, then $q^{\star}(\nu_1)<q^{\star}(\nu_2)$ follows by local monotonicity and continuity of $f_2$. Moreover, if $l(\nu_1)<U^{\prime}(L_1)$ and $l(\nu_2)\geq U^{\prime}(L_1)$, then $q^{\star}(\nu_1)<U^{\prime}(L_1)\leq q^{\star}(\nu_2)$.

Thus, it remains the case of $\min\{l(\nu_1),l(\nu_2)\}\geq U^{\prime}(L_1)$. In this case, if $l(\nu_1)<q^{\star\star}(\nu_1)$ and $l(\nu_2)\geq q^{\star\star}(\nu_2)$, then $q^{\star}(\nu_1)<q^{\star\star}(\nu_1)$ and $q^{\star}(\nu_2)\geq q^{\star\star}(\nu_2)$ and by monotonicity of $q^{\star\star}$ we obtain that
\[
    q^{\star}(\nu_1)<q^{\star\star}(\nu_1)<q^{\star\star}(\nu_2)\leq q^{\star}(\nu_2).
\]
Next, consider the case of $l(\nu_1)\geq q^{\star\star}(\nu_1)$ and $l(\nu_2)< q^{\star\star}(\nu_2)$. Then, by continuity of $l$ and $q^{\star\star}$, there exists $\nu_0\in(\nu_1,\nu_2]$ such that $l(\nu_0)=q^{\star\star}(\nu_0)$. But then $H_2(q^{\star\star}) = -\nu_0$, where $H_2(q) = U(L_2)-qL_1-\big[U(I(q))-qI(q)\big]$. Indeed, recall that 
\[
    H_2(q)-H_1(q) = U(L_2)-U(L_1) - q(L_2-L_1),
\]
where $H_1(q) = U(L_1)-qL_1-\big[U(I(q))-qI(q)\big]$. Then,
\begin{align*}
    H_2(q^{\star\star}(\nu_0)) = -\nu\alpha_1  + U(L_2)-U(L_1) - q^{\star\star}(\nu_0)(L_2-L_1) = -\nu,
\end{align*}
where we used $H_1(q^{\star\star}(\nu_0)) = -\nu\alpha_1$ and $q^{\star\star}(\nu_0) = l(\nu_0)$ to obtain the last equality.

Hence, by uniqueness of the root of $H_2+\nu$ on $(U^{\prime}(L_1),\infty)$, it holds that
\[
    x^{\star}(\nu_0) = q^{\star\star}(\nu_0) = l(\nu_0),
\]
where $x^{\star}(\nu_0)$ solves $H(x^{\star}(\nu_0)) = -\nu$. Then, since $x^{\star}(\nu)$ is increasing in $\nu$ and $\nu_1<\nu_0$, we have $q^{\star}(\nu_1)=x^{\star}(\nu_0)<x^{\star}(\nu_0) = l(\nu_0)$. Moreover, by $l$ being strictly increasing, it follows
\[
    q^{\star}(\nu_1)<l(\nu_0)\leq l(\nu_2) = q^{\star}(\nu_2),
\]
i.e.,~$q^{\star}(\nu_1)<q^{\star}(\nu_2)$. The remaining cases are clear and we obtain that $q^{\star}$ is strictly increasing.

Furthermore, it holds that
\[
    \lim\limits_{\nu\downarrow 0}q^{\star}(\nu) = U^{\prime}(L_2),\quad \lim\limits_{\nu\downarrow 0}q^{\star\star}(\nu) = U^{\prime}(L_1)\quad \text{and}\quad  \lim\limits_{\nu\rightarrow \infty}q^{\star\star}(\nu)=  \lim\limits_{\nu\rightarrow \infty}q^{\star}(\nu) = \infty. 
\]

Then, for $\nu>0$ we define the function
\[
    h_\nu:[0,\infty)\rightarrow (0,\infty],\lambda\mapsto \mathbb{E}\big[\xi_T X_T^{\star}\big(\lambda,q^{\star}(\nu),q^{\star\star}(\nu)\big)\big],
\]
where 
\begin{align*}
    X_T^\star(\lambda,q^{\star},q^{\star\star})=
        \begin{cases}
            I(\lambda \xi_T), & \xi_T \leq \frac{U^{\prime}(L_2)}{\lambda},\\
            L_{2}, & \frac{U^{\prime}(L_2)}{\lambda} <  \xi_T \leq \frac{q^{\star}}{\lambda},\\
            I(\lambda \xi_T), & \frac{q^{\star}}{\lambda}<\xi_T \leq \frac{U^{\prime}(L_1)}{\lambda},\\
            L_{1}, &\frac{\max\{U^{\prime}(L_1),q^{\star}\}}{\lambda} < \xi_T\leq \frac{q^{\star\star}}{\lambda},\\
            I(\lambda \xi_T),  &  \xi_T >\frac{\max\{q^{\star},q^{\star\star}\}}{\lambda}. 
         \end{cases}
\end{align*}

The function $h_\nu$ is continuous. Indeed, first note that 
\[
    \liminf_{\lambda\downarrow 0} X_T^{\star}\big(\lambda,q^{\star}(\nu),q^{\star\star}(\nu)\big) = \infty,\quad \mathbb{Q}\text{-a.s.}
\]
So, by Fatou's lemma it holds that
\begin{align*}
    \lim\limits_{\lambda\downarrow 0} h_\nu(\lambda) &= e^{-rT}\liminf_{\lambda\downarrow 0} \mathbb{E}_{\mathbb{Q}}\big[ X_T^{\star}\big(\lambda,q^{\star}(\nu),q^{\star\star}(\nu)\big)\big]\\
    &\geq e^{-rT}\mathbb{E}_{\mathbb{Q}}\big[\liminf_{\lambda\downarrow 0} X_T^{\star}\big(\lambda,q^{\star}(\nu),q^{\star\star}(\nu)\big)\big] = \infty.
\end{align*}
Furthermore, for some $\Tilde{\lambda}>0$, note that for all $\lambda>\tilde{\lambda}$ it holds that 
\[
    0 = \lim_{\kappa\rightarrow\infty}X_T^\star(\kappa,q^{\star}(\nu),q^{\star\star}(\nu)) \leq X_T^\star(\lambda,q^{\star}(\nu),q^{\star\star}(\nu))\leq X_T^\star\big(\tilde{\lambda},q^{\star}(\nu),q^{\star\star}(\nu)\big),\quad \mathbb{Q}\text{-a.s.}
\]
So, by Lebesgue's dominated convergence theorem we obtain
\[
    \lim_{\lambda\rightarrow \infty}h_{\nu}(\lambda) = e^{-rT}\lim_{\lambda\rightarrow \infty}\mathbb{E}_{\mathbb{Q}}\big[ X_T^{\star}\big(\lambda,q^{\star}(\nu),q^{\star\star}(\nu)\big)\big] = e^{-rT}\mathbb{E}_{\mathbb{Q}}\big[\lim_{\lambda\rightarrow \infty} X_T^{\star}\big(\lambda,q^{\star}(\nu),q^{\star\star}(\nu)\big)\big] = 0.
\]
Then, let $\tilde{\lambda}\in(0,\infty)$. Again by dominated convergence, we obtain
\[
    \lim_{\lambda\uparrow \tilde{\lambda}} h_\nu(\lambda) = \lim_{\lambda\downarrow \tilde{\lambda}} h_\nu(\lambda)  = h_\nu\big(\tilde{\lambda}\big),
\]
which shows that $h_\nu$ is continuous. Finally, let $\lambda_2>0$ and $\lambda_1\in[0,\lambda_2)$. Then, $X_T^\star(\lambda_1,q^{\star}(\nu),q^{\star\star}(\nu))\geq X_T^\star\big(\lambda_2,q^{\star}(\nu),q^{\star\star}(\nu)\big)$ $\mathbb{Q}$-a.s., with strict inequality for all $\omega\in\left\{\xi_T<\frac{U^{\prime}(L_2)}{\lambda_2}\right\}$ (which is a $\mathbb{Q}$-non-null set). Hence, 
\[
    h_{\nu}(\lambda_1) = e^{-rT}\mathbb{E}_{\mathbb{Q}}\big[ X_T^{\star}\big(\lambda_1,q^{\star}(\nu),q^{\star\star}(\nu)\big)\big]>e^{-rT}\mathbb{E}_{\mathbb{Q}}\big[ X_T^{\star}\big(\lambda_2,q^{\star}(\nu),q^{\star\star}(\nu)\big)\big] = h_{\nu}(\lambda_2). 
\]
Thus, $h_\nu$ is strictly decreasing. All of the derived properties of $h_\nu$ together imply then that there exists a unique $\lambda(\nu)\in(0,\infty)$ such that $h_\nu(\lambda(\nu)) = x_0$. 

Before we proceed to prove the existence of $\nu>0$, we state properties of the map $\lambda(\nu)$. By the monotonicity of $q^{\star}$ and $q^{\star\star}$ in dependence of $\nu$, it is easy to see that $\nu\mapsto h_\nu$ is strictly increasing with respect to the pointwise order, i.e.,~$\nu_1<\nu_2$ implies $h_{\nu_1}<h_{\nu_2}$, which in turn gives us that $\lambda(\nu_1)< \lambda(\nu_2)$. 

Moreover, note that $h:(\nu,\lambda)\mapsto h_\nu(\lambda)$ is continuously differentiable in $(\nu,\lambda(\nu))$ for an arbitrary $\nu>0$. Hence,
\begin{align*}
    \frac{\partial h_\nu}{\partial \lambda}(\lambda(\nu)) &= e^{-rT} \frac{\partial}{\partial\lambda}\bigg|_{\lambda=\lambda(\nu)} \mathbb{E}_{\mathbb{Q}}\Big[X_T^{\star}(\lambda,q^{\star}(\nu),q^{\star\star}(\nu))\Big]\\
    &=e^{-rT}  \mathbb{E}_{\mathbb{Q}}\Big[\frac{\partial}{\partial\lambda}\bigg|_{\lambda=\lambda(\nu)}X_T^{\star}(\lambda,q^{\star}(\nu),q^{\star\star}(\nu))\Big]\\
    &=\mathbb{E}\Big[I^{\prime}\big(\lambda(\nu)\xi_T\big)\xi_T^2\mathbf{1}_A\big(\lambda(\nu)\xi_T\big)\Big] <0,
\end{align*}
where
\[
    A = \left\{\xi_T\leq \frac{U^{\prime}(L_2)}{\lambda(\nu)}\right\}\bigcup \left\{\frac{q^{\star}(\nu)}{\lambda(\nu)}<\xi_T \leq \frac{U^{\prime}(L_1)}{\lambda(\nu)}\right\}\bigcup \left\{\xi_T >  \frac{\max\{q^{\star}(\nu),q^{\star\star}(\nu)\}}{\lambda(\nu)}\right\},
\]
and note that $I^{\prime}(x)<0$ for all $x>0$. Hence, applying the implicit function theorem for all $(\nu,\lambda(\nu))$, we obtain that $\lambda(\nu)$ is continuous (even differentiable).

Then, note that $\lambda(\nu)\downarrow \lambda^{M}$ when $\nu\downarrow 0$, i.e.,~$\lambda(\nu)$ converges to the Lagrange multiplier of the classical Merton payoff. Then, $X_T^{\star}\big(\lambda(\nu),q^{\star}(\nu),q^{\star\star}(\nu)\big)\downarrow X_T^{\star}(\lambda^M,U^{\prime}(L_2),U^{\prime}(L_1)) = I(\lambda^{M}\xi_T)$ a.s.~as $\nu\downarrow 0$ and for the map
\begin{align}\label{eq:functionToDetermineSecondLagrange}
    \tilde{h}:\nu\mapsto \alpha_1 \P\Big( X_T^{\star}\big(\lambda(\nu),q^{\star}(\nu),q^{\star\star}(\nu)\big)\geq L_1\Big) +\alpha_2 \P\Big( X_T^{\star}\big(\lambda(\nu),q^{\star}(\nu),q^{\star\star}(\nu)\big)\geq L_2\Big),
\end{align}
by dominated convergence, we get
\[
    \lim_{\nu\downarrow 0} \tilde{h}(\nu) = \alpha_1 \P\Big(I\big(\lambda^{M}\xi_T\big)\geq L_1\Big) + \alpha_2 \P\Big(I\big(\lambda^{M}\xi_T\big)\geq L_2\Big) = a^{M}<a.
\]
For the limit as $\nu\rightarrow\infty$, the following holds:
\begin{itemize}
    \item \underline{$e^{-rT}L_2<x_0$:} By way of contradiction, assume $\lim\limits_{\nu\rightarrow\infty}\lambda(\nu)=\infty$, which implies 
    \[
        \lim_{\nu\rightarrow\infty}X_T^{\star}\big(\lambda(\nu),q^{\star}(\nu),q^{\star\star}(\nu)\big)\leq L_2\quad \text{a.s.}
    \]
    Hence, by applying dominated convergence with payoff bound 
    \[
        Y = I(\lambda\xi_T)\mathbf{1}_{\left(0,\frac{U^{\prime}(L_2)}{\tilde{\lambda}}\right]}(\xi_T)+ L_2 \mathbf{1}_{\left(\frac{U^{\prime}(L_2)}{\tilde{\lambda}},\infty\right)}(\xi_T),
    \]
    for some $\lambda>0$, it holds that
    \begin{align*}
        \lim_{\nu\rightarrow\infty}\mathbb{E}_{\mathbb{Q}}\Big[e^{-rT}X_T^{\star}\big(\lambda(\nu),q^{\star}(\nu),q^{\star\star}(\nu)\big)\Big] &= \mathbb{E}_{\mathbb{Q}}\Big[e^{-rT}\lim_{\nu\rightarrow\infty}X_T^{\star}\big(\lambda(\nu),q^{\star}(\nu),q^{\star\star}(\nu)\big)\Big]\\
        &\leq e^{-rT}L_2<x_0,
    \end{align*}
    contradicting the assumption that $X_T^{\star}\big(\lambda(\nu),q^{\star}(\nu),q^{\star\star}(\nu)\big)$ satisfies the budget constraint for all $\nu>0$. So, it has to hold that $\lim\limits_{\nu\rightarrow\infty}\lambda(\nu)\in(0,\infty)$ and consequently
    \[
        \lim_{\nu\rightarrow\infty}X_T^{\star}\big(\lambda(\nu),q^{\star}(\nu),q^{\star\star}(\nu)\big) = \begin{cases}
            I\Big(\xi_T\lim\limits_{\nu\rightarrow\infty}\lambda(\nu)\Big), & \xi_T\leq \frac{U^{\prime}(L_2)}{\lim\limits_{\nu\rightarrow\infty}\lambda(\nu)},\\
            L_2,&\xi_T> \frac{U^{\prime}(L_2)}{\lim\limits_{\nu\rightarrow\infty}\lambda(\nu)}.        \end{cases}
    \]
    By Proposition~\ref{prop:riskMaximizer}, we see that this limit has to be equal to the payoff in~\eqref{prop:riskMaximizer} and hence, $
    \lim\limits_{\nu\rightarrow \infty} \tilde{h}(\nu) = \alpha_1+\alpha_2>a$. 
    \item \underline{$e^{-rT}L_2=x_0$:} Analogously to the previous case, we can assume that either $\lim\limits_{\nu\rightarrow\infty}\lambda(\nu)\in(0,\infty)$ or $\lim\limits_{\nu\rightarrow\infty}\lambda(\nu)=\infty$ with $\lim\limits_{\nu\rightarrow\infty}\frac{q^{\star}(\nu)}{\lambda(\nu)} < \infty$ and show that these cases contradict the budget constraint. For brevity, we skip the detailed argumentation. Then, it has to hold that $\lim\limits_{\nu\rightarrow\infty}\lambda(\nu) = \infty$ and $\lim\limits_{\nu\rightarrow\infty}\frac{q^{\star}(\nu)}{\lambda(\nu)} = \infty$, which implies that
    \[
        \lim_{\nu\rightarrow\infty}X_T^{\star}\big(\lambda(\nu),q^{\star}(\nu),q^{\star\star}(\nu)\big) = L_2\quad \text{a.s.}
    \]
    This implies that $
    \lim\limits_{\nu\rightarrow \infty} \tilde{h}(\nu) = \alpha_1+\alpha_2>a$.
    \item \underline{$e^{-rT}L_2>x_0$:}  Similar to before, one can show that this case is equivalent to $\lim\limits_{\nu\rightarrow\infty}\lambda(\nu)=\infty$ and $\lim\limits_{\nu\rightarrow\infty}\frac{q^{\star}(\nu)}{\lambda(\nu)} < \infty$, otherwise a contradiction to the budget constraint occurs.

    For brevity and to emphasize the parameter $\nu$, let us  denote by $f_1(.;\nu)$ and $f_2(.;\nu)$ the functions $f_1$ and $f_2$. Then, $x^{\star}(\nu)<q^{\star\star}(\nu)$ holds if and only if the intersection point $y(\nu) = \frac{U(L_2)-U(L_1)+\nu\alpha_2}{L_2-L_1}$ of the curves $f_2(.;\nu)$ and $f_1(.;\nu)$ satisfies $f_2(y(\nu))>0$. Now, note that 
    \begin{align*}
        \lim_{\nu\rightarrow \infty} f_2(y(\nu)) = \lim_{\nu\rightarrow \infty} y(\nu) \Big(-L_2+\frac{\nu}{y(\nu)}\Big)>0
    \end{align*}
    holds iff $\alpha_1>\frac{L_1}{L_2}$. So, we perform a case-by-case study, in which we differ the two cases of $\alpha_1> \frac{L_1}{L_2}$ and $\alpha_1\leq \frac{L_1}{L_2}$. We start with the case of $\alpha_1> \frac{L_1}{L_2}$ and assume always that $\nu$ is large enough such that $x^{\star}(\nu)<q^{\star\star}(\nu)$ holds. Then, recall that $q^{\star\star}(\nu)$ satisfies the following equation:
    \begin{align}\label{eq:auxiliaryExistenceLagrange}
        1 = \frac{\alpha_1}{L_1}\cdot \frac{\nu}{q^{\star\star}(\nu)} + \underbrace{\frac{U(L_1)}{L_1 q^{\star\star}(\nu)}}_{\xrightarrow{\nu\rightarrow \infty}0} - \frac{U(I(q^{\star\star}(v)))-q^{\star\star}(\nu)I(q^{\star\star}(\nu))}{L_1q^{\star\star}(\nu)}. 
    \end{align}
    Then, by applying Lemma~\ref{lem:Inada_2}, we obtain that
    \[
        \lim_{\nu\rightarrow\infty}\frac{U(I(q^{\star\star}(\nu)))-q^{\star\star}(\nu)I(q^{\star\star}(\nu))}{q^{\star\star}(\nu)} = \lim_{x\rightarrow\infty}\frac{U(I(x))-xI(x)}{x}  = 0.
    \]
    Together with~\eqref{eq:auxiliaryExistenceLagrange} we obtain the following asymptotic behavior of $q^{\star\star}(\nu)$:
    \[
        q^{\star\star}(\nu)\sim \frac{\alpha_1}{L_1} \nu,\quad \nu\rightarrow \infty.
    \]
    Hence, 
    \begin{align}\label{eq:asymtotic_q_doubleStar}
        (q^{\star\star})^{-1}(y)\sim \frac{L_1}{\alpha_1} y,\quad y\rightarrow \infty.
    \end{align}
    Based on this result, we can calculate the following limit
    \begin{align}\label{eq:limits_q_stars}
        \begin{split}
            \lim\limits_{\nu\rightarrow\infty}\frac{q^{\star\star}(\nu)}{q^{\star}(\nu)} = \frac{L_2-L_1}{L_1}\lim\limits_{\nu\rightarrow\infty}\Bigg(\frac{\alpha_1}{\alpha_2+\frac{U(L_2)-U(L_1)}{\nu}} &+ \frac{U(L_1)}{\nu\alpha_2+U(L_2)-U(L_1)}\\
            &-\frac{U(I(q^{\star\star}(\nu)))-q^{\star\star}(\nu) I(q^{\star\star}(\nu))}{\nu\alpha_2+U(L_2)-U(L_1)}\Bigg).
        \end{split}
    \end{align}
    
    Using the asymptotic from~\eqref{eq:asymtotic_q_doubleStar} we obtain in the case of 
    \begin{align*}
        \lim_{\nu\rightarrow\infty}\frac{U(I(q^{\star\star}(\nu)))-q^{\star\star}(\nu) I(q^{\star\star}(\nu))}{\nu\alpha_2+U(L_2)-U(L_1)} &= \lim_{y\downarrow 0}\frac{U(y)-U^{\prime}(y)y}{\alpha_2 (q^{\star\star})^{-1}(U^{\prime}(y))+U(L_2)-U(L_1)}\\
        &= \lim_{y\downarrow 0}\frac{U(y)-U^{\prime}(y)y}{\frac{L_1\alpha_2}{\alpha_1}U^{\prime}(y)}\\
        &= \frac{\alpha_1}{L_1\alpha_2}\lim_{y\downarrow 0}\left(\frac{U(y)}{U^{\prime}(y)}-y\right)= 0. 
    \end{align*}
    Together with~\eqref{eq:limits_q_stars} we have
    \[
        \lim\limits_{\nu\rightarrow\infty}\frac{q^{\star\star}(\nu)}{q^{\star}(\nu)} = \frac{L_2-L_1}{L_1}\cdot\frac{\alpha_1}{\alpha_2}. 
    \]
    
    For brevity, we set $c = \lim\limits_{\nu\rightarrow\infty}\frac{q^{\star}(\nu)}{\lambda(\nu)}$. Since $\alpha_1> \frac{L_1}{L_2}$, we have $\frac{L_2-L_1}{L_1}\cdot\frac{\alpha_1}{\alpha_2}> 1$ and the limiting payoff has the following form:
    \[
        \lim_{\nu\rightarrow\infty}X_T^{\star}\big(\lambda(\nu),q^{\star}(\nu),q^{\star\star}(\nu)\big) = \begin{cases}
            L_2,&\xi_T\leq c,\\
            L_1, & c<\xi_T\leq \frac{L_2-L_1}{L_1}\cdot\frac{\alpha_1}{\alpha_2}\cdot c,\\
            0, & \xi_T>c.\end{cases}
    \]
    A comparison with the budget constraint shows that the limiting payoff is the one in Proposition~\ref{prop:optimalPayoffBoundaryCasePI} (ii), i.e.,~$c = \frac{\alpha_2}{\lambda^{\star}(L_2-L_1)}$, where $\lambda^{\star}$ is calculated as in Proposition~\ref{prop:optimalPayoffBoundaryCasePI} (ii). Hence, 
    \[
        \lim_{\nu\rightarrow\infty}\tilde{h}(\nu) = \overline{a}>a.
    \]

    In the remaining case, i.e.,~$\alpha_1\leq \frac{L_1}{L_2}$, it holds that $q^{\star}(\nu) = x^{\star}(\nu)\geq q^{\star\star}(\nu)$ for $\nu$ large enough. Hence, we directly see that the limiting payoff has the following form:
    \begin{align}\label{eq:payoffProofLimit_nu}
        \lim_{\nu\rightarrow\infty}X_T^{\star}\big(\lambda(\nu),q^{\star}(\nu),q^{\star\star}(\nu)\big) = \begin{cases}
            L_2,&\xi_T\leq c,\\
            0, & \xi_T>c.\end{cases}
    \end{align}
    Then, by Proposition~\ref{prop:optimalPayoffBoundaryCasePI} (i) we know that there exists a unique value $c$ such that the limiting payoff satisfies the budget constraint and this value is $c = \frac{\alpha_1+\alpha_2}{\lambda^{\star}L_2} = \frac{1}{\lambda^{\star}L_2}$, where $\lambda^{\star}$ is chosen as in Proposition~\ref{prop:optimalPayoffBoundaryCasePI} (i). Hence, the limiting payoff is exactly the one from Proposition~\ref{prop:optimalPayoffBoundaryCasePI} (i) and therefore,
    \[
        \lim_{\nu\rightarrow\infty}\tilde{h}(\nu) = \overline{a}>a. 
    \]
\end{itemize}

Continuity of $\tilde{h}$ can be shown with the help of dominated convergence, analogously to earlier considerations in the proof. This ensures the existence of $\nu$ such that $\tilde{h}(\nu) = a$.
\end{proof}

\section{Numerical procedure} \label{sec:numerical}

This numerical scheme is used to obtain the results in Section~\ref{sec:numericalStudy}. In addition, it serves as an overview for the reader and is intended to clarify the interplay between the various theoretical results in Section~\ref{sec:solution}. For simplicity, let us recall our assumptions, which are important to choose permitted model parameters. It has to hold that $a\in(0,1)$, $\alpha_1+\alpha_2 = 1$ with $\min\{\alpha_1,\alpha_2\}>0$ and $0< L_1<L_2$. Also recall that we need to distinguish three cases, namely $a\leq a^{M}$, $a^{M}<a<\bar{a}$ and $a\geq \bar{a}$. To be able to check which of the conditions apply, we have to perform the following steps:
\begin{enumerate}[align = left, leftmargin = 1.25cm]
    \item[(Step 1a)] Calculate $\lambda^{M}>0$ as the root of the function $\lambda\mapsto \mathbb{E}[\xi_T I(\lambda\xi_T)] - x_0$, for instance via a bisection root-search algorithm. The expectation can then be calculated numerically. 
    \item[(Step 1b)] Calculate $a^{M}$ by~\eqref{eq:riskConstraintMerton}, i.e.,~set
    \[
        a^{M} = \alpha_1\P\left(\xi_T<\frac{U^{\prime}(L_1)}{\lambda^{M}}\right) + \alpha_2\P\left(\xi_T<\frac{U^{\prime}(L_2)}{\lambda^{M}}\right).
    \]
    \item[(Step 2a)] Calculate $\lambda^{\star}>0$ as the root of the following function 
    \[
        \lambda\mapsto \begin{cases}
            e^{-rT} L_2\mathbb{Q}\left(\lambda\xi_T\leq \frac{1}{L_2}\right) - x_0, & \alpha_1\leq \frac{L_1}{L_2},\\
            e^{-rT} L_2\mathbb{Q}\left(\lambda\xi_T\leq \frac{\alpha_2}{L_2-L_1}\right) + e^{-rT}L_1\mathbb{Q}\left(\frac{\alpha_2}{L_2-L_1}<\lambda\xi_T\leq \frac{\alpha_1}{L_1}\right) - x_0, & \text{otherwise},
        \end{cases}
    \]
    for instance as in (Step 1a) via a bisection root-search algorithm.
    \item[(Step 2b)] If $e^{-rT}L_2\leq x_0$, then $\bar{a} = 1$. Otherwise, following  Proposition~\ref{prop:optimalPayoffBoundaryCasePI}, set 
    \[
        \bar{a} = \begin{cases}
            \P\left(\lambda^{\star}\xi_T\leq \frac{1}{L_2}\right), & \alpha_1\leq \frac{L_1}{L_2},\\
            \alpha_1\P\left(\lambda^{\star}\xi_T\leq \frac{\alpha_1}{L_1}\right)+\alpha_2 \P\left(\lambda^{\star}\xi_T\leq \frac{\alpha_2}{L_2-L_1}\right), & \text{otherwise}.
        \end{cases}
    \]
\end{enumerate}

Now we are in the position to compare the three cases mentioned above. 
\begin{enumerate}[align = left, leftmargin = 1.25cm]
    \item[(Step 3)] If $a\leq a^{M}$, then perform the following steps:
    \item[(Step 3a)] Use $\lambda^{M}$ from (Step 1a) and set the optimal terminal payoff to $X_T^{\star} = \sum_{i=1}^n
    	\beta_i^{1/\gamma_i}(\lambda^M\xi_T)^{-1/\gamma_i}$. 
    \item[(Step 3b)] The optimal trading strategy is given by $\pi_t^\star
    	=
    	\frac{\mu-r}{\sigma^2}
    	\sum_{i=1}^n
    	\frac{1}{\gamma_i} \beta_i^{1/\gamma_i}(\lambda^M\xi_t)^{-1/\gamma_i}$.
    \item[(Step 4)] If $a\geq \bar{a}$, then perform the following steps:
    \item[(Step 4a)] If $a>\bar{a}$, by Corollary~\ref{cor:benchmark-cases} no solution exists.
    \item[(Step 4b)] If $a=\bar{a}$, then use $\lambda^{\star}$ from (Step 2a) and by Proposition~\ref{prop:optimalPayoffBoundaryCasePI} set the optimal payoff to
    \[
        X_T^{\star} = \begin{cases}
            L_2\mathbf{1}_{\left\{\lambda^{\star}\xi_T\leq \frac{1}{L_2}\right\}}, & \alpha_1\leq \frac{L_1}{L_2},\\
            L_2\mathbf{1}_{\left\{\lambda^{\star}\xi_T\leq \frac{\alpha_2}{L_2-L_1}\right\}} + L_1\mathbf{1}_{\left\{\frac{\alpha_2}{L_2-L_1}<\lambda^{\star}\xi_T\leq \frac{\alpha_1}{L_1}\right\}}, & \text{otherwise}.
            \end{cases}
    \]
    \item[(Step 4c)] If $a=\bar{a}$, then the optimal trading strategy is given by 
    \[
        \pi_t^\star
    	=
    	\begin{cases}
            \frac{e^{-r\tau}}{\sigma\sqrt{\tau}}L_2\varphi\big(d_{\infty}(t,\xi_t;b_2)\big), & \alpha_1\leq \frac{L_1}{L_2},\\
            \frac{e^{-r\tau}}{\sigma\sqrt{\tau}}\Big((L_2-L_1)\varphi\big(d_{\infty}(t,\xi_t;b_2)\big) +  L_1\varphi\big(d_{\infty}(t,\xi_t;b_1)\big)\Big), & \text{otherwise}.
        \end{cases}
    \]
\end{enumerate}

For the intermediate case $a^{M}<a<\bar{a}$, the proof of Theorem~\ref{thm:2bounds} is already constructive in the sense that it provides a procedure to calculate the Lagrange multipliers. This leads to the following procedure:
\begin{enumerate}[align = left, leftmargin = 1.25cm]
    \item[(Step 5)] If $a^{M}<a< \bar{a}$, then perform the following steps: 
    \item[(Step 5a)] Determine the parameters $\lambda$, $q^{\star}$, $q^{\star\star}$ and $\nu$ by performing a bisection root-search algorithm for function $\tilde{h}-a$ from~\eqref{eq:functionToDetermineSecondLagrange} in the proof of Theorem~\ref{thm:2bounds}, see also the explanations in Remark~\ref{rem:calculateLagrangeParameters}.
    \item[(Step 5b)] Set the optimal payoff to~\eqref{eq:optimalPayoffTwoBounds} from Theorem~\ref{thm:2bounds}.
    \item[(Step 5c)] Determine the optimal trading strategy via Equation~\eqref{eq:tradingTwoBounds}. 
\end{enumerate}

\section{Auxiliary figures and tables}\label{sec:auxiliaryFiguresTables}

\begin{figure}[h]
\centering
\hspace{1cm}
\begin{subfigure}{0.425\textwidth}
\centering
\includegraphics[width=\linewidth]{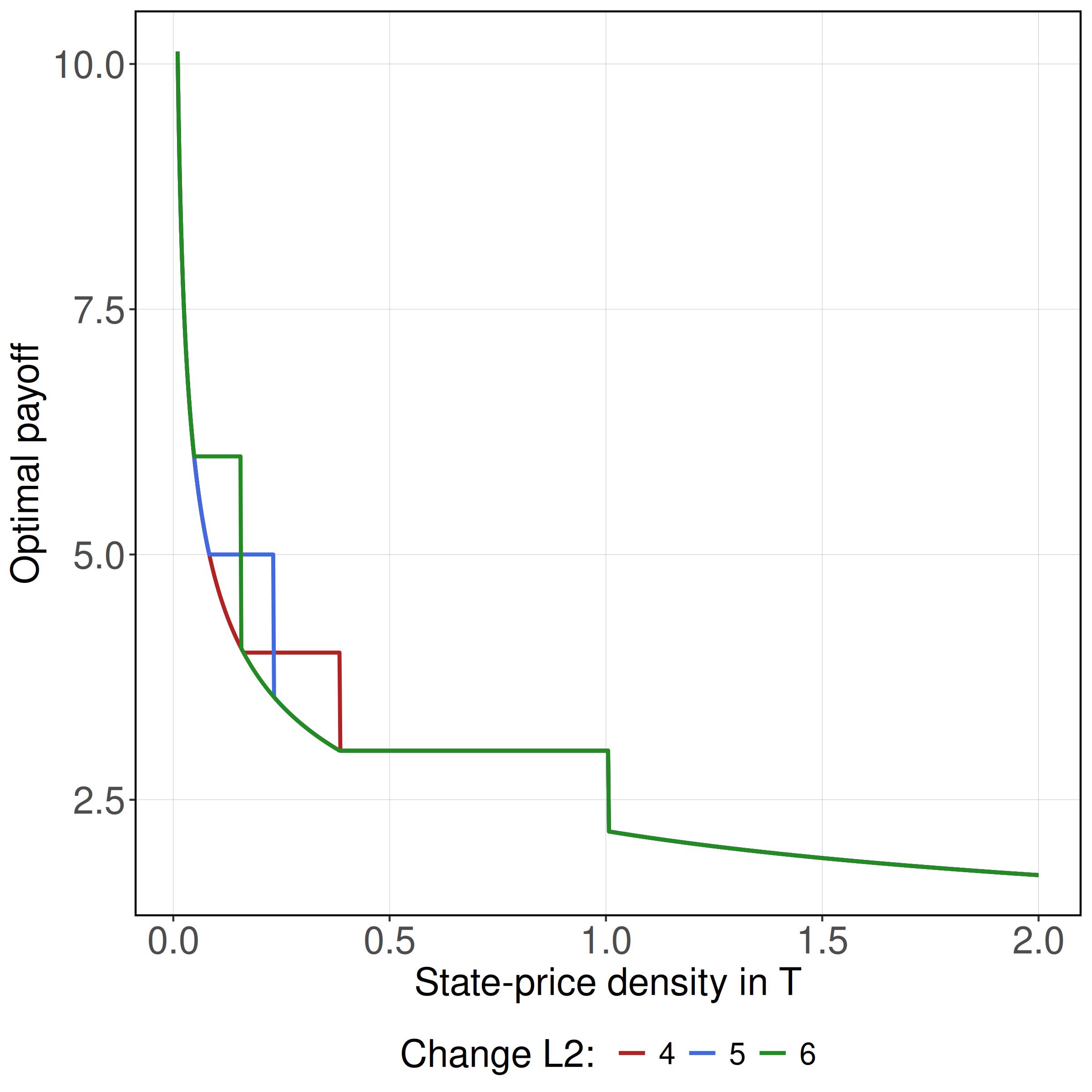}
\caption{Change $L_2$}
\end{subfigure}
\hspace{1cm}
\begin{subfigure}{0.425\textwidth}
\centering
\includegraphics[width=\linewidth]{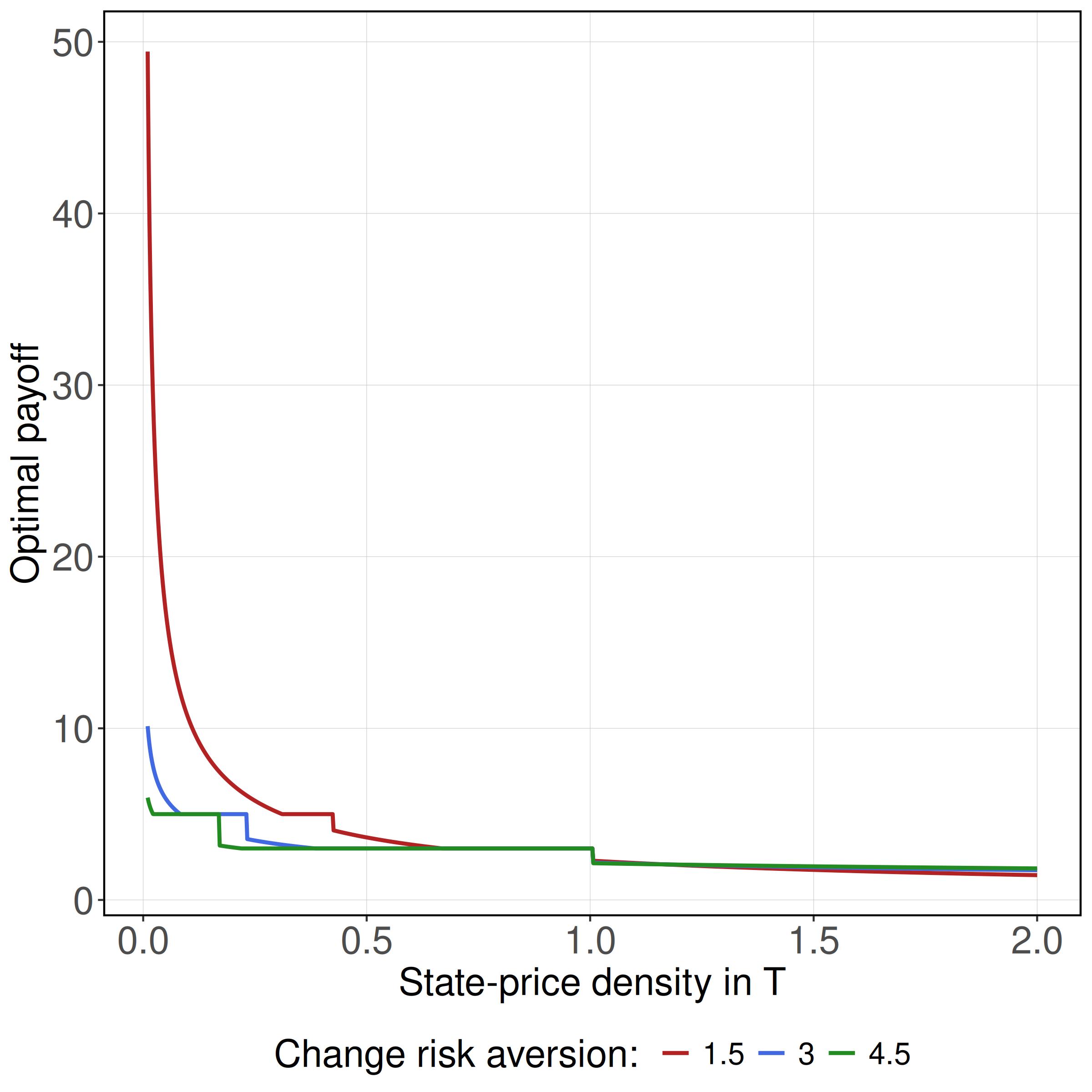}
\caption{Change $\gamma$}
\end{subfigure}
\caption{Optimal payoff $X_T^{\star}$ depending on the state-price density
$\xi_T$ by varying either the second threshold $L_2$ or the risk aversion parameter $\gamma$.}
\label{fig:optimal_payoff_auxiliary}
\end{figure}

To complement Figure~\ref{fig:optimal_payoff}, we present in Figure~\ref{fig:optimal_payoff_auxiliary} the optimal payoff structure in dependence of the state-price density for varying either $L_2$ or $x_0$. In contrast to variations in $L_1$, changing $L_2$ has only a limited effect, see panel~(b). This is caused by two facts. First, the upper threshold has the smaller weight
$\alpha_2=0.3$. Second, reaching $L_2$ is relatively costly given the
initial wealth $x_0=2.5$. 

Panel~(b) varies the risk aversion parameter $\gamma$. Lower risk aversion leads to much
larger payoffs in good states, as the investor is more willing to accept
dispersion in terminal wealth. However, these high-payoff states occur with
small probability. Higher risk aversion reduces the upside exposure and
increases the relative importance of the threshold plateaus. In other words, a more risk-averse investor reduces the shortfall in bad states which is compensated by smaller returns in good states. Concluding, risk
aversion affects the constrained solution through two channels: it changes the
Merton branch of the payoff, and it changes the opportunity cost of allocating
wealth to the regulatory plateaus.

However, as the values in Table~\ref{tab:certainty_equivalents_auxiliary} shows, risk aversion has a moderate effect on the welfare loss. The relative
loss increases from $0.73\%$ for $\gamma=1.5$ to $4.93\%$ for $\gamma=4.5$.
A more risk-averse investor dislikes the payoff distortions caused by the
average compliance constraint more strongly.

\begin{table}[ht]
\centering
\begin{tabular}{lrrr}
  \hline
\textbf{Parameter} & \multicolumn{1}{c}{\textbf{$CE^M$}} & \multicolumn{1}{c}{\textbf{$CE^{\star}$}} & \multicolumn{1}{c}{\textbf{$rc$}} \\ 
  \hline
  $L_2 = 4.0$ & 2.5423 & 2.4731 & 2.72\% \\ 
  $L_2 = 5.0$ & 2.5423 & 2.4731 & 2.72\% \\ 
  $L_2 = 6.0$ & 2.5423 & 2.4731 & 2.72\% \\ 
  \hline
  $\gamma = 1.5$ & 2.5590 & 2.5403 & 0.73\% \\ 
  $\gamma = 3.0$ & 2.5423 & 2.4731 & 2.72\% \\ 
  $\gamma = 4.5$ & 2.5355 & 2.4105 & 4.93\% \\ 
   \hline
\end{tabular}
\caption{Certainty equivalents (CE) and resulting welfare loss $rc$ for changes in the second threshold $L_2$ and the risk aversion parameter $\gamma$ in the single investor case.} 
\label{tab:certainty_equivalents_auxiliary}
\end{table}
       
\end{document}